\documentclass[10pt]{article}

\usepackage{newtxtext,newtxmath}
\usepackage{graphicx}
\usepackage[letterpaper,margin=1in]{geometry}
\renewenvironment{abstract}
	{\quotation}
	{\endquotation}
\date{}

\makeatletter
\renewcommand{\fnum@figure}{\textbf{Figure \thefigure}}
\renewcommand{\fnum@table}{\textbf{Table \thetable}}
\makeatother

\usepackage{scicite}
\usepackage{url}

\usepackage{amsmath}
\usepackage{siunitx}
\usepackage{booktabs}
\usepackage{upgreek}
\usepackage{float}
\usepackage{placeins}
\usepackage{threeparttable}
\usepackage{xcolor}
\usepackage{hyperref}
\usepackage{tabularx}
\usepackage{setspace}
\usepackage{caption}

\newcolumntype{Y}{>{\raggedright\arraybackslash}X}

\let\oldeqref\eqref
\renewcommand{\eqref}[1]{Eq.~\oldeqref{#1}}

\newcommand{\SeeSI}[0]{\emph{Supplementary Text}}
\newcommand{\SeeMM}[0]{\emph{Materials and Methods}}

\def\scititle{
	Following the thread: surface and bulk solvent migration in silicone elastomers from local volumetric swelling
}
\title{\bfseries \boldmath \scititle}

\author{
	Chenzhuo~Li$^{1}$,
	Tom~Beyeler$^{1}$,
	Marc~Antonio~Chalhoub$^{1}$,
	John~M.~Kolinski$^{1\ast}$\and
	\small$^{1}$Institute of Mechanical Engineering, School of Engineering, EPFL, 1015 Lausanne, Switzerland.\and
	\small$^\ast$Corresponding author. Email: john.kolinski@epfl.ch
}

\begin{document}

\maketitle

\begin{abstract} \bfseries \boldmath
% limit 250 words; 0217 --> 244 words
Poroelastic materials, consisting of a permeable solid matrix infiltrated with fluid, are ubiquitous in natural and engineering contexts. In poroelastic polymer solids, the elastic matrix swells to equilibrium when immersed in a solvent bath; thus, the network elasticity couples to the solvent transport. Despite the ubiquity and importance of poroelastic theory in describing phenomena as diverse as earthquakes and biological tissues, there is a paucity of experimental data that probe the local network response to controlled stress and solvent boundary conditions. Here, we first probe the baseline diffusion kinetics of a polymeric solvent during free swelling of a polydimethylsiloxane (PDMS) network with well-characterized silicone oils. \textit{In situ} 3D spatiotemporal measurements identify a flux-limited interfacial boundary condition, contradicting the canonical fully drained assumption. This correction eliminates an order-of-magnitude underestimation of diffusivity in standard bulk analysis. The swelling equilibrium is accurately captured by a Flory--Rehner theory that requires modification to include the effective finite extensibility of the filled network. Solvent migration is then studied using a bending configuration for three material preparations: as-prepared, mobile-phase-free, and fully swollen in silicone oils. The as-prepared and mobile-phase-free beams show no discernible volumetric change or force relaxation, whereas local \textit{in situ} measurements directly resolve tensile-side dilation and compressive-side contraction, yielding the effective diffusivities in agreement with the force-relaxation data. These measurements rigorously benchmark solvent diffusivity in polymer networks, underscoring the importance of unambiguous interfacial boundary conditions and shedding light on mechanics and engineering across poroelastic polymers and geomaterials.

% By leveraging \textit{in situ} 3D spatiotemporally resolved swelling fields, we show that solvent uptake is limited by interfacial flux, overturning the standard fully drained assumption. This correction eliminates an order-of-magnitude underestimation of diffusivity in bulk analysis.
\end{abstract}

\begin{figure}[!htbp]
\centering
\includegraphics[width=0.9\textwidth]{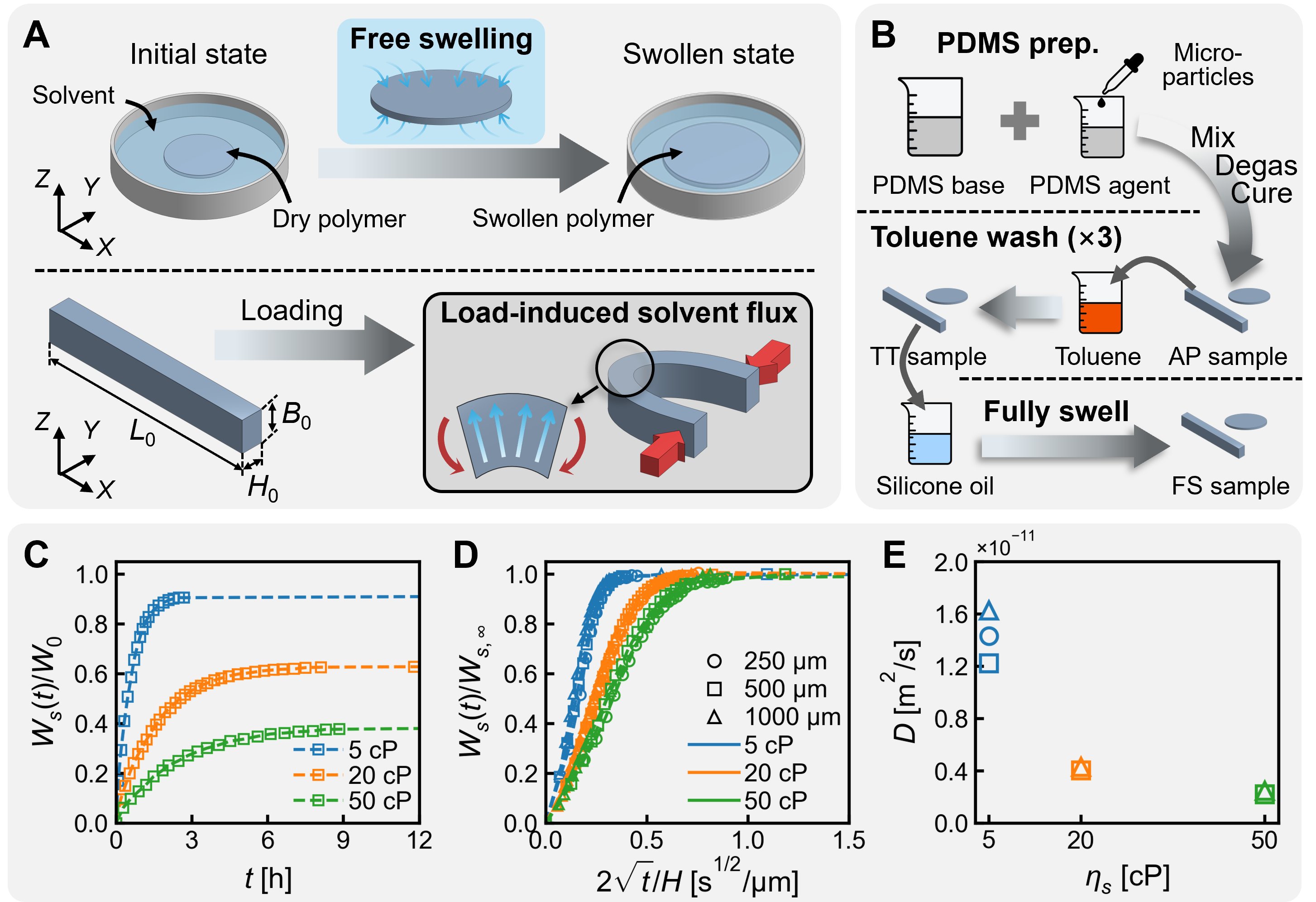}
\caption{(A) Two different boundary conditions are used to drive solvent transport in polymer networks. \textit{Concentration-driven solvent flux} (e.g., free swelling, upper panel): an initially solvent-poor polymer (typically dry polymer) is immersed in a large solvent bath without any constraint. The bath-polymer contrast in concentration establishes a chemical potential gradient and drives solvent uptake until thermodynamic equilibrium is reached. \textit{Load-induced solvent flux} (e.g., bent beam, lower panel): in a pre-swollen sample with initially uniform solvent concentration, solvent flux will occur when an external load is applied. External load generates spatial gradients in the chemical potential, driving internal solvent flow; in the bent-beam example, under applied force and local moment (red arrows), solvent flows from the compressed to the tensile side (blue arrows).
(B) Sample preparation protocol. PDMS base and curing agent (10:1~w/w) are first mixed, degassed, and thermally cured per documentation for as-prepared (AP) samples. The residual uncrosslinked sol fraction is removed by rinsing AP samples in toluene three times and then dried thoroughly, yielding mobile-phase-free toluene-treated (TT) samples. TT samples are immersed in silicone oil of the desired viscosity $\eta_s$, resulting in fully swollen (FS) samples once free-swelling equilibrium is reached. See \SeeMM\ for further details.
(C) Solvent mass uptake normalized by the initial weight during the free swelling of TT samples in silicone oils of viscosity $\eta_s=\numlist{5;20;50}\,\unit{\centi P}$ with sample thickness $H_0=\SI{500}{\micro m}$. Different uptake rates and equilibrium ratios are measured across the viscosities.
(D) The solvent mass is normalized by its equilibrium value, and time is scaled with the sample initial half-thickness ($H_0=\numlist{250;500;1000}\,\unit{\micro m}$; donated by the open symbols). The scaling collapses the data across these geometries for each $\eta_s$, consistent with Fickian diffusion.
(E) Effective diffusion coefficients $D$ are estimated from linear fits to the early-time uptake and graphed as a function of $\eta_s$. $D$ decreases with increasing $\eta_s$.}
\label{fig:bulk_swelling}
\end{figure}

\noindent Porous materials are ubiquitous in natural and engineering contexts, ranging from geomaterials and biological tissues to engineered applications involving artificial hydrogels and elastomers~\cite{wang2000theory, mow1980cartilage, cowin1999bone, liu2025intercellular, porous_gel_review, hu2011indentPDMS}. The mechanical response of poroelastic solids emerges from the coupling of the elastic deformation of the permeable solid skeleton and the pore fluid solvent~\cite{biot1941, cheng2016poroelasticity}. Polymeric solids exemplify this biphasic architecture, generally comprising a solid polymer network and an interstitial solvent~\cite{doi2009, rubinstein2003}; this unique composite structure underpins the versatility of polymers in a vast array of modern technologies, such as microfluidics~\cite{beebe2000, microfluidic2004, microfluidic2023, microfluidic2025pdms}, drug delivery~\cite{delivery1992, delivery2010, delivery2010, biomedical2021microneedle}, and biomedical devices~\cite{biomedical2021, biomedical2023, biomedical2023expander}. At free swelling equilibrium~\cite{flory1943, flory1953}, the resulting polymer volume fraction not only dictates the polymer's initial mechanical properties~\cite{G_phi1994, treloar1975physics, tang2017fatigue, yamamoto2022scaling, swellingPAAm}, but also strongly influences its effective diffusivity that governs diffusion within the polymer network~\cite{scherer1992bending, hu2010indentation, yoon2010, bouklas2012}. Furthermore, in many practical applications, the swollen polymers are subjected to external loads~\cite{wei2019plasticiser, seals2018, OA2025, wound2021} or environmental stresses~\cite{Spores2023, Spores2024, Monica2021, polotsky2013}. Such stimuli create gradients in the local chemical potential, driving solvent migration until a new thermodynamic equilibrium is reached, as exemplified in Fig.~\ref{fig:bulk_swelling}A. This solvent transport is not confined to swollen polymers; it can also emerge in as-prepared polymers with a native mobile phase from  uncrosslinked sol fraction~\cite{PNAS_Vikram, Holmes2015PVS, wang2021programmable, marcilla2004plasticizers}.

In poroelastic solids, the solvent phase intrinsically diffuses; it is thus modeled using similar physics to those used to describe the consolidation of saturated soils~\cite{biot1941}. Recently, this theory has been adapted to analyze solvent migration in polymer networks~\cite{tanaka1979, durning1993, scherer1989, hong2008theory, Monica2025} and applied to free swelling~\cite{yoon2010, bouklas2012} as well as to more complex configurations including cracks~\cite{bouklas2015crack, RHuang_crack2018, RHuang_crack2020, yang2022, plummer2024}. 
In this framework, effective diffusivity is the key governing parameter, and has been measured with numerous techniques~\cite{vesely2008review}, ranging from simple mass-/volume-uptake protocols~\cite{microchannel, sotiri2018tunability, yoon2010, tanaka1979, uptake2001, uptake2015, uptake2023} to relaxation-based methods~\cite{hu2010indentation, hu2011indentPDMS, hui2006contact, hu2012indentation, hu2012thinlayer, scherer1992bending, shape_recovery}. The effective diffusion constant $D$ depends on the solvent, the network, and the coupling between the two; $D$ can span a wide range from very small $\approx10^{-11}\,\unit{m^2/s}$ (silicone V100 in polydimethylsiloxane (PDMS)~\cite{microchannel}) to very large $\approx10^{-9}\,\unit{m^2/s}$ (water in hydrogel~\cite{hu2010indentation} or Decane in PDMS~\cite{hu2011indentPDMS}). However, these measurements typically ignore deformation or stress in the bulk and lack the resolution to determine interfacial boundary conditions, whereas diffusion is inherently local and the estimate of $D$ is sensitive to boundary transport kinetics; indeed, stresses can generate nonuniform chemical-potential gradients that drive heterogeneous solvent fluxes~\cite{IJF2024}. Recent \textit{in-situ}, spatially resolved measurements of rubbery networks have uncovered a remarkably high $D=1.8\times10^{-6}\,\unit{m^2/s}$ and counter-intuitive volumetric responses~\cite{PNAS_Vikram, hartquist2024local}. Although the measured $D$ largely exceeds all classical measurements in polymers, it may be physically plausible: shear thinning of the solvent can lower the solvent viscosity~\cite{shearthinning} and strong stress fields can align and reorient solvent molecules so that their migration becomes anisotropic~\cite{viovy2000electrophoresis, Monica1900Reorientation}. To rigorously probe these dynamics and establish a robust bound on $D$, independent spatially-resolved \textit{in-situ} measurements with a controlled solvent are therefore essential.

%  barrier-lowering via delocalized, percolating pathways (akin to ‘colloidal metallicity’~\cite{Monica2019colloidalmetallicity}) can further elevate effective mobility~\cite{Monica2021D}

We study solvent-diffusion dynamics during both free swelling and load-induced migration in a PDMS-based system comprising three distinct sample states: as-prepared (AP), toluene-treated (TT), and fully-swollen (FS). The fabrication workflow is sketched in Fig.~\ref{fig:bulk_swelling}B, with details in \SeeMM. AP samples contain a sol fraction of $\approx4\,\text{wt}\%$ uncrosslinked PDMS chains (Fig.~\ref{figS:toluene}, \SeeSI); toluene washing removes this mobile phase, yielding TT samples that serve as a pure network control. To prescribe the viscosity of the pore fluid, $\eta_s$, TT samples are immersed in silicone oil solvents with $\eta_s=\numlist{5;20;50}\,\unit{\centi P}$, producing FS states. We quantify $D$ during free swelling by global and local modalities: gravimetric mass uptake and volumetric strain measurement via 3D particle tracking of embedded micro-particles; load-induced migration in a bending configuration is probed using the same particle tracking technique and validated with force-relaxation kinetics.

\subsection*{Free swelling of PDMS network in viscosity-controlled silicone oils}

We first investigate the free swelling dynamics by measuring the global mass uptake of thin TT samples (thickness $H_0 = \numlist{250;500;1000}\,\unit{\um}$) during free swelling in silicone oils with prescribed viscosities ($\eta_s = \numlist{5;20;50}\,\unit{\centi P}$). The experimental details are given in \SeeMM. As shown for $\num{500}\,\unit{\um}$-thick samples in Fig.~\ref{fig:bulk_swelling}C, the uptake of solvent mass normalized by the initial sample mass, $W_s(t)/W_0$, reveals that both the swelling rate and the equilibrium $W_s(\infty)/W_0$ vary across the solvent viscosities. Normalizing the transient mass uptake $W_s(t)$ by its equilibrium value $W_s(\infty)$ and plotting it as a function of normalized time, $2\sqrt{t}/H$, we collapse the swelling curves for different sample thicknesses onto a master curve for each $\eta_s$ (Fig.~\ref{fig:bulk_swelling}D). This collapse confirms that the swelling is governed by a Fickian diffusion. $D$ is extracted by measuring the slope of the normalized mass-uptake for each silicone oil (Sec.~\ref{secS:FS_solution}, \SeeSI), and is plotted in Fig.~\ref{fig:bulk_swelling}E. The fitted values for $D$ fall in a narrow range of $10^{-12}\text{--}10^{-11}\,\unit{m\squared\per\second}$; furthermore, as expected, $D$ is independent of sample geometry, and decreases with $\eta_s$.

\begin{figure}[!htbp]
\centering
\includegraphics[width=0.45\textwidth]{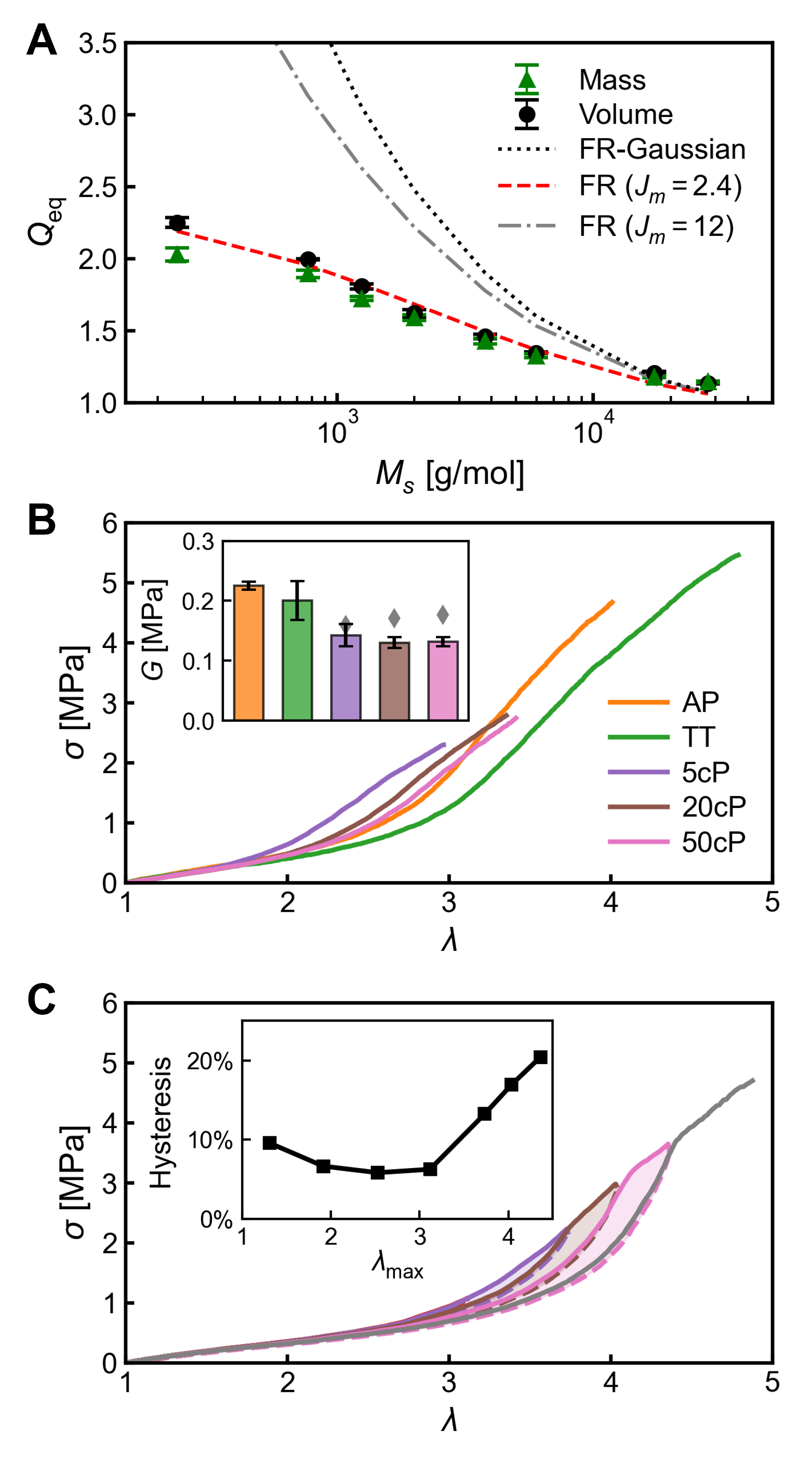}
\caption{(A) The equilibrium swelling ratio $Q_{\mathrm{eq}}$ versus silicone oil molar mass $M_s$. The classic Flory--Rehner prediction overestimates swelling at small $M_s$ (dotted line). 
Including finite extensibility via the Gent model and fitting the locking parameter $J_m$ captures the equilibrium swelling ratio (red dashed line). However, setting $J_m=12$ as obtained from uniaxial tensile tests (Fig.~\ref{figS:tensile}B, \SeeSI) still leaves a discrepancy, which we attribute to the different stress states---triaxial in swelling vs. uniaxial in tension.
(B) Uniaxial stress--stretch curves for PDMS samples. Inset: small-strain shear moduli (mean$\pm$s.d., $n=3$) with a classic affine prediction $G=G_{\text{dry}}/Q_{\text{eq}}^{1/3}$ (gray diamonds) for FS samples. AP and TT samples show similar shear moduli, while for FS samples, the measured moduli are consistently smaller than the theoretical prediction from the affine theory.
(C) Cyclic stress--stretch curves with incremental maximum loading, illustrating the Mullins effect~\cite{Mullins} in a TT sample. Inset: the hysteresis is defined as the fractional energy lost for each loading--unloading cycle. Even at small maximum stretch, a non-negligible loss is observed.}
\label{fig:uniaxial}
\end{figure}

While solvent viscosity governs the diffusion rate, it cannot account for the observed equilibrium swelling ratios. For silicone oils, $\eta_s$ varies with the molar mass $M_s$; as $M_s$ increases, the equilibrium swelling ratio $Q_\text{eq}$ decreases as predicted by Flory-Rehner (FR) theory~\cite{flory1943, flory1953} and explained in Sec.~\ref{secS:FR_equil}, \SeeSI. To test this dependence directly, we measure $Q_\text{eq}$ for an expanded range of silicone oil molar mass ($\eta_s=1\text{ to }\num{1000}\,\unit{\centi P}$; $\eta_s$--$M_s$ mapping in Fig.~\ref{figS:tensile}A, \SeeSI) and compare the measured values with the FR prediction in Fig.~\ref{fig:uniaxial}A. The classic FR model significantly over-predicts the observed $Q_\text{eq}$, particularly for low-$M_s$ silicone oils. We considered whether this discrepancy could be explained by the non-swellable silica nanoparticles present in Sylgard~184~\cite{clough2016covalent, suriboot2021amphiphilic}. However, explicitly correcting $Q_\text{eq}$ for this excluded volume only marginally shifts the inferred swelling of the PDMS matrix and does not fundamentally resolve the discrepancy (Fig.~\ref{figS:filler_correction}, \SeeSI). This indicates that fillers do not merely act as inert volume; rather, filler--polymer interactions and associated constraints further suppress equilibrium swelling. Notably, for a low $M_s$, the network stretch dramatically increases to accommodate the extra solvent volume; however, the deformation cannot increase indefinitely, and at large stretch, filler-induced constraints and the finite extensibility of polymer chains become salient~\cite{Zhao2023Lock, Suo2014gent, okumura2018gentAB, zhou2020hydrolysis}. To model this effect, we modify the FR framework using the phenomenological Gent model~\cite{gent1996} to incorporate the strain stiffening of the silica-filled network. This modified FR-Gent model successfully captures the experimental swelling behavior across the entire range of the tested $M_s$, and provides an effective chain-locking parameter $J_m=2.4$.

To further probe the mechanical response of the materials, uniaxial tensile tests are performed on AP, TT, and FS ($\eta_s=\numlist{5;20;50}\,\unit{\centi P}$) dog-bone samples, with loading curves shown in Fig.~\ref{fig:uniaxial}B. Experimental details can be found in \SeeMM. Each of the FS samples fails at a lower stretch than either the AP or TT samples; the considerable reduction in strength and stretch at failure is consistent with a significant pre-stretch in the swollen network prior to tensile loading. Notably, the stretch at failure of the FS samples is ordered inversely with $\eta_s$, and consequently $M_s$; thus, the FS sample with the greatest $Q_\text{eq}$ had the lowest stretch at failure. The small-strain shear moduli are extracted from the initial slopes (inset of Fig.~\ref{fig:uniaxial}B) and compared with the classical affine prediction, $G = G_\text{dry}\phi_\text{eq}^{1/3}$~\cite{G_phi1994, treloar1975physics, tang2017fatigue}, where $G_\text{dry}=G_\text{TT}$ and the solid fraction $\phi_\text{eq}=1/Q_\text{eq}$. The affine theory overestimates the modulus for all FS conditions, indicating departures from affine Gaussian elasticity, possibly due to entanglement defects and non-affine kinematics~\cite{yamamoto2022scaling}, and elastically ineffective strands~\cite{kim2022polyacrylamide}.

Fitting the TT stress--stretch curve with Gent model gives an effective locking parameter $J_m\approx12$ (Fig.~\ref{figS:tensile}B, \SeeSI), which is significantly larger than $J_m=2.4$ evaluated by fitting the equilibrium swelling data. This discrepancy arises from the fundamentally distinct deformation states probed by the two protocols in a silica-filled PDMS: equilibrium swelling is an isotropic expansion at modest stretch, where the response is tightly limited by the filler-mediated constraints and the shortest polymer strands. In contrast, uniaxial tension imposes large deviatoric stretches that can overstrain and irreversibly damage the filler--polymer constraints and short strands; this damage relieves constraints and allows longer strands to realign, manifesting as a delayed locking stretch (higher effective $J_m$). Indeed, incremental increases in subsequent loading cycles reveals hysteresis and softening even at the modest strains, as shown in Fig.~\ref{fig:uniaxial}C. We attribute this behavior to irreversible damage to the filler--polymer network, having ruled out viscoelastic effects with loading rate tests (Fig.~\ref{figS:tensile}C) and recovery-reloading tests (Fig.~\ref{figS:tensile}D). These results confirm the distinct constraint modes active in the two loading states, reconcile the larger tension-derived $J_m$, and validate the swelling-derived $J_m$ as an effective description of the silica-filled network's finite extensibility governing swelling equilibrium. With this parameterization, the FR-Gent model offers a high-fidelity prediction of $Q_{\mathrm{eq}}$ as a function of $M_s$.

\begin{figure}[!htbp]
\centering
\includegraphics[width=0.9\textwidth]{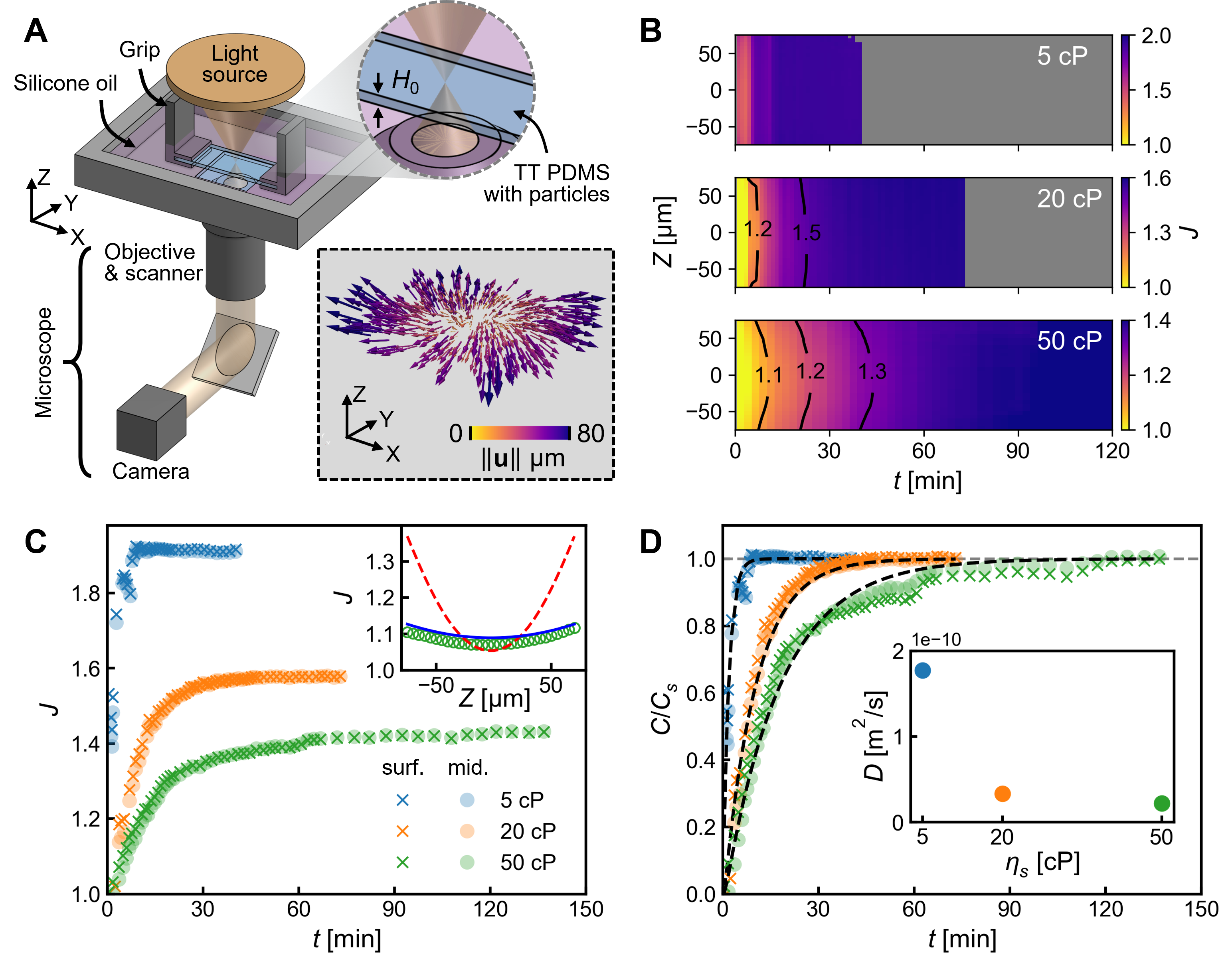}
\caption{Validation of 3D particle tracking in local free swelling and effective diffusivity measurement. (A) Experimental setup. A TT sample (initial thickness $H_0=\SI{160}{\micro m}$) with \SI{1.1}{\micro m} embedded microparticles is gently clamped on the grip and silicone oil of viscosity $\eta_s=\numlist{5;20;50}\,\unit{\centi P}$ is introduced at $t=0$. The TT PDMS swells freely; only minimal far-field tension is applied to maintain the sample flatness for imaging. The sample is illuminated from the above and imaged through a microscope. The objective is mounted on a fast $Z$-scanner, enabling time-resolved 3D image stacks. 3D particle trajectories are retrieved as described in the \SeeMM. Inset: particle displacement after reaching the free swelling equilibrium in $20\,\rm cP$ silicone oil; color encodes the displacement magnitude $\|u\|$.
(B) From 3D particle trajectories, the deformation gradient tensor $\mathbf{F}(\mathbf{X},t)$ is estimated via a local least-squares routine, and its determinant $J=\det(\mathbf{F})$ is computed to quantify the local volumetric swelling. The field of $J(Z,t)$ reveals the solvent diffusion across the thickness over time. A penetration front can be seen in the curvature of the iso-contours of $J$ for $\eta_s=50\,\rm cP$.
(C) Time traces of $J(Z,t)$ at the mid-plane ($Z=\SI{0}{\micro m}$, filled circles) and near the surface ($Z=\pm \SI{75}{\micro m}$, crosses). Three viscosities exhibit different swelling rate and equilibrium ratios, consistent with bulk measurements in Fig.~\ref{fig:bulk_swelling}C. Inset: a representative through-thickness profile $J(Z)$ at $t\approx5\,\rm min$ for $\eta_s=50\,\rm cP$ (green circles), compared with the fully drained Dirichlet prediction (red dashed; $D$ fitted in Fig.~\ref{figS:FS_sheet_Compare}A) and the limited-flux Robin prediction (blue solid; $(D,\mathrm{Bi})$ fitted in panel~(D)).
% Inset: enlarged view of the early-time swelling, where $J$ near the surface is consistently higher than at the mid-plane.
(D) Solvent concentration normalized with the equilibrium $C/C_s$, converted from $J$, assuming incompressible constituents. For each $\eta_s$, the Robin finite-slab solution (\eqref{eq:FS_robin_C_series}, \SeeSI) is fit to the full spatiotemporal field $C(Z,t)/C_s$, providing best-fit parameters $(D,\mathrm{Bi})$. The fitted $\mathrm{Bi}$ is summarized in Fig.~\ref{figS:FS_sheet_Compare}B, and the fitted $D$ is reported here in the inset. The dashed curves show the corresponding best-fit prediction evaluated at the mid-plane ($Z=0$). A comparison between Dirichlet- and Robin-based diffusivities is shown in Fig.~\ref{figS:FS_sheet_Compare}C.}
\label{fig:local-free-swelling}
\end{figure}

% \subsection*{Direct free-swelling measurement via 3D particle tracking}\label{sec:free_swelling_local}
To complement the global mass-uptake measurements, we implement a local 3D measurement of swelling kinetics \textit{in situ} by tracking tracer particles embedded in TT PDMS strips ($H_0=\SI{160}{\micro m}$). Particles become sterically bound during polymerization, and remain within the network after toluene treatment. The TT strips are immersed in silicone oils and imaged with a microscope as they freely swell (Fig.~\ref{fig:local-free-swelling}A; details in \SeeMM). A time series of 3D image stacks is acquired by scanning an objective along the optical axis, thereby sweeping the focal plane through the sample thickness throughout the experiment. Particle trajectories (inset of Fig.~\ref{fig:local-free-swelling}A) are retrieved in 3D from these stacks and used to compute the deformation gradient tensor $\textbf{F}$~\cite{EM2023}. Its determinant, $J=\det(\textbf{F})$, directly quantifies the local volumetric change and by incompressibility, the equivalent swelling ratio.

The spatiotemporal evolution of swelling is shown in Fig.~\ref{fig:local-free-swelling}B, where the swelling ratio $J$ across the sample thickness ($Z$) is plotted versus time. As expected, the overall swelling increases over time, until saturating at the final equilibrium state. Iso-$J$ contours are plotted atop the kymograph, and show a slight surface-to-center lag that is most evident for the high-viscosity oils; for $\eta_s=5\,\unit{cP}$, diffusion is so rapid that the field becomes nearly uniform, approaching our temporal and spatial resolution. $J$ is extracted near the free surface and at the mid-plane and shown in Fig.~\ref{fig:local-free-swelling}C. Both the swelling rate and equilibrium value exhibit a clear dependence on the solvent viscosity, with minimal $Z$-dependence. For $\eta_s=50\,\rm cP$, we extract the through-thickness profile $J(Z)$ at $t=5\,\rm min$ (Fig.~\ref{fig:local-free-swelling}C inset), highlighting the small but measurable gradient and confirming the spatial fidelity of our particle-tracking measurement.

Assuming incompressible molecules, we derive the local solvent concentration $C$ from $J$ and normalize it by the equilibrium value $C_s$ as $C/C_s=(J-1)/(J_s-1)$. The time evolution of $C/C_s$ at the mid-plane is first compared with the canonical fully drained Dirichlet solution (\eqref{eq:S5:C_Cs}, \SeeSI), as shown in Fig.~\ref{figS:FS_sheet_Compare}A. While this fit matches the temporal evolution and the fitted $D$ is consistent with the independent bulk mass-uptake measurement in Fig.~\ref{fig:bulk_swelling}E, a spatial comparison reveals that the Dirichlet prediction significantly overestimates the near-surface swelling at early times, as shown in the inset of Fig.~\ref{fig:local-free-swelling}C and Fig.~\ref{figS:FS_sheet_BC}. This discrepancy indicates that uptake at the free surfaces is not instantaneous; instead, finite interfacial transfer limits the flux into the sheet. Accordingly, we replace the Dirichlet condition with a limited-flux Robin boundary condition (parameterized by a Biot number $\mathrm{Bi}$; in the limit $\mathrm{Bi}\to\infty$, the Robin condition reduces to Dirichlet), and fit the full spatiotemporal field $C(Z,t)/C_s$ to the Robin finite-slab solution (\eqref{eq:FS_robin_C_series}, \SeeSI). This analysis well captures both the temporal evolution at the mid-plane (Fig.~\ref{fig:local-free-swelling}D) and the spatial through-thickness profiles (inset of Fig.~\ref{fig:local-free-swelling}C and Fig.~\ref{figS:FS_sheet_BC}). The finite values of the fitted $\mathrm{Bi}$, in the range 0.3 to 0.4, confirm a flux-limited interface under our experimental conditions (Fig.~\ref{figS:FS_sheet_Compare}B). This interfacial resistance is neither a casting artifact associated with the glass-contact surface (Fig.~\ref{figS:FS_Edge_Comparison}, \SeeSI) nor a consequence of formulating the diffusion problem in terms of $C$ rather than the chemical potential (Fig.~\ref{figS:FS_Numerical}, \SeeSI). With this boundary condition, the obtained $D$ are $\approx 5 \times$ larger than those inferred under the inappropriate Dirichlet assumption (Fig.~\ref{fig:local-free-swelling}D and Fig.~\ref{figS:FS_sheet_Compare}C).

This finding motivates us to re-assess the bulk mass-uptake measurements in Fig.~\ref{fig:bulk_swelling}C--E, which were performed under the same experimental conditions and therefore should obey the same limited-flux boundary condition. Using the interfacial $\mathrm{Bi}$ identified in Fig.~\ref{figS:FS_sheet_Compare}B, we refit the bulk uptake curves and obtain good agreement (Fig.~\ref{figS:FS_bulk_Compare}A--C). The resulting $D$ are consistent with the local measurements and correct the underestimated Dirichlet-based values by approximately one order of magnitude (Fig.~\ref{figS:FS_bulk_Compare}D). 
This comparison reveals a fundamental limitation of standard bulk characterization---intrinsic transport generally cannot be decoupled from boundary kinetics. It further underscores the essential role of 3D spatiotemporally resolved measurements for resolving this ambiguity and extracting an unbiased estimate of $D$.

\subsection*{Solvent migration in bent PDMS beams with controlled solvent-viscosity}

We next investigate solvent migration under external loading using a bent-beam geometry. Bending imposes an antisymmetric through-thickness strain field---tension on the outer surface and compression on the inner surface---thereby generating a chemical-potential gradient that drives the internal solvent fluxes. We use a ‘buckling’ configuration---rather than the conventional three- or four-point bending configurations---to attain a large deformation gradient and chemical-potential contrast, while providing a simple and stable boundary condition for long-duration experiments, with clear optical access.

% \subsection*{Direct measurement of load-induced solvent migration via 3D particle tracking}

\begin{figure}[!htbp]
\centering
\includegraphics[width=0.9\textwidth]{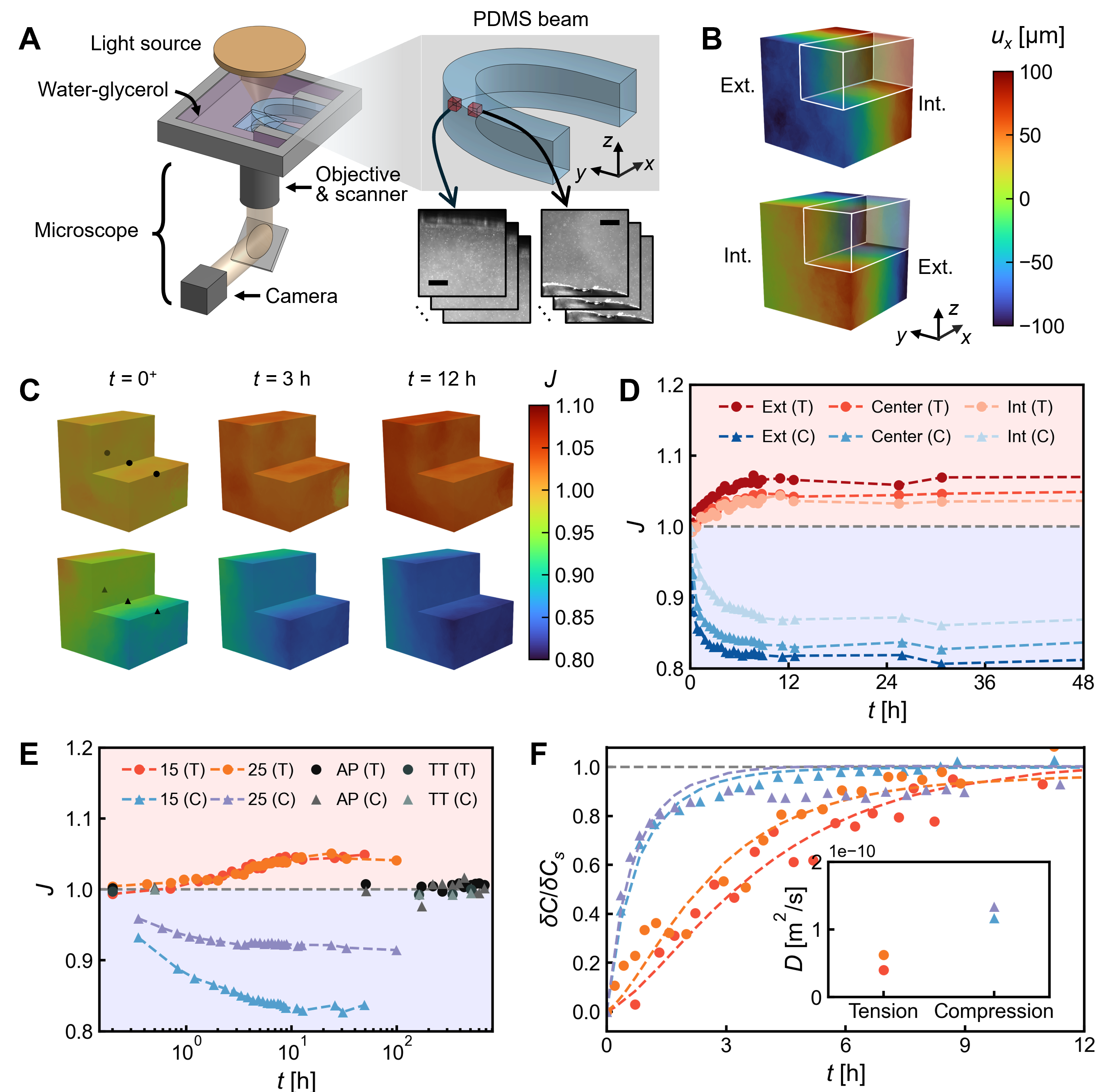}
\caption{Quantification of local volumetric change in bent PDMS via 3D particle tracking.
(A) Experimental setup: the bent PDMS beam is immersed in a glycerol--water bath with refractive index matched to PDMS, and tracer particles are imaged with scattered light. Upper-right inset: two volumes of interest (VOIs, red) at mid-span---one on the tensile side and one on the compressive side--- are imaged over time. Lower-right inset: a representative 3D stack for a bent beam; the scale bar is $200\,\unit{\um}$. 
% (A) inset image stacks are from for a FS-\SI{5}{\centi P} beam bent at span $15\,\unit{\mm}$
(B) Reconstructed 3D displacement fields at $t=0^+$ in the tensile (top) and compressive (bottom) VOIs for the FS-\SI{5}{\centi P}, \SI{15}{\mm}-span case. “Int.” and “Ext.” mark the VOI surfaces adjacent to the neutral plane and to the free surface, respectively. The cubic VOI measures $550\,\unit{\um}$ on each side. A quadrant of each VOI is rendered semi-transparent to reveal the interior displacement fields along the beam axis.
(C) Volumetric change is measured by $J=\det(\mathbf{F})$ at $t=0^+, 3,\,\text{and}\,12 \,\unit{h}$ in the tensile (top) and compressive (bottom) VOIs of the FS-\SI{5}{\centi P}, \SI{15}{\mm}-span beam. Dilation ($J>1$) appears on the tensile side and contraction ($J<1$) on the compressive side; the $J$ fields for both VOIs evolve over time. Three markers at $t=0^+$ indicate the centerline sampling locations shown in (D).
(D) Time traces $J(t)$ from the three sampling locations extracted from (C)---exterior surface, mid-thickness, and interior surface---share a re-equilibration time of $\approx\SI{10}{\hour}$. However, the re-equilibration amplitudes are asymmetric; on the tensile-side, dilation is modest ($J\approx1.05$) with little depthwise spread, whereas on the compressive side, contraction is stronger ($J\approx0.8-0.9$).
(E) Comparison of $J(t)$ in AP and TT beams (span \SI{15}{\mm}) and in an FS–\SI{5}{\centi P} beam at reduced curvature (span \SI{25}{\mm}). Over a one-month timescale, AP and TT traces remain flat ($J\approx1$) on both sides. Lower curvature (span \SI{25}{\mm}) mainly changes the compressive-side contraction (higher plateau, slower approach), with limited effect on the tensile-side plateau or re-equilibration time.
(F) Normalized solvent concentration change, $\delta C / \delta C_s$, derived from the $J$ traces in (E). Dashed lines show fits to solution of a 1D poroelastic diffusion problem (Sec.~\ref{secS:beam_solution},  \SeeSI); the resulting $D$ are summarized in the inset.
}
\label{fig:bending_local}
\end{figure}

We implement a 3D particle-tracking technique to quantify the local volumetric change in AP, TT, and FS ($\eta_s=5\,\unit{cP}$) beams with the bending configuration, using the experimental setup and the volume of interest (VOI), shown with typical image stacks in Fig.~\ref{fig:bending_local}A. The bent PDMS beam is immersed in a refractive-index-matched water-glycerol mixture to mitigate optical distortions from the curvature of the beam's surface. The particle tracking procedure follows that of the free swelling experiments: a 3D image stack is acquired; particles are tracked, and $J$ is estimated from the particle kinematics; additional experimental details are provided in \SeeMM.

The particle trajectories provide a full-field, 3D displacement map, as shown in Fig.~\ref{fig:bending_local}B for a representative FS-\SI{5}{\centi P} beam bent to a $15\,\unit{\mm}$ span. The smoothness of these fields confirms the tracking accuracy and underscores the robustness of the linking algorithm, even when the particle displacements far exceed the inter-particle spacing. The spatiotemporal $J$ fields are calculated at $t=0^+, 3,\,\text{and}\,12 \,\unit{h}$ and are shown in Fig.~\ref{fig:bending_local}C. Immediately after the bending step, the tensile VOI (top) remains near $J\approx1$ without discernible dilation or contraction, whereas the compressive VOI (bottom) exhibits an incipient contraction ($J<1$) with a pronounced through-thickness gradient; the contraction is strongest at the exterior surface and decreases toward the neutral plane. Over time, the volumetric strain evolves: the tensile VOI develops measurable dilation, and the compressive VOI increasingly contracts, indicating solvent migration from the compressed side of the beam toward the stretched side.

To quantitatively analyze the local evolution of $J$, we extract $J$ at three through-thickness locations along each VOI’s centerline, labeled at $t=0^+$ in Fig.~\ref{fig:bending_local}C, and plot the corresponding time traces in Fig.~\ref{fig:bending_local}D. The traces share a re-equilibration time of $\approx10\,\unit{h}$, whereas neither the rate of change nor the limiting value of $J$ is the same on the compressive or tensile side of the beam: on the tensile side, $J$ rises modestly to $\approx1.05$ with only minor depthwise variation in the asymptote, whereas on the compressive side, $J$ steeply drops to $0.8-0.9$ with a higher rate closer to the exterior surface, and then re-equilibrates with greater spread as a function of distance to the neutral plane. 

The same bending experiment is performed on AP and TT beams (span \SI{15}{\mm}), and the temporal evolution of $J$ at the centers of the VOIs is extracted and compared with the FS beams in Fig.~\ref{fig:bending_local}E. In sharp contrast to FS beams, the AP and TT traces remain essentially flat with $J$ indistinguishable from 1 on both the tensile and compressive sides over a one-month long timescale, indicating negligible solvent transport. For FS beams, reducing the bending amplitude (span $15\,\unit{\mm}\to25\,\unit{\mm}$) does not significantly change the re-equilibration rate or the asymptotic value on the tensile-side; however, the compressive-side response is reduced: the approach to re-equilibrium becomes slower and the $J$ asymptote shifts closer to $1$. Despite these bending amplitude-dependent differences, all FS cases re-equilibrate on a comparable timescale of a few hours, consistent with a diffusion length governed predominantly by the beam thickness.

% ----------------------------- 1201 from here
From the $J$ traces of FS-\SI{5}{\centi P} beams, we derived the change of solvent concentration normalized by saturation, $\delta C / \delta C_s$, as shown in Fig.~\ref{fig:bending_local}F. These $\delta C / \delta C_s$ curves are fitted to the solution of a 1D poroelastic diffusion problem across the beam thickness, assuming an impermeable (no-flux) tensile surface and a drained compressive surface with an attached oil-coating (Fig.~\ref{figS:bending_BC}, \SeeSI). The resulting calculations agree with the measured temporal evolution of $J$; the fitted $D$ values are consistently on the order of $10^{-10}\unit{m^2/s}$, with the compressive-side $D$ roughly twice that on the tensile side, and no systematic trend with bending amplitude, as shown in the inset; notably, $D$ measured from bending is roughly an order of magnitude larger than that measured from free swelling.

% --------

\begin{figure}[!htbp]
{\centering
\includegraphics[width=0.9\textwidth]{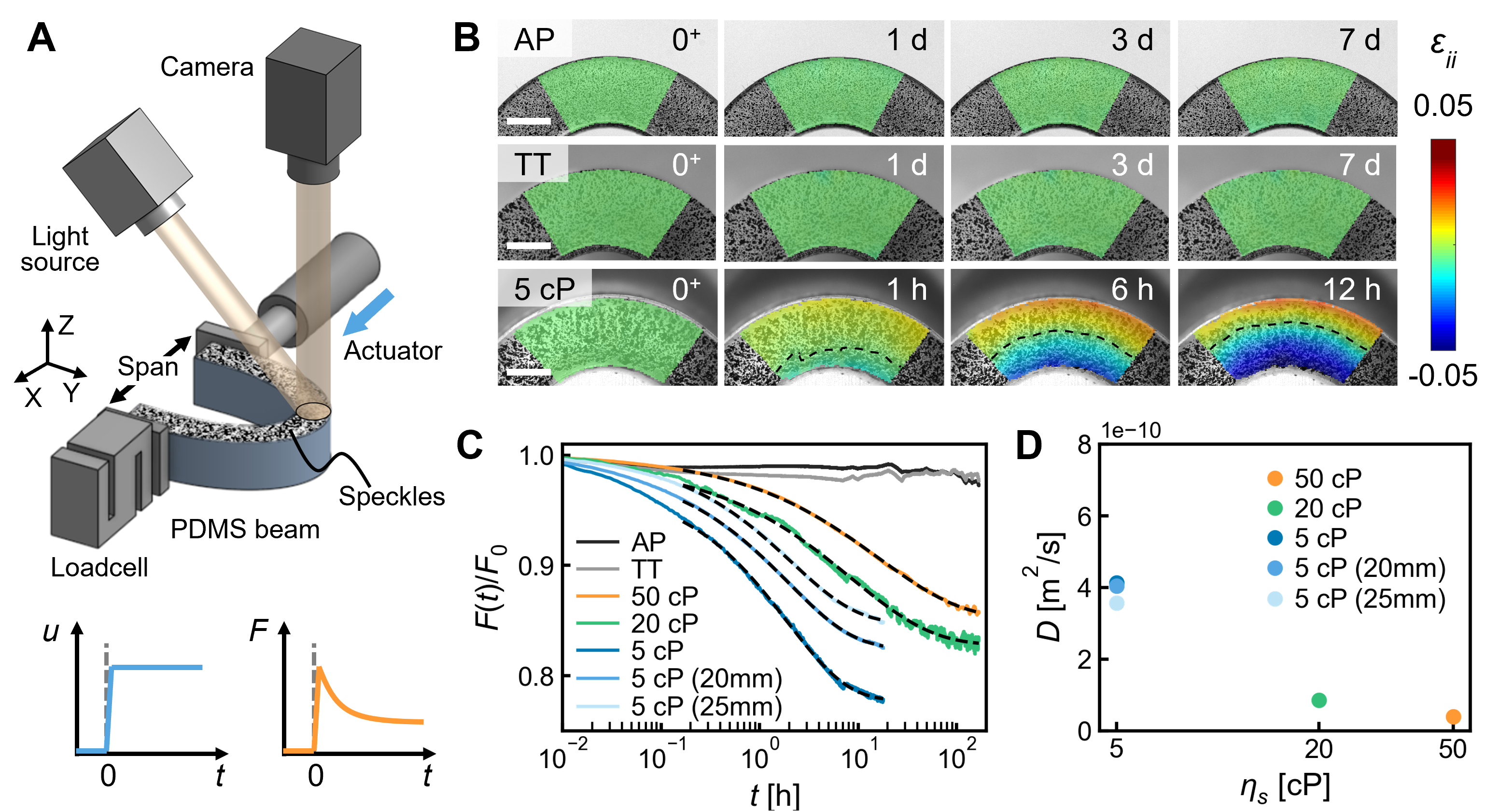}
\caption[Force relaxation and in-plane area change in bent PDMS beams]{
Force relaxation and in-plane area change in bent PDMS beams.
(A) Experimental setup. AP, TT, and FS ($\eta_s=\numlist{5;20;50}\,\unit{\centi P}$) PDMS beams are spray-painted with speckle patterns on one face. At $t=0$, a motorized actuator rapidly imposes a displacement step by setting the span to $15\,\unit{mm}$ (unless noted; corresponding curvatures in Fig.~\ref{figS:bending_global}). Schematic of the relaxation protocol is shown in the lower inset: after the displacement step $u$, the actuator position is held fixed while the reaction force and a time series of mid-span images for digital image correlation (DIC) are continuously recorded.
(B) Time-lapse snapshots and DIC strain fields for AP, TT, and FS-\SI{5}{\centi P} beams. The trace of the 2D strain tensor, $\varepsilon_{ii}=\varepsilon_{xx}+\varepsilon_{yy}$, measures the local area change and serves as a proxy for the local volumetric change. The dashed curve marks the iso-contour of $\varepsilon_{ii}=0$ and is used to track the migration front. The scale bar is $2\,\unit{\mm}$.
(C) Reaction-force evolution during relaxation, normalized by the initial force $F_0$. Apart from a brief transient viscoelastic relaxation at $t=0^+$ (Fig.~\ref{figS:bending_global}, \SeeSI), AP and TT beams show negligible decay over one week\textsuperscript{*}, whereas all fully swollen beams relax markedly, with faster decay for lower $\eta_s$ and for higher curvature. These curves are fit with a stretched-exponential function, from which we extract the characteristic diffusion timescale $\tau$; the fits (black dashed lines) show good agreement with the measurements.
(D) $D$ is derived from the fitted relaxation time $\tau$ and the beam half-thickness $H_0/2$ via $D\approx(H_0/2)^2/\tau$. The resulting $D$ values agree well with the local particle tracking measurements in Fig.~\ref{fig:bending_local}F. The measured $D$ is essentially independent of loading amplitude, and decreases as $\eta_s$ increases.
}
\label{fig:bending_global}\par}

\vspace{1ex}
\noindent
\begin{minipage}{\textwidth}
\raggedright
\setlength{\parindent}{0pt}
\footnotesize
\textsuperscript{*}The apparent correlation between the AP and TT force traces is not an intrinsic material effect but reflects shared environmental fluctuations, as the two experiments are conducted simultaneously on parallel setups.
\end{minipage}
\end{figure}
% \footnotetext{The apparent correlation between the AP and TT force traces is not an intrinsic material effect but reflects shared environmental fluctuations, as the two experiments are conducted simultaneously on parallel setups.}

While 3D particle tracking resolves the local volumetric fields associated with stress-driven solvent migration, it provides only kinematic information, and cannot directly access the conjugate forces that complete the poroelastic coupling. To close this mechanical loop and rigorously confirm that the observed local volumetric changes are indeed stress-driven, we conduct a macroscopic bending experiment consisting of long-duration reaction force recording and 2D digital image correlation (DIC), as shown schematically in Fig.~\ref{fig:bending_global}A. The stability of this measurement modality enables recording of FS samples with higher viscosities, and thus we can not only cross-validate $D$ for the reference case ($\eta_s=\SI{5}{\centi P}$), but also extend our stress-driven study to higher viscosities ($\eta_s = 20$ and $\SI{50}{\centi P}$). Additional experimental details are provided in \SeeMM.
% high-viscosity silicone oils, due to their longer diffusion timescales and reduced fractions, are prone to compromise 3D image-stack fidelity and undermine particle tracking, whereas force-relaxation measurements remain sensitive and robust.

2D DIC is implemented to probe the in-plane deformations induced by solvent migration. To isolate the strain arising from solvent migration, we compute kinematic fields using the image acquired immediately \emph{after} the displacement step ($t=0^{+}$) as a reference, thereby removing the large background strain from the initial bending. We interrogate the in-plane dilatation and contraction using $\varepsilon_{ii} = \varepsilon_{xx} + \varepsilon_{yy}$ as a 2D proxy for the real volumetric change for the AP, TT, and FS-\SI{5}{\centi P} beams bent to a $15\,\unit{\mm}$ span, as shown in Fig.~\ref{fig:bending_global}B. Over a period exceeding one week, no dilatation or contraction is observed in either the AP or the TT beams, consistent with the local volumetric measurement in Fig.~\ref{fig:bending_local}E; in contrast, a dilatation band on the tensile side and a contraction band on the compressive side is visible in the FS-\SI{5}{\centi P} beam, after only $1\,\unit{h}$ is elapsed from the application of the deformation. Within the field of view, the poroelastic strain field is approximately uniform along the neutral axis and exhibits a pronounced asymmetry across the thickness: the neutral-dilation iso-contour ($\varepsilon_{ii} = 0$) originates closer to the compressive side and subsequently propagates toward the tensile side, indicating the direction of the solvent flux and a greater flux on the compressive side, consistent with the local measurements show in Fig.~\ref{fig:bending_local}.

A signature of the solvent transport is directly captured by the simultaneous force-relaxation measurements shown in Fig.~\ref{fig:bending_global}C, where the reaction force is normalized by its initial value, $F_0$. Consistent with the DIC observations, AP and TT beams show negligible force decay over the one-week experiment, apart from an initial viscoelastic transient (Fig.~\ref{figS:bending_global}B). In contrast, the force relaxation is clearly visible in all FS beams: beams with lower $\eta_s$ relax faster, consistent with diffusion-controlled transport where the characteristic timescale $\tau \propto 1/D$ with $D \propto 1/\eta_s$; for FS-\SI{5}{\centi P} beams, increasing the bending amplitude accelerates the decay and lowers the long-time plateau. Because the global force relaxation reflects stress-induced solvent migration with a strong stress gradient, the functional form of the solution cannot be easily expressed in terms of simple algebraic relations of transcendental functions. The solution to such diffusion problems can be expressed by an infinite sum of functions of $Dt/H_0^2$. We estimate $D$ by fitting the curves---excluding the initial $10\,\unit{min}$ to avoid viscoelastic effects---to a phenomenological stretched-exponential form~\cite{KWW1, KWW2, hu2010indentation, berry2020poroelastic}, $F(t)/F_0=C_0+C_1e^{-(t/\tau)^\beta}$, where $\tau$ is the characteristic poroelastic timescale, $\beta\in(0,1]$ is a stretching exponent that captures the complex nature of the relaxation, and $C_0$ and $C_1$ are the long-term plateau and relaxation amplitude, respectively. This function accurately captures the force relaxation for all FS beams over several decades in time. The key fitting parameter, $\tau$, is used to approximate $D$ via the scaling relation with the beam thickness $D\approx (H_0/2)^2/\tau$; these values are plotted in Fig.~\ref{fig:bending_global}D.

\subsection*{Discussion}
\begin{figure}[!htbp]
\centering
\includegraphics[width=0.48\textwidth]{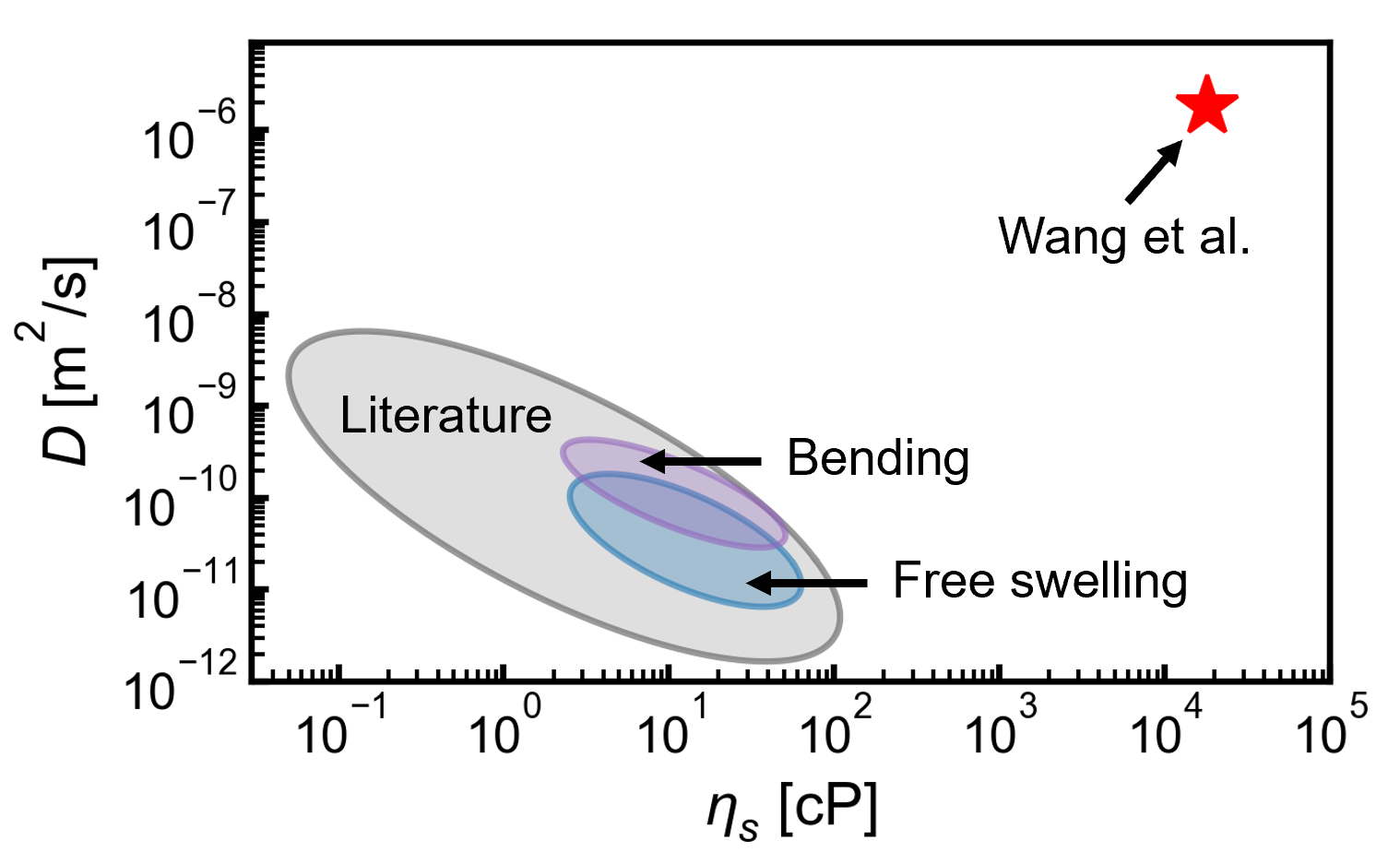}
\caption{Summary of solvent diffusivity $D$ versus solvent viscosity $\eta_s$ in PDMS and related silicone-rubber networks. Colored regions indicate the ranges measured here in free swelling (purple) and bending-driven migration (blue); the gray region summarizes literature values (compiled in Table~\ref{tabS:D_values}, \SeeSI). The red star marks the recently reported value for a similar silicone system~\cite{PNAS_Vikram}.
}
\label{fig:all_D}
\end{figure}

Our measured $D$ values from both free swelling and bending-induced solvent migration are compared with reported solvent diffusivities in PDMS and related systems in Figure~\ref{fig:all_D}; the corresponding literature data are summarized in Table~\ref{tabS:D_values}, \SeeSI. While our measurements fall within the range of reported values, they consistently lie near the upper bound of this established region. This distribution is expected because literature values are typically inferred from bulk measurements using theoretical models and calibrated simulations that assume fully drained boundary conditions. However, as demonstrated in our spatiotemporal measurements, these interfaces can be flux-limited. Because interfacial resistance cannot be distinguished from intrinsic diffusion in bulk measurements, imposing a fully drained boundary condition lumps interfacial resistance into an apparent diffusivity and thereby systematically underestimate the true $D$. Our experimental framework provides a practical route to eliminate this interfacial boundary-condition ambiguity via local full-field kinematics and, when integrated with conventional mechanical tests such as indentation~\cite{hu2010indentation}, elevates a purely global force--relaxation measurement into a closed-loop bulk--local, kinematic--mechanical characterization.

Our measurements enable a direct comparison with recent findings in similar rubbery solids~\cite{PNAS_Vikram}, which reported a remarkably high diffusivity $D^* \approx 1.8 \times 10^{-6}\,\unit{m^2/s}$---more than three orders of magnitude higher than our benchmarks---along with counter-intuitive local volumetric responses. While solvent shear-thinning was proposed to explain this rapid transport, our AP beams containing $\approx 4\,\text{wt}\%$ uncrosslinked sol fraction show no measurable volumetric change or force relaxation over weeks, suggesting that shear-thinning alone is insufficient to explain such a high value for $D^*$. Furthermore, our results demonstrate that the volumetric expansion goes hand-in-hand with solvent migration, in striking contrast to these recent measurements. Further studies may be required to elucidate the physical origin of these discrepancies.

We have established a quantitative framework for silicone-oil diffusion in a crosslinked PDMS network by experimentally probing the swelling equilibrium and swelling kinetics under free and mechanically loaded conditions. We first demonstrate that the equilibrium swelling ratios are accurately predicted only when the classic Flory--Rehner theory is extended to include the finite extensibility of the silica-filled network through the Gent model. Turning to transport kinetics, we leverage 3D spatiotemporally resolved particle tracking to rigorously determine the transport boundary conditions and rationalize global mass uptake data. This combination, while underscoring the critical role of spatially resolved data in eliminating boundary condition ambiguity, establishes high-fidelity benchmark measurements of $D$ in the range of $10^{-12}\text{--}10^{-10}\,\unit{m^2/s}$; $D$ systematically decreases with increasing $\eta_s$. 
In a bent-beam configuration, local particle-tracking measurements reveal dilation on the tensile side and contraction on the compressive side, but only in FS beams; in AP and TT beams, volumetric change is not detected over a long measurement timescale, consistent with independent force relaxation recording. For FS beams, $D$ is of the order of $10^{-10}\,\unit{m^2/s}$, essentially independent of bending amplitude and decreasing with increasing $\eta_s$, sharing a viscosity dependence with the free-swelling case. Our results present a quantitative local assessment of poroelastic solvent migration in polymer systems that falls within the range of all prior bulk diffusivity measurements in silicone rubbers~\cite{hu2011indentPDMS, hu2012thinlayer, microchannel, fierro2011experimental, randall2005PNAS, uptake2001, uptake2015}. These measurements are in striking contrast to the recent local measurements in similar polymer systems~\cite{PNAS_Vikram}; resolving the origin of this discrepancy remains an important direction for future studies and will sharpen our understanding of fundamental polymer transport physics. Nevertheless, despite their differing conclusions, both studies underscore the distinct value of 3D spatiotemporally resolved measurements for probing inherently local processes such as diffusion. By providing direct access to the evolving local chemo-mechanical state, such measurements not only enable unambiguous interpretation of experimental results, but also establish benchmark datasets for building and validating high-fidelity poroelastic models in polymers and, by extension, other fluid-infiltrated solids.

%%%%%%%%%%%%%%%% MAIN TEXT FIGURES %%%%%%%%%%%%%%%

%%%%%%%%%%%%%%%% REFERENCES %%%%%%%%%%%%%%%

\clearpage
\bibliography{references}
\bibliographystyle{sciencemag}

%%%%%%%%%%%%%%%% ACKNOWLEDGEMENTS %%%%%%%%%%%%%%%

\section*{Acknowledgments}
We acknowledge support from the Swiss National Science Foundation (Grant No. 200021\_19716). We are grateful to Prof. Pedro Reis for raising this problem to our attention.
\paragraph*{Funding:}
Swiss National Science Foundation Grant No. 200021\_19716.
\paragraph*{Author contributions:}
C.L. and J.M.K. designed research; C.L., T.B., and M.A.C. performed research; C.L., T.B., M.A.C., and J.M.K. analyzed data; and C.L. and J.M.K. wrote the paper.
\paragraph*{Competing interests:}
There are no competing interests to declare.
\paragraph*{Data and materials availability:}
All data needed to evaluate the conclusions in the paper are present in the paper and/or the Supplementary Materials.

%%%%%%%%%%%%%%%% SUPPLEMENT LIST %%%%%%%%%%%%%%%

\subsection*{Supplementary materials}
Materials and Methods\\
Supplementary Text\\
Figs. S1 to S15\\
Table S1\\
References \textit{(1-\arabic{enumiv})}

%%%%%%%%%%%%%%%% END OF MAIN TEXT %%%%%%%%%%%%%%%

\newpage

%%%%%%%%%%%%%%%% START OF SUPPLEMENT %%%%%%%%%%%%%%%

\renewcommand{\thefigure}{S\arabic{figure}}
\renewcommand{\thetable}{S\arabic{table}}
\renewcommand{\thepage}{S\arabic{page}}
\renewcommand{\thesection}{S\arabic{section}}
\renewcommand{\thesubsection}{\thesection\Alph{subsection}}
\setcounter{figure}{0}
\setcounter{table}{0}
\setcounter{equation}{0}
\setcounter{section}{0}
\setcounter{page}{1}

%%%%%%%%%%%%%%%% SUPPLEMENT TITLE PAGE %%%%%%%%%%%%%%%

\begin{center}
\section*{Supplementary Materials for\\ \scititle}

Chenzhuo~Li$^{\ast}$,
Tom~Beyeler,
Marc~Antonio~Chalhoub,
John~M.~Kolinski$^{\ast}$\\
\small$^\ast$Corresponding author. Email: john.kolinski@epfl.ch
\end{center}

\subsubsection*{This PDF file includes:}
Materials and Methods\\
Supplementary Text\\
Figures S1 to S15\\
Table S1

\newpage

%%%%%%%%%%%%%%%% MATERIALS AND METHODS %%%%%%%%%%%%%%%

\section*{Materials and Methods}
\subsubsection*{Fabrication of as-prepared, toluene-treated, and fully swollen PDMS samples}

To enable 3D particle tracking, polystyrene microparticles ($\SI{1.1}{\micro m}$ diameter, sigma-aldrich) were embedded into PDMS prepared from Dow Sylgard~184. The as-received particle suspension (aqueous form, 10\% solid fraction) was first mixed with the curing agent at a volumetric-to-mass ratio of \SI{1}{\micro\liter}:\SI{2}{\gram}. The particle-laden curing agent was vortex-mixed for \SI{5}{\minute} and sonicated for \SI{5}{\minute} to break agglomerates and achieve a uniform dispersion. The Sylgard~184 based was then mixed with this particle-laden curing agent at a \SI{10}{:1} (w/w) ratio, hand-mixed thoroughly for \SI{5}{\minute}, and degassed under vacuum for \SI{30}{\minute} to completely remove entrapped air bubbles. The degassed mixture was then cast between glass plates separated by precision spacers of thickness $H_0\in\{0.25, 0.5, 1.0, 5.0\}\,\unit{\mm}$ and cured in an oven at $\SI{65}{\celsius}$ for $\SI{2}{\hour}$. After curing, PDMS sheets/slabs were demolded and cut to the required geometries for subsequent experiments. These freshly cured samples are referred to as as-prepared (AP). Note that nanoparticles are often added in these commercial PDMS to enhance the mechanical properties~\cite{clough2016covalent, suriboot2021amphiphilic}.

\begin{figure}[H]
\centering
\includegraphics[width=0.8\linewidth]{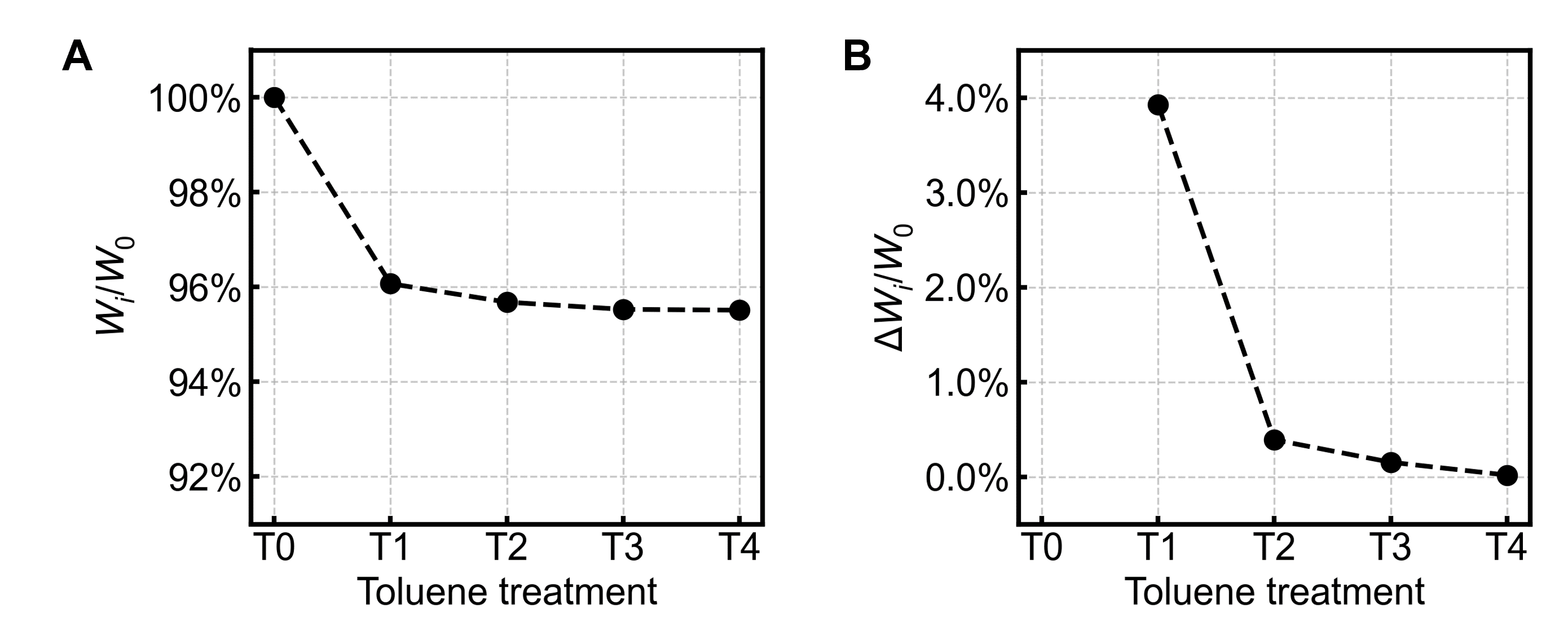}
\caption{Removal of uncrosslinked PDMS sol fraction by sequential toluene treatments. (A) Remaining mass fraction of a thin as-prepared PDMS sample after each wash $W_i/W_0$. where $W_0$ is the pre-wash mass (T0), and $i=1,...,4$ denote successive washes (T1–T4). Each wash used fresh toluene at $>10\times$ the sample volume for $\geq \SI{24}{\hour}$, followed by drying for $\geq \SI{24}{\hour}$ before weighing.(B) Normalized mass removed per wash, $\Delta W_i/W_0=(W_i-W_{i-1})/W_0$. The first wash removes $\approx4\%$ of the initial mass; subsequent washes each remove $<1\%$ and the mass approaches a plateau after the 3 repeated washes, indicating that the material is effectively solvent-free.}
\label{figS:toluene}
\end{figure}

To obtain a mobile-phase–free reference network, AP PDMS sheets/slabs were washed exhaustively with toluene. They were immersed in sealed glass containers filled with fresh toluene (bath volume $>10\times$ the PDMS volume) and left $\SI{24}{\hour}$ in the fume hood. This wash was repeated three times with complete toluene replacement each cycle. Following the final extraction, the sheets/slabs were briefly rinsed in fresh toluene, gently blotted with lint-free wipes, and  dried in a vacuum chamber for $\SI{24}{\hour}$ to remove residual solvent. This protocol removes uncrosslinked PDMS prepolymer, leaving only the crosslinked network without interstitial solvent (Fig.~\ref{figS:toluene}, \SeeSI). The dried PDMS were cut to samples with specific dimensions; these samples serve as the mobile-phase–free reference and are referred to as toluene-treated (TT).

TT PDMS sheets/slabs were fully immersed in silicone oils of prescribed kinetic viscosity $\eta_s=\numlist{5;20;50}\,\unit{\centi P}$ at room temperature. This free swelling process lasted for at least one week (for sheets, $H_0 \leq \SI{1}{\mm}$) or two weeks (for slabs, $H_0 = \SI{3}{\mm}$) to ensure a stable, fully-equilibrium state. Following equilibration, the PDMS were cut to samples with specified geometries, referred to as fully swollen (FS).

\subsubsection*{Tensile tests of AP, TT, and FS PDMS samples}
\begin{figure}[H]
\centering
\includegraphics[width=0.5\linewidth]{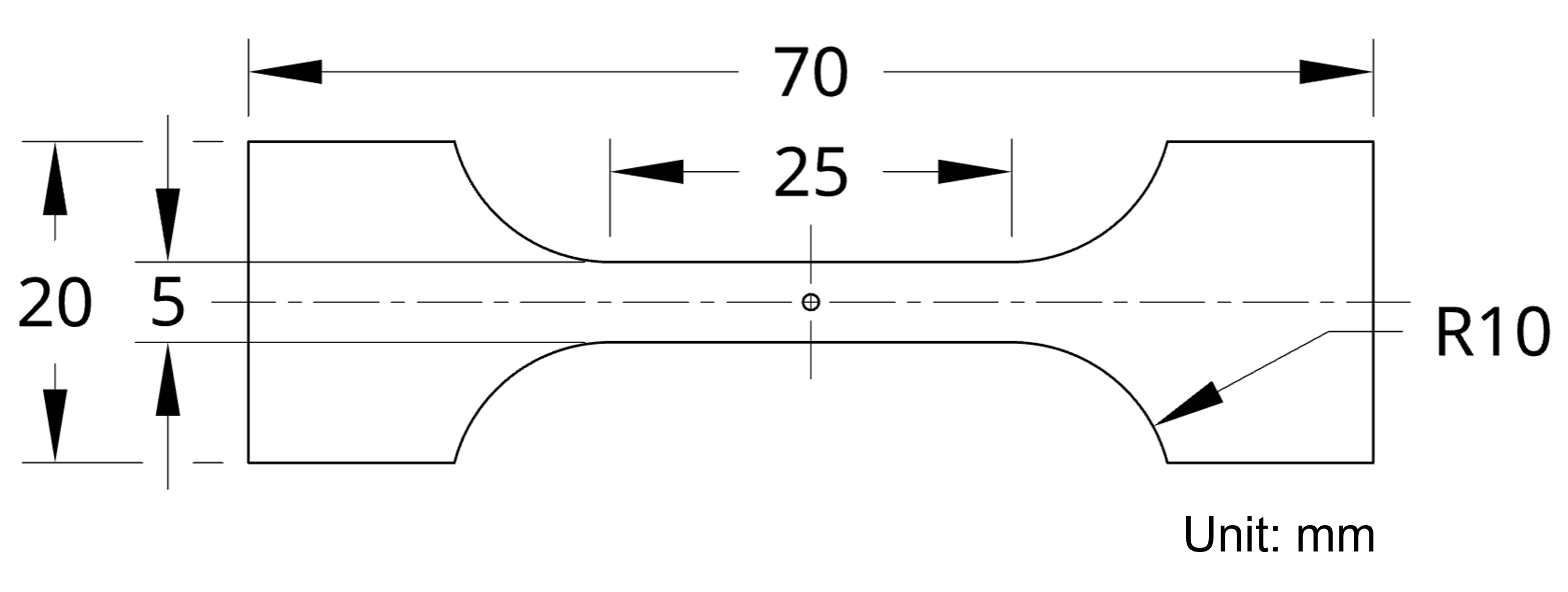}
\caption{Planform geometry of laser-cut dog-bone samples from as-prepared PDMS sheets for uniaxial tensile testing. For each test, the actual gauge length and gauge width were obtained from the measured swelling ratios of the samples at the test state, and the thickness was measured with a caliper. All dimensions are in millimeters.}
\label{figS:dogbone}
\end{figure}

Dog-bone samples were laser-cut from \SI{0.25}{\mm}-thick as-prepared (AP) PDMS sheets with a nominal gauge length ($L_0$) of \SI{25}{mm} and width of \SI{5}{mm} (detailed geometry in Fig.~\ref{figS:dogbone}, \SeeSI). AP samples were tested as cut, while TT and FS samples were prepared as described in Fig.~\ref{fig:bulk_swelling}B. Prior to testing, in-plane gauge dimensions (length and width) for TT and FS states were corrected by the corresponding equilibrium swelling ratio extracted from Fig.~\ref{fig:uniaxial}A. The through-thickness dimension was measured directly at equilibrium using a caliper at three positions within the gauge section and averaged. The initial cross-sectional area $A_0$ was computed from the corrected width and the measured thickness.

Uniaxial tension was performed on a customized bench-top testing frame instrumented with a \SI{10}{N} load cell (HBM S2M/10N, connected to HBM ClipX amplifier) and a linear variable differential transformer (LVDT, HBM WA/100mm) for displacement. Force ($F$) and displacement signals $\Delta L$ were recorded synchronously on a oscilloscope (Tektronix MDO3054) and logged for post-processing. Unless otherwise noted, samples were stretched at a constant engineering strain rate $\dot\varepsilon=\SI{0.05}{s^{-1}}$ (crosshead speed set from the corrected gauge length) under ambient laboratory conditions.

Engineering stress was defined as $\sigma=F/A_0$ and engineering stretch as $\lambda=\Delta L/L_0 +1$. The small-strain tensile modulus ($E$) was obtained by a linear least-squares fit over the initial linear regime (typically $\lambda\in[1,1.05]$), and small-strain shear modulus $G$ was derived by setting Poisson's ratio $\nu=0.5$. For each material state (AP, TT, and each FS solvent condition), at least three independent samples were tested; results are reported as mean $\pm$ standard deviation.

\subsubsection*{Mass uptake measurement during free swelling of thin TT samples in silicone oils}
Mass uptake during free swelling was quantified on TT PDMS thin samples in silicone oils with viscosities of $\eta_s=\numlist{5;20;50}\,\unit{\centi P}$. Samples were cut from TT PDMS sheets using a punching tool and had initial thicknesses of $H_0\in\{0.25, 0.5, 1.0\}\,\unit{\mm}$ with an in-plane diameter $D_0=\SI{20}{\mm}$. This geometry ($D_0 \gg H_0$) approximates 1D diffusion across the thickness and suppresses edge effects.

Prior to immersion, the initial weight of each sample was measured with a precision balance (resolution \SI{0.0001}{g}). The sample was then fully immersed in a bath of silicone oil at the prescribed viscosity, and the time-stamp at first contact was recorded. After an immersion interval (\SIrange{5}{30}{\minute}, depending on sample thickness and oil viscosity), the sample was removed from the oil bath and the superficial residual oil was gently blotted with lint-free wipes. The sample was then briefly dipped in a sonicating isopropanol bath for \SI{10}{\second}, gently blotted again, and dried with nitrogen to remove residual isopropanol; the mass was then recorded with the time stamp, and the sample was returned to the oil bath. This cycle was repeated until the mass increase between successive measurements was less than 1\%. To ensure an accurate equilibrium weight, an additional measurement was taken $\SI{1}{\day}$ after the last measurement cycle. Each entire cleaning and weighing cycle took approximately $\SI{30}{\second}$.

For the equilibrium swelling ratio measurement, samples were immersed in silicone oil baths with an extended viscosity range ($\eta_s=\numlist{1;5;10;20;50;100;500;1000}\,\unit{\centi P}$). For each viscosity, three TT samples (nominal $D_0=\SI{20}{\mm}$, $H_0=\SI{0.5}{\mm}$) from different batches were tested. The initial weight ($W_0$) and diameter ($D_0$) of each sample were measured with a precision balance and a caliper prior to immersion. After two weeks of free swelling, the samples were removed from the silicone oil baths, cleaned using the procedure described above, and measured again for their fully swollen weight ($W_\infty$) and diameter ($D_\infty$). The equilibrium mass swelling ratio was defined as $W_\infty/W_0$ and the the equilibrium volume swelling ratio as $(D_\infty/D_0)^3$.

% \subsubsection*{Local quantification of swelling ratios via 3D particle tracking}
\subsubsection*{Quantification of local swelling ratios during free swelling of TT PDMS in silicone oils}
A TT PDMS strip (thickness $H_0=\SI{160}{\micro\meter}$, width $\approx\SI{10}{\mm}$, and length $\approx\SI{30}{\mm}$) with embedded \SI{1.1}{\micro\meter} tracer particles was used in this experiment. Before immersion, a strip was placed flat on a cover glass, and a reference 3D image stack was acquired at its central region using a Nikon Eclipse Ti microscope (objective: Nikon Plan Fluor, 10x water-immersion, $\mathrm{NA}=0.3$; camera: Hamamatsu C13440, $2048\times2048$ pixel, 16 bit) equipped with a $Z$-piezo stage (Nano-ZL300-M, Mad City Labs Inc.). Such a typical image stack spanned \SI{300}{\micro\meter} in $Z$ with 254 slices and required a \SI{3}{\second} acquisition time. The sample was then transferred and clamped between the grips shown in Fig.~\ref{fig:local-free-swelling}A, and a minimal tensile preload was applied via a precision servo motor to keep the strip flat. After centering the same region within the field of view, silicone oil of a prescribed viscosity was gently poured into the bath, and the immersion time $t=0$ was recorded. As swelling progressed, the grips were manually adjusted as needed to keep the field of view centered on the same region and to avoid out-of-plane wrinkling from sample elongation. In the meanwhile, time-stamped 3D stacks spanning the full thickness were recorded through the cover-glass window at the bottom of the bath at approximately \SI{1}{\minute} intervals with imaging parameters held fixed.

\begin{figure}[H]
\centering
\includegraphics[width=0.7\linewidth]{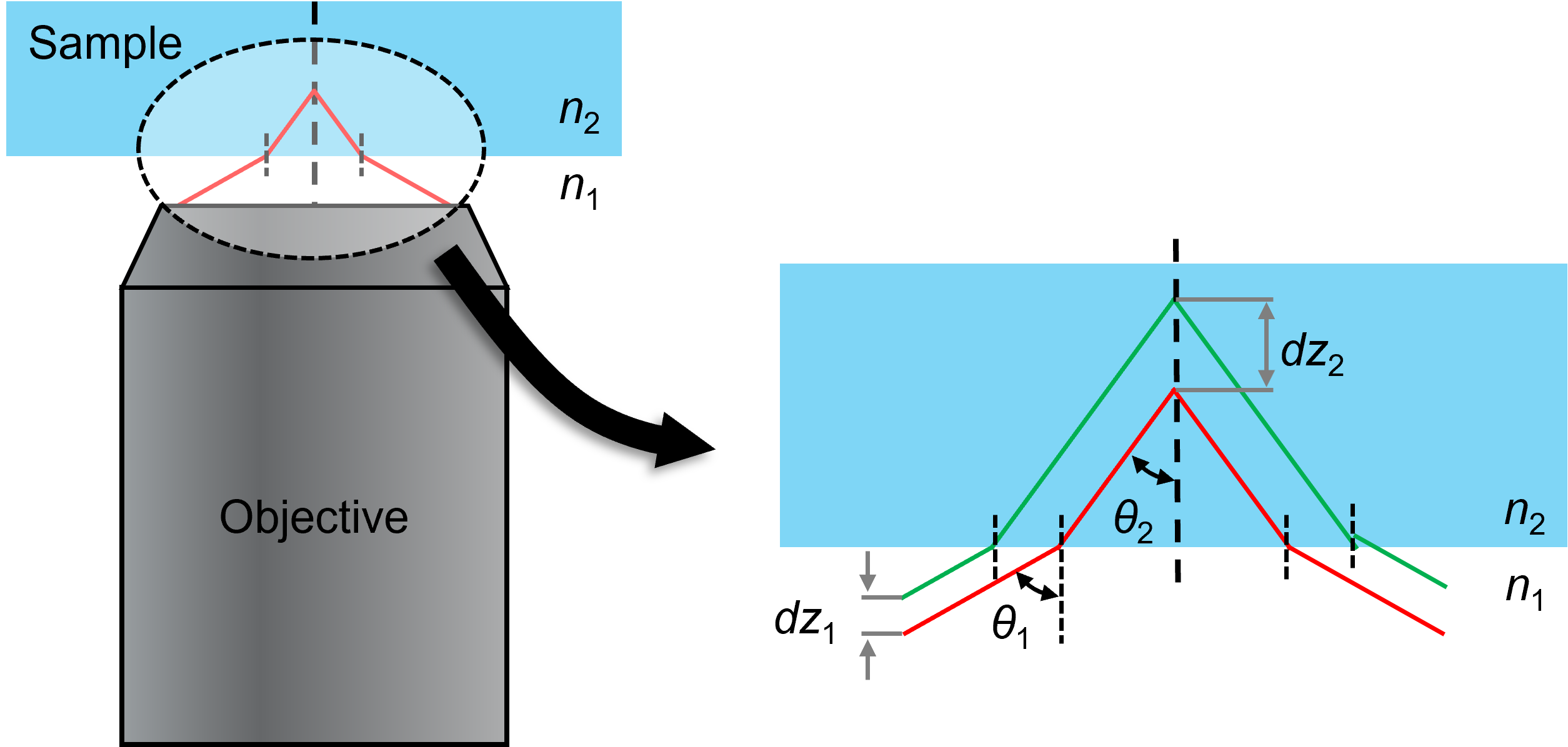}
\caption{Geometric correction of the axial step size when imaging across a refractive-index interface. 
For a commanded objective step $dz_1$ in the immersion medium ($n_1$), the true axial step in the sample ($n_2$) is $dz_2/dz_1=\tan\theta_1/\tan\theta_2$, where $\theta_1$ and $\theta_2$ are the chief-ray angles relative to the optical axis in media $n_1$ and $n_2$, respectively. 
Using $\mathrm{NA}=n_1\sin\theta_1$ and Snell’s law $n_1\sin\theta_1=n_2\sin\theta_2$, we obtain $\sin\theta_1=\mathrm{NA}/n_1$ and $\sin\theta_2=\mathrm{NA}/n_2$. In our setup---10$\times$ water-immersion objective ($\mathrm{NA}=0.30$, $n_1=1.33$) imaging PDMS with measured $n_2=1.4125$---we obtain $\sin\theta_1=0.30$, $\sin\theta_2=0.212$, and $dz_2/dz_1=1.063$. Accordingly, reconstructed $z$-positions are scaled by $1.063$.}
\label{figS:Z_correction}
\end{figure}

Reference (pre-immersion) and current (during swelling) particle locations were obtained in each 3D stack using the open-source package TrackPy~\cite{trackpy_crocker, trackpy_code}. Pixel coordinates were converted to physical units using careful stage calibrations, including a $Z$-step size correction (Fig.~\ref{figS:Z_correction}, \SeeSI). Particles were then linked into time-resolved trajectories using a RAFT-based algorithm~\cite{RAFT} modified with an updating predictor. From the retrieved 3D trajectories, the deformation gradient tensor $\mathbf{F}$ was estimated at each particle location via a local least-squares routine~\cite{EM2023}, and the local volumetric change was subsequently quantified as $J=\det(\mathbf{F})$. Based on the incompressible polymer and solvent molecules, $J$ equals the local swelling ratio and provides an accurate measure of the solvent concentration field.

\subsubsection*{Quantification of local solvent migration in bent beams}

PDMS beams were cut from AP, TT, and FS ($\eta_s=\SI{5}{cP}$) slabs to dimensions of height $B_0\approx\SI{5}{\mm}$, thickness $H_0\approx\SI{3.3}{\mm}$, and length $L_0\approx\SI{40}{\mm}$, as annotated in Fig.~\ref{fig:bulk_swelling}A. Before bending, the beam was placed flat on a large cover glass (thickness \SI{160}{\micro m}), and reference 3D image stacks were acquired near the midpoint on each side of the beam using the same microscope setup as in the fully swollen experiment, with the $Z$-piezo stage replaced by an objective scanner to achieve a larger scanning range. A typical image stack spanned \SI{600}{\micro\meter} in $Z$ with a step size of \SI{2}{\micro m} and required \SI{3}{\minute} to acquire. The beam was then rapidly bent by hand at $t=0$ and inserted into a 3D-printed fixture with a prescribed span (Fig.~\ref{fig:bending_local}A). A water/glycerol mixture was subsequently added to the bath to match the refractive index of PDMS ($n=1.4125$), thereby minimizing optical distortions from the curved bottom surface induced by the Poisson's effect. Image stacks of the reference regions on both sides were then acquired at intervals of approximately \SIrange{20}{30}{\minute} for FS samples over more than two days, and every \SIrange{2}{3}{\day} for AP and TT samples for four weeks. These time-lapse stacks were processed using the same pipeline as for the local free-swelling measurements to track particle trajectories and to estimate the local deformation gradient tensor $\mathbf{F}$ and volumetric swelling $J=\det(\mathbf{F})$.

\subsubsection*{Force relaxation with simultaneous in-plane kinematic measurement for bent beams}

PDMS beams were cut from AP, TT, and FS ($\eta_s=\numlist{5;20;50}\,\unit{\centi P}$) slabs to dimensions ($B_0\approx\SI{5}{\mm}$, $H_0\approx\SI{3.3}{\mm}$, and $L_0\approx\SI{40}{\mm}$) as annotated in Fig.~\ref{fig:bulk_swelling}A. The beams were spray-painted with a single-layer black speckle pattern on one surface for DIC analysis,, and an imaging system was focused on a midspan region of interest (ROI). To increase measurement throughput, two imaging configurations were used in parallel: a Nikon AZ100 macroscope with a Hamamatsu camera ($2048\times2048$ pixel, 16 bit), and a telescope-lens-equipped CMOS camera (Thorlabs CS2100M). Force was measured using a load cell (HBM S2M/10N, connected to HBM ClipX amplifier) and logged with a digital multimeter (Keysight 34465A) at \SI{12}{\second} intervals. Prior to loading, a reference image of the ROI was captured, and force logging (voltage-calibrated) was initiated to establish a baseline. To bend the beam, the actuator (HBM TRB25CC) was driven at \SI{25}{\mm\per\second} to reduce the span to a prescribed value (Fig.~\ref{fig:bending_global}A); a slight lateral perturbation was applied before actuation to ensure the bending direction. Once the target span was reached, the actuator was held fixed and an image of the ROI was captured immediately ($t=0^+$). During the subsequent relaxation, force logging continued and ROI images were acquired every $\SI{12}{\minute}$, with all imaging parameters held constant.

DIC analysis was performed using the open-source software Ncorr~\cite{Ncorr}. To isolate the time-dependent relaxation strain from the background bending field, the image acquired at $t=0^+$ was used as the reference. The subset radius was set between 20 and 30 pixels, with the subset spacing (step size) $\approx1/3$ of the chosen radius and a strain calculation window between 8 and 10.

%%%%%%%%%%%%%%%% SUPPLEMENTARY TEXT %%%%%%%%%%%%%%%
\newpage
\section*{Supplementary Text}
\section{Supplementary figures and tables}

% \subsection{Toluene extraction of residual uncrosslinked PDMS chains}\mbox{}\par

% \subsection{Mapping silicone-oil viscosity to molecular weight and additional tensile tests of PDMS samples}\mbox{}\par

\begin{figure}[H]
\centering
\includegraphics[width=0.7\linewidth]{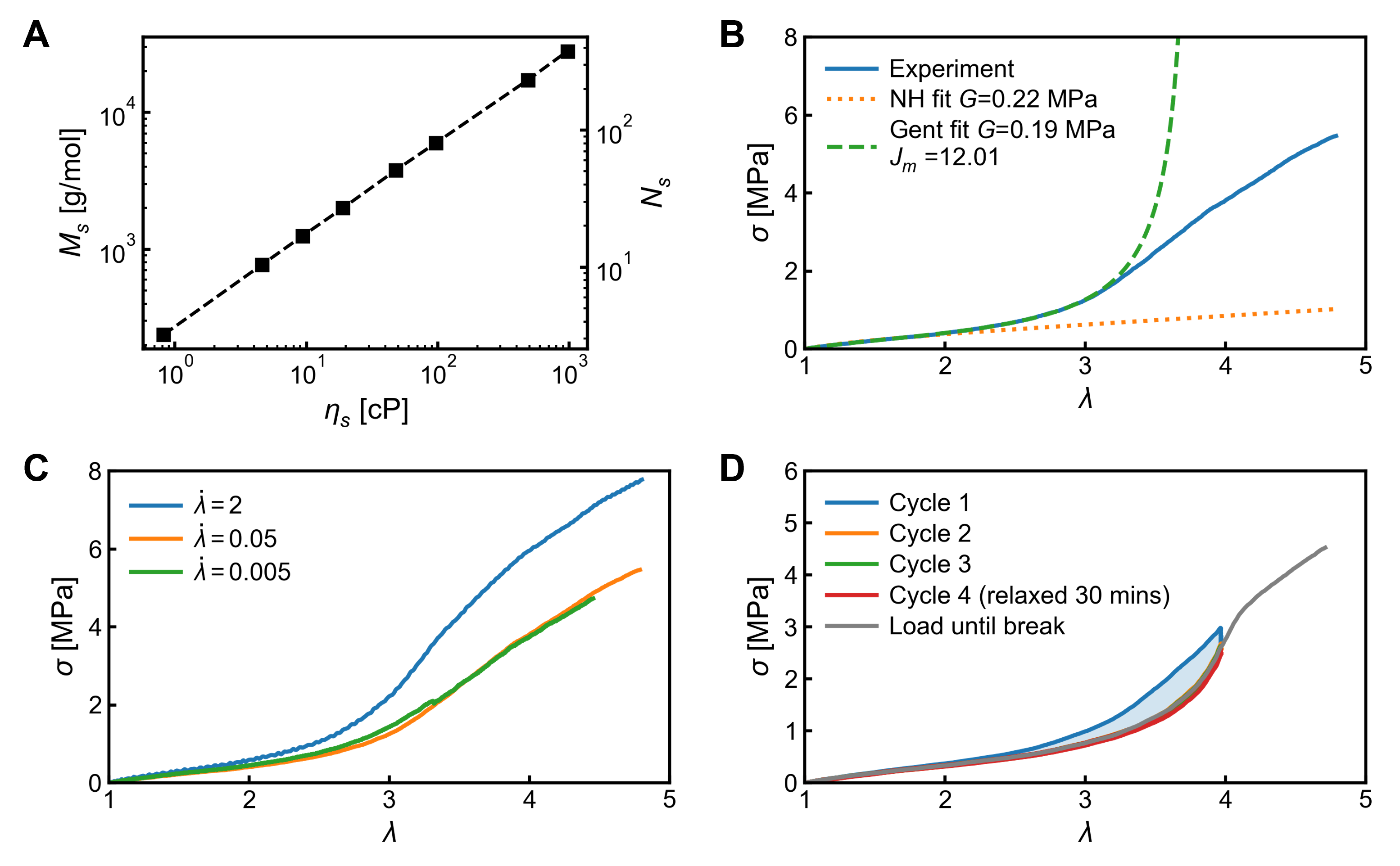}
\caption{(A) Mapping between silicone oil viscosity $\eta_s$ and molecular weight $M_s$, and chain length $N_s$. Data sourced from Gelest Technical Library~\cite{GelestSiliconeFluids}. (B) Uniaxial tensile stress--stretch curve of TT sample at $\dot{\lambda}=0.05~\mathrm{s^{-1}}$, with neo-Hookean (NH) fitting ($1\leq\lambda\leq2$) and Gent fitting ($1\leq\lambda\leq3$). (C) Nominal stress--stretch curves of a TT sample at different stretch rates. The rate used in the main text, $\dot{\lambda}=0.05~\mathrm{s^{-1}}$, collapses with the tenfold lower rate $\dot{\lambda}=0.005~\mathrm{s^{-1}}$, indicating negligible rate dependence at $\dot{\lambda}=0.05~\mathrm{s^{-1}}$.
(D) Cyclic loading--recovery--reloading of a TT sample at $\dot{\lambda}=0.05~\mathrm{s^{-1}}$. The sample was cycled three times to $\lambda_{\max}=4$, fully unloaded and allowed to recover for \SI{30}{\minute}, then reloaded to $\lambda_{\max}=4$ and finally to failure. The first cycle shows pronounced hysteresis (blue shading), while the reloading closely retraces the prior unloading with minimal additional dissipation. The persistence of stress softening relative to the virgin curve after recovery indicates network damages (or other irreversible processes), rather than viscoelastic relaxation, as the dominant source of hysteresis.}
\label{figS:tensile}
\end{figure}

% \newpage
% \subsection{Effect of non-swellable silica filler on the equilibrium swelling of Sylgard~184.}\mbox{}\par
\begin{figure}[H]
\centering
\includegraphics[width=0.5\linewidth]{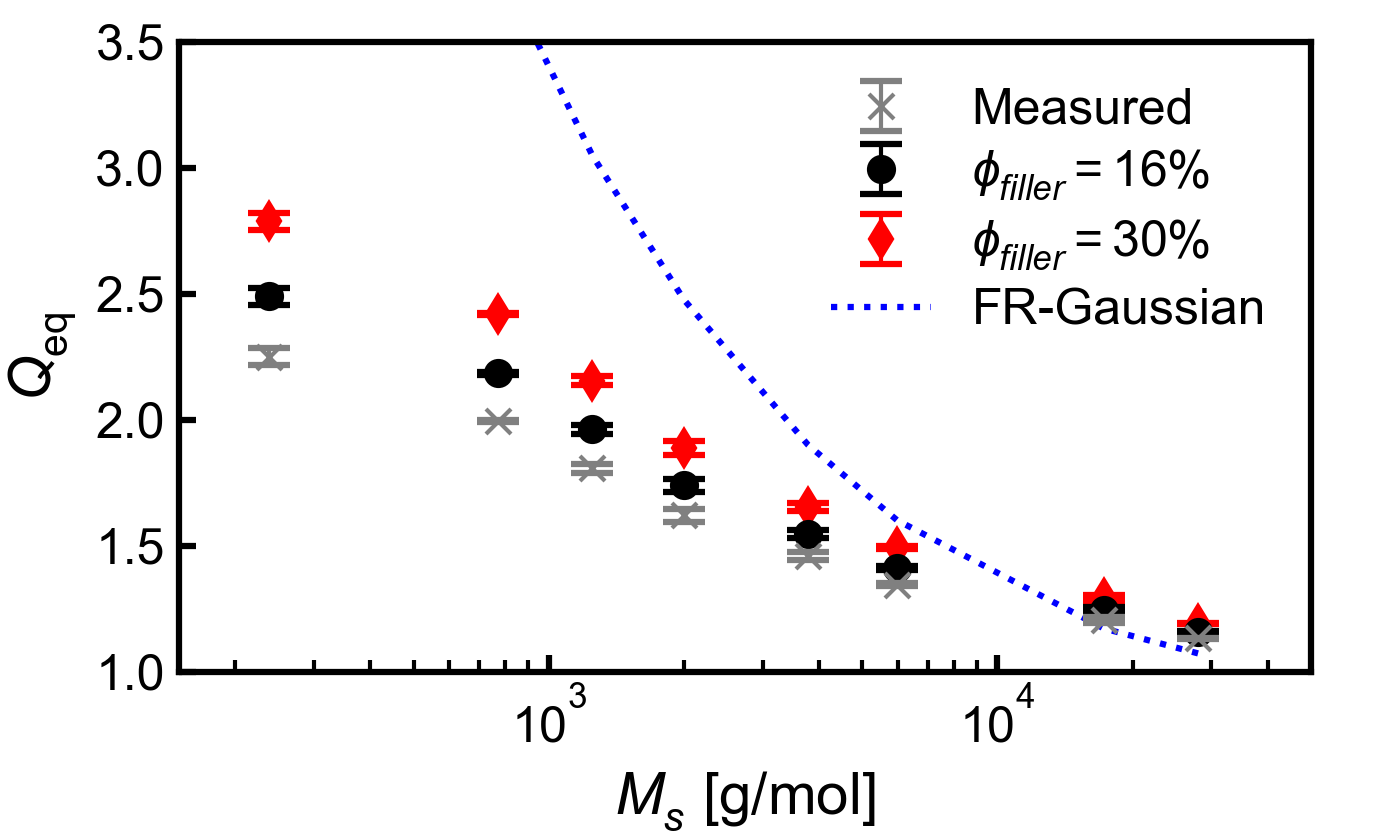}
\caption{Equilibrium volume swelling ratio $Q_{\mathrm{eq}}$ as a function of solvent molar mass $M_s$. Gray crosses are the measured swelling ratios computed using the total dry volume (polymer + filler). Black circles and red diamonds show the swelling ratios of the PDMS matrix corrected for a non-swellable filler volume fraction $\phi_{\mathrm{filler}}=16\%$ and $30\%$, respectively~\cite{clough2016covalent, suriboot2021amphiphilic}, via $Q_{\mathrm{corr}}=(Q_{\mathrm{meas}}-\phi_{\mathrm{filler}})/(1-\phi_{\mathrm{filler}})$. The blue dotted curve is the classic Flory--Rehner prediction as defined in the main text.
}
\label{figS:filler_correction}
\end{figure}
% \newpage

% \subsection{Dog-bone sample geometry for tensile tests}\mbox{}\par

% \newpage
% \subsection[Axial z-step correction]{Axial $z$-step correction for refractive-index mismatch}\mbox{}\par

% \newpage
% \subsection{Effect of boundary conditions on the fitted $D$ in free swelling of thin sheets}\mbox{}\par
\begin{figure}[H]
\centering
\includegraphics[width=0.8\linewidth]{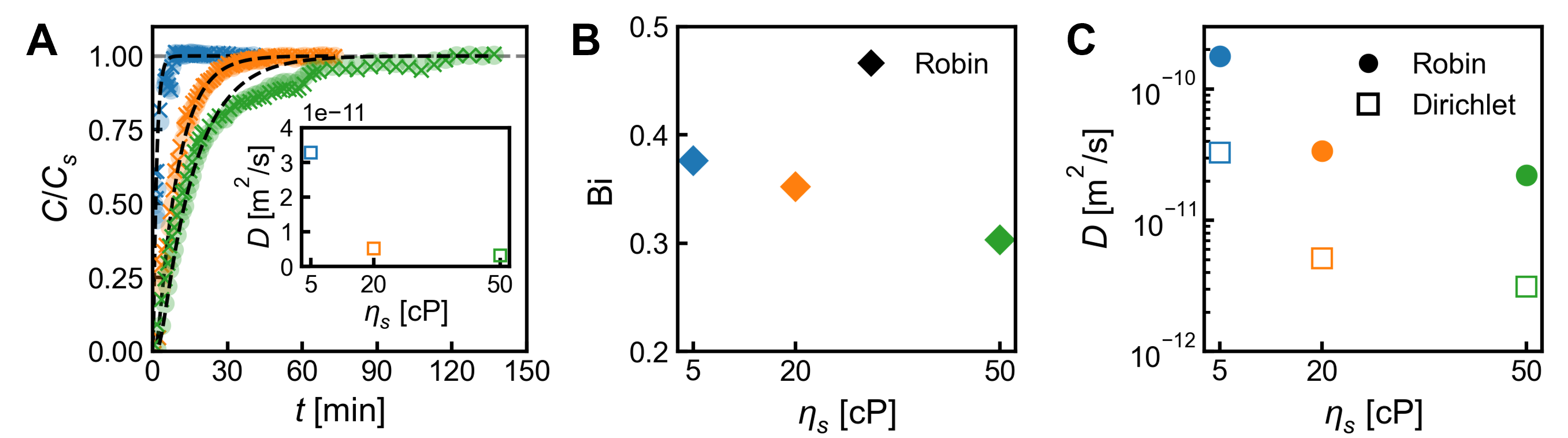}
\caption{Local free-swelling analysis under fully drained (Dirichlet) and limited-flux (Robin) boundary conditions.
(A) Normalized concentration $C/C_s$ extracted at mid-thickness (open cross) and near the surface (filled circle) as a function of time. Dashed curves show best fits of the Dirichlet finite-slab solution to the mid-thickness $C/C_s$ data; the inset summarizes the fitted diffusivity $D$ for each $\eta_s$.
(B) Biot number $\mathrm{Bi}$ obtained from fits of the Robin finite-slab solution (\eqref{eq:FS_robin_C_series}); the corresponding fitted $D$ values are reported in Fig.~\ref{fig:local-free-swelling}D in the main text.
(C) Comparison of $D$ inferred from Dirichlet and Robin fits. Accounting for limited interfacial flux increases the fitted $D$ by approximately a factor of five.
}
\label{figS:FS_sheet_Compare}
\end{figure}

% \newpage
% \subsection{Through-thickness $J$ profiles during free swelling of thin sheets}\mbox{}\par
\begin{figure}[H]
\centering
\includegraphics[width=0.8\linewidth]{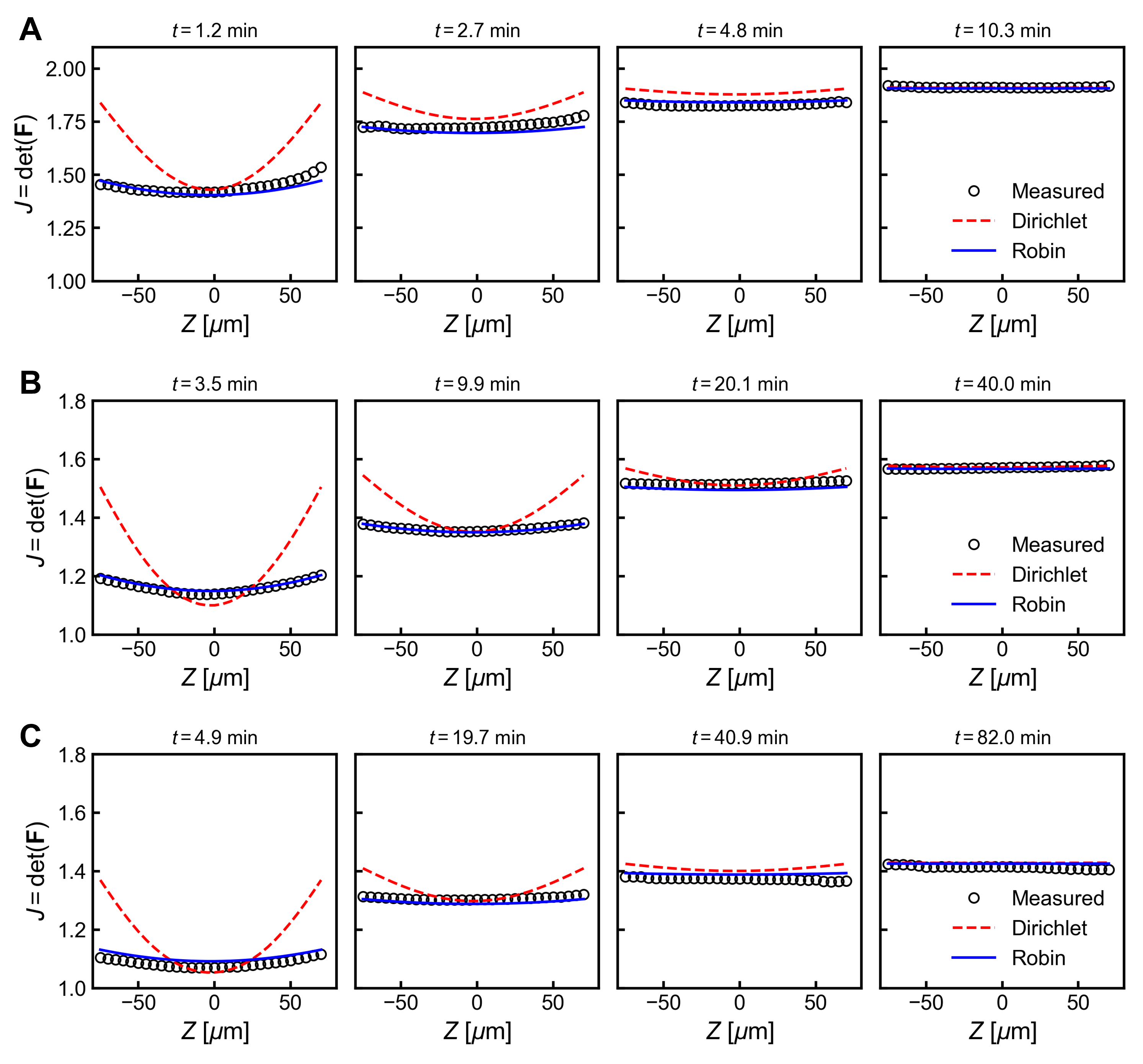}
\caption{
Through-thickness $J$ profiles during free swelling of thin sheets in silicone-oil baths of viscosity (A) 5~cP, (B) 20~cP, and (C) 50~cP. Open symbols show the measured volumetric swelling $J(z,t)=\det(\mathbf{F})$ at the indicated times. Curves compare two solutions of the diffusion equation: a fully drained (Dirichlet) boundary condition (\eqref{eq:S3:ICBCb}, dashed red) and a limited-flux (Robin) boundary condition (\eqref{eq:robin_bc_FS}, solid blue).
}
\label{figS:FS_sheet_BC}
\end{figure}

% \newpage
% \subsection{Diffusion through a freshly cut PDMS surface}\mbox{}\par
\begin{figure}[H]
\centering
\includegraphics[width=0.9\linewidth]{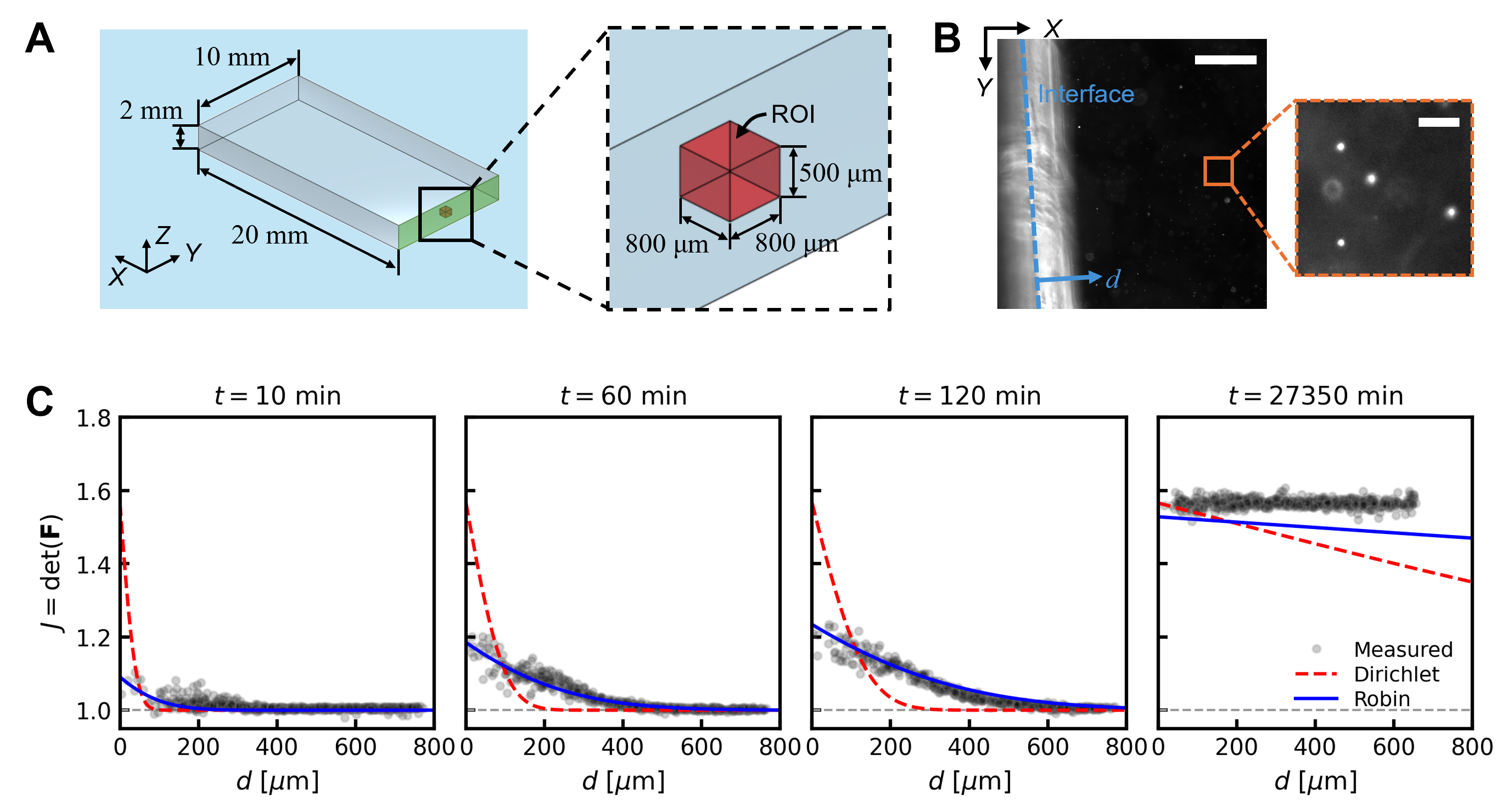}
\caption{Finite interfacial transfer is observed during diffusion through a freshly cut PDMS surface.
(A) A \SI{2}{mm}-thick long-strip TT PDMS sample was immersed in 50~cP silicone oil, and swelling was measured at a mid-thickness region near a freshly cut surface (light green). Using the same $Z$-scanner-equipped microscope as in the local free-swelling experiments, we focused on an $800 \times 800 \times 500~\mu\mathrm{m}^3$ region of interest (ROI) adjacent to this fresh surface, as shown in the inset (light red). This geometry probes diffusion through a surface that was not formed by casting against glass, and also provides an in-plane measurement direction with higher spatial resolution than the through-thickness $Z$ direction used in Fig.~\ref{fig:local-free-swelling}. 
(B) Representative slice from the acquired 3D image stack. The PDMS--oil interface is defined by the blue dashed line, and particle locations are referenced by the in-plane distance $d$ from this interface. The inset shows a contrast-adjusted, zoomed-in view of the tracer particles, which remain well resolved during imaging. By tracking the embedded tracer particles in 3D over time, we reconstruct the deformation gradient tensor $\mathbf{F}$ and quantify the local volumetric swelling as $J=\det(\mathbf{F})$. The scale bars are \SI{200}{\micro m} and \SI{20}{\micro m} in the inset. 
(C) Swelling profiles $J(d,t)$ calculated from the particle trajectories, compared with semi-infinite one-dimensional diffusion models with either Dirichlet or Robin boundary conditions. Both models were fitted to the same early-time data, $t \lesssim \SI{120}{min}$, for which the diffusion length remains smaller than the distance from the ROI to the upper and lower sample surfaces; over this time window, transport within the ROI can therefore be approximated as one-dimensional and driven primarily through the freshly cut surface. The Dirichlet condition does not capture the measured profiles and gives $D=7.86\times10^{-13}~\mathrm{m^2/s}$. The Robin condition, which allows finite interfacial transfer, provides a substantially better fit, with $D=1.14\times10^{-11}~\mathrm{m^2/s}$, $k=2.22\times10^{-8}~\mathrm{m/s}$, and $\mathrm{Bi}=1.56$ based on $L_\mathrm{ref}=800~\mu\mathrm{m}$. The observation of finite interfacial transfer at a freshly cut edge, consistent with the measurement in Fig.~\ref{fig:local-free-swelling}, indicates that contact with glass during casting is not responsible for the interfacial resistance. At long times, the ROI approaches equilibrium swelling, whereas the semi-infinite model does not capture this saturation, as expected for a finite sample geometry.
}
\label{figS:FS_Edge_Comparison}
\end{figure}

% \newpage
% \subsection{Comparison of numerical simulation and local measurement for free swelling}\mbox{}\par
\begin{figure}[H]
\centering
\includegraphics[width=0.9\linewidth]{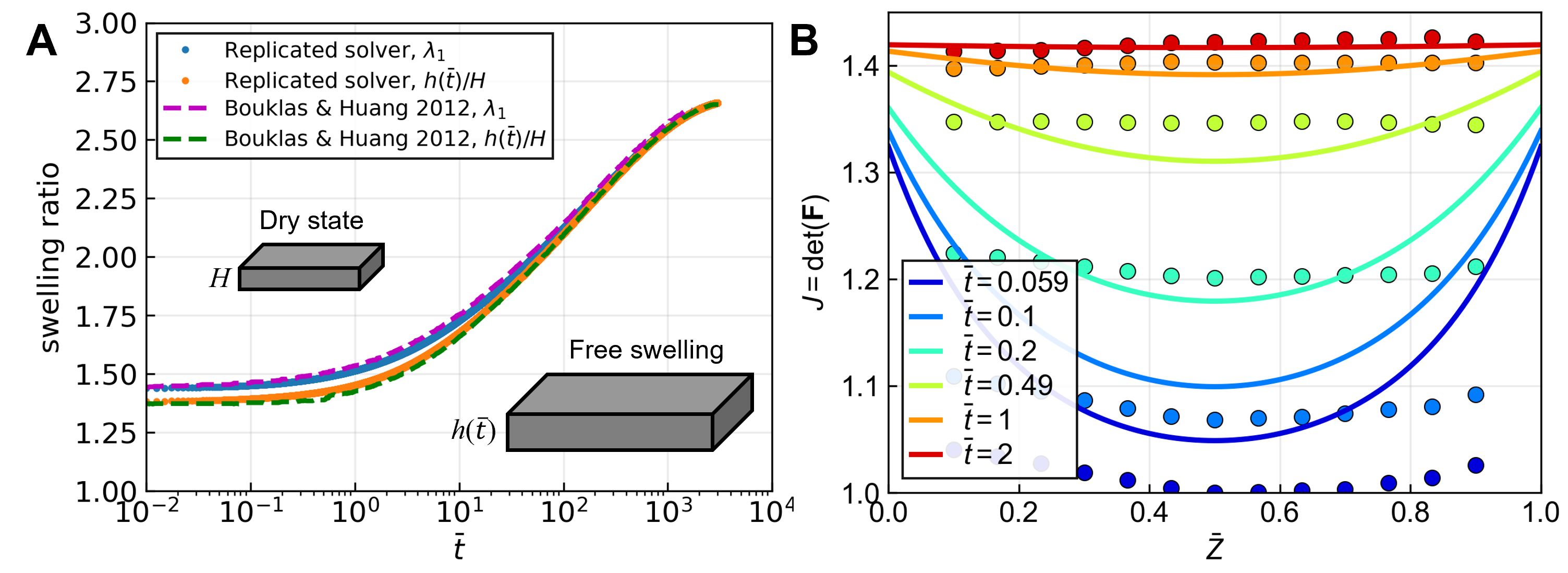}
\caption{Comparison between free-swelling measurements and numerical simulation with a Dirichlet chemical-potential boundary condition.
(A) The free-swelling solver was replicated from \cite{bouklas2012} and benchmarked against the reported solution for a thin sample swelling from the initial dry thickness $H$ to the transient thickness $h(\bar{t})$, where $\bar{t}=Dt/H^2$ is the dimensionless time, with $D$ the diffusivity and $t$ the physical time. The in-plane swelling ratio, $\lambda_1$, and through-thickness swelling ratio, $h(\bar{t})/H$, agree closely with the reference results over the full swelling process.
(B) The solver was then applied to free swelling of a TT sample in 50~cP silicone oil. The effective interaction parameter, $\chi_\mathrm{eff}$, was determined from the final equilibrium swelling ratio using the classical Flory--Rehner relation. The dimensionless network parameter was calculated as $N\Omega = G\Omega/(RT)$, where $N$ is the molar density of elastically active network chains, $\Omega$ is the solvent molar volume, $G$ is the measured shear modulus, $R$ is the gas constant, and $T$ is the absolute temperature. A small non-zero initial swelling ratio of 1.001 was used to avoid the singularity associated with the dry reference state. Fitting the full spatiotemporal swelling measurement in Fig.~\ref{fig:local-free-swelling} to the numerical solution gives $D=5.26\times10^{-12}~\mathrm{m^2\,s^{-1}}$. However, the predicted through-thickness profiles of $J=\det(\mathbf{F})$ as a function of normalized thickness coordinate $\bar{Z}=Z/H$ deviate substantially from the measurements, with the simulation substantially over-predicting swelling near the free surfaces. This discrepancy indicates that a prescribed chemical-potential boundary condition alone does not capture the transient diffusion process, and suggests that interfacial resistance, represented by a Robin boundary condition, should be considered at these interfaces.
}
\label{figS:FS_Numerical}
\end{figure}

% \newpage
% \subsection{Revised $D$ from bulk mass uptake accounting for limited interfacial flux}\mbox{}\par
\begin{figure}[H]
\centering
\includegraphics[width=0.69\linewidth]{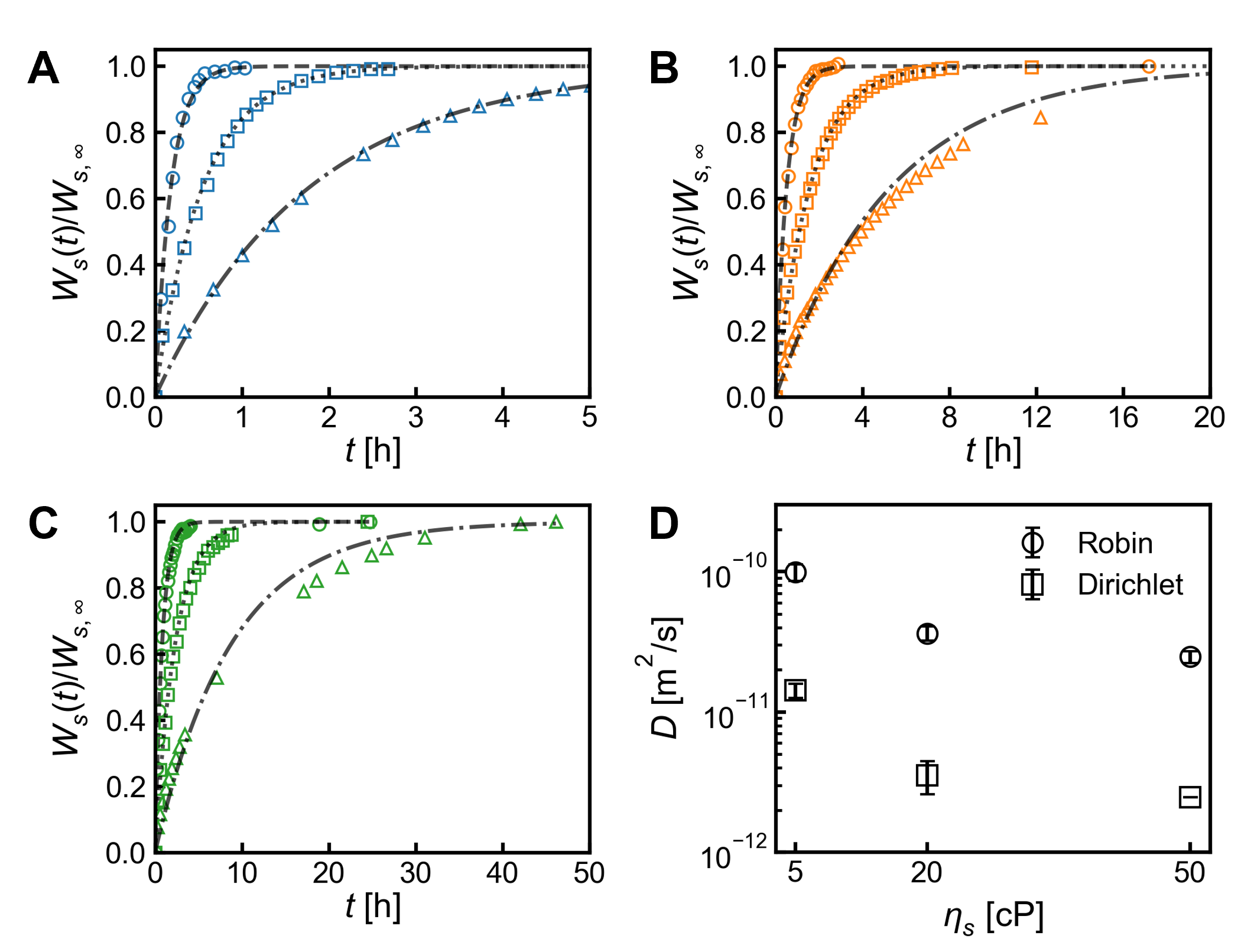}
\caption{Mass-uptake kinetics in free swelling and refits with limited interfacial flux. Normalized mass uptake $W_s(t)/W_{s,\infty}$ for three initial thicknesses $H_0=$ 250~$\mu$m (circles), 500~$\mu$m (squares), and 1000~$\mu$m (triangles) in silicone oils of viscosity (A) 5~cP, (B) 20~cP, and (C) 50~cP. Curves show best fits of the finite-slab diffusion solution with Robin boundary conditions at both surfaces (\eqref{eq:FS_robin_W_series}); the Biot number $\mathrm{Bi}$ is fixed to the value inferred from the local measurements (Fig.~\ref{figS:FS_sheet_Compare}B), and only $D$ is fitted. Line styles correspond to thickness (dashed: 250~$\mu$m; dotted: 500~$\mu$m; dash-dotted: 1000~$\mu$m). (D) $D$ calculated from the same mass-uptake data using Robin fits (circles) and Dirichlet fits (squares); points denote the mean over the three thicknesses and error bars indicate the standard deviation. Switching from Dirichlet to Robin boundary conditions increases the inferred $D$ by approximately one order of magnitude.
}
\label{figS:FS_bulk_Compare}
\end{figure}

% \newpage
% \subsection{Curvatures for prescribed buckling spans and brief viscoelastic relaxation in AP and TT beams}\mbox{}\par
\begin{figure}[H]
\centering
\includegraphics[width=0.69\linewidth]{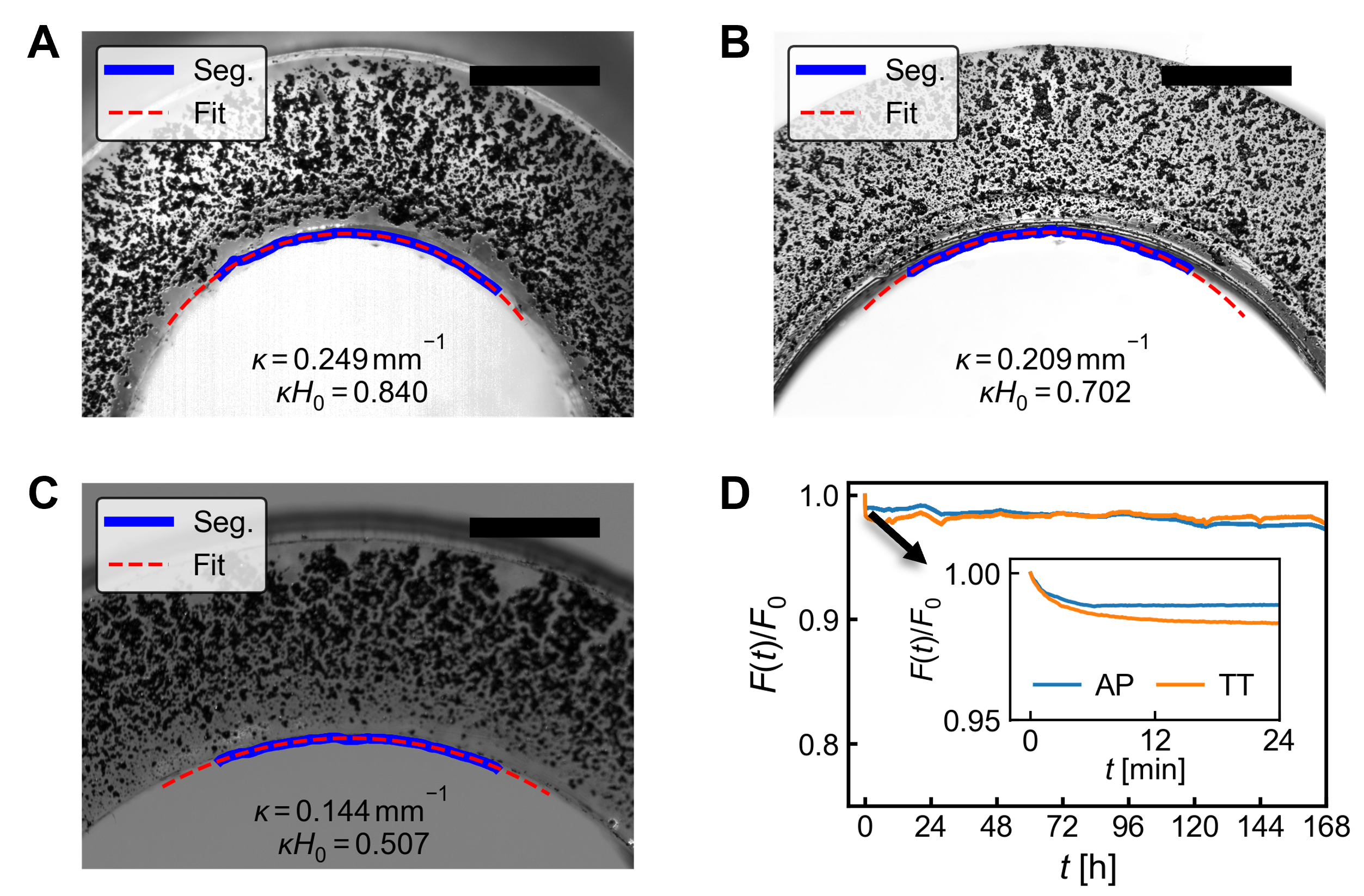}
\caption{Curvatures of PDMS beams bent to prescribed spans: (A) $\SI{15}{\mm}$, (B) $\SI{20}{\mm}$, and (C) $\SI{25}{\mm}$. Blue curves: segmented inner-arc profiles; red dashed curves: circular-arc fits used to extract the curvature $\kappa$ and the thickness-normalized curvature $\kappa H_0$ ($H_0$ is the beam thickness), as annotated. The scale bars are \SI{2}{\mm}. (D) Normalized force relaxation $F/F_0$ for AP and TT beams. Inset: zoom-in view of the first \SI{24}{\minute}, highlighting a fast viscoelastic decay that is orders of magnitude shorter than typical poroelastic relaxation timescales for FS beams.}
\label{figS:bending_global}
\end{figure}

% \newpage
% \subsection{Literature diffusivities used for comparison}\mbox{}\par
\begin{table}[H]
\centering
\begin{threeparttable}
\caption{Literature diffusivities $D$ of solvents in PDMS and related silicone-rubber networks used for comparison in Fig.~\ref{fig:all_D}.}
\label{tabS:D_values}
\begin{tabularx}{\textwidth}{@{} Y Y r r c @{}}
\toprule
Polymer network & Solvent & {$\eta_s$ [cP]} & {$D$ [\si{\meter\squared\per\second}]} & Reference \\
\midrule
PDMS 10:1 (Essex Brownell) & Pentane      & $0.23$   & $\left(5.9 \pm 0.1\right)\times 10^{-9}$    & \cite{hu2011indentPDMS} \\
PDMS 10:1 (Essex Brownell) & Heptane      & $0.386$  & $\left(2.3 \pm 0.03\right)\times 10^{-9}$   & \cite{hu2011indentPDMS} \\
PDMS 10:1 (Essex Brownell) & Decane       & $0.92$   & $\left(1.1 \pm 0.08\right)\times 10^{-9}$   & \cite{hu2011indentPDMS} \\
PDMS 10:1 (Essex Brownell) & Cyclohexane  & $0.9$    & $\left(1.2 \pm 0.03\right)\times 10^{-9}$   & \cite{hu2011indentPDMS} \\
% \addlinespace
PDMS & Heptane      & $0.386$  & $\left(3.1 \pm 0.2\right)\times 10^{-9}$   & \cite{hu2012thinlayer} \\
% \addlinespace
PDMS 10:1 (Sylgard~184) & Silicone oil (V100) & $96.0$   & $\left(1.0 \pm 0.1\right)\times 10^{-11}$  & \cite{microchannel} \\
PDMS 10:1 (Sylgard~184) & Hexadecane          & $3.03$   & $\left(4.4 \pm 0.4\right)\times 10^{-11}$  & \cite{microchannel} \\
% \addlinespace
Silicone rubber & D4                  & $2.20$   & $3.2\times 10^{-11}$  & \cite{uptake2001} \\
Silicone rubber & Silicone oil        & $48.0$   & $1.76\times 10^{-12}$ & \cite{uptake2001} \\
% \addlinespace
Filler-free PDMS (Dehesive 994) & Hexane        & $0.295$  & $9.0\times 10^{-11}$  & \cite{uptake2015} \\
Filler-free PDMS (Dehesive 994) & Dodecane      & $1.36$   & $1.0\times 10^{-11}$  & \cite{uptake2015} \\
Filler-free PDMS (Dehesive 994) & Hexadecane    & $3.03$   & $2.0\times 10^{-11}$  & \cite{uptake2015} \\
Filler-free PDMS (Dehesive 994) & Acetone       & $0.302$  & $8.5\times 10^{-11}$  & \cite{uptake2015} \\
Filler-free PDMS (Dehesive 994) & Water         & $0.890$  & $5.5\times 10^{-11}$  & \cite{uptake2015} \\
Filler-free PDMS (Dehesive 994) & Ethyl acetate & $0.455$  & $3.0\times 10^{-11}$  & \cite{uptake2015} \\
Filler-free PDMS (Dehesive 994) & Ethanol       & $1.074$  & $6.0\times 10^{-11}$  & \cite{uptake2015} \\
% \addlinespace
PDMS & Methyl ethyl ketone (MEK) & $0.4$ & $\left(2.2 \pm 0.2\right)\times 10^{-9}$  & \cite{fierro2011experimental} \\
PDMS & Diethyl ketone (DEK)    & $0.48$  & $\left(2.0 \pm 0.2\right)\times 10^{-9}$  & \cite{fierro2011experimental} \\
PDMS & Tetraethylene glycol dimethyl ether (TEGDME) & $2.5$  & $\left(2.7 \pm 0.3\right)\times 10^{-10}$ & \cite{fierro2011experimental} \\
PDMS & Poly(ethylene glycol) dimethyl ether (PEGDME) & $6.0$  & $\left(1.3 \pm 0.1\right)\times 10^{-10}$ & \cite{fierro2011experimental} \\
% \addlinespace
PDMS 10:1 (Sylgard~184) & Water  & $0.890$ & $8.5\times 10^{-10}$ & \cite{randall2005PNAS} \\
% \addlinespace
RTV silicone rubber (Polycraft GP3481-F) & Uncrosslinked base & $18000$ & $\left(1.8 \pm 0.0\right)\times 10^{-6}$ & \cite{PNAS_Vikram} \\
\bottomrule
\end{tabularx}
\begin{tablenotes}[flushleft]
\footnotesize
\item \textit{Note. Some of the $\eta_s$ values are taken from Wikipedia.}
\end{tablenotes}

\end{threeparttable}
\end{table}

\FloatBarrier
\section{Linear poroelastic diffusion during free swelling: analytical solutions}\label{secS:FS_solution}

Based on linear poroelastic theoretical development for polymeric gels~\cite{yoon2010, bouklas2012}, the coupled transport–deformation problem reduces to a Fickian diffusion equation with an effective poroelastic diffusivity $D$:
\begin{equation}
\partial_t C = D \nabla^{2} C,
\label{eq:diff_eq}
\end{equation}
where $C(\mathbf{x},t)$ is the local solvent concentration, $\partial_t$ the time derivative, and $\nabla^2$ the spatial Laplacian. This equation is developed in the linear poroelastic regime based on an isotropic network with homogeneous reference state, individually incompressible phases, and a Newtonian solvent obeying Darcy flow.

Note that a rigorous poroelastic treatment of free swelling (e.g., in~\cite{yoon2010}) shows that even a freestanding layer does not admit an exactly self-similar concentration profile. Because the main focus of this work is the solvent diffusion under external load, in this section, we nevertheless adopt the classical Fickian slab solution as a phenomenological description of swelling kinetics. This formulation is widely used in the literature~\cite{crank1979mathematics, uptake2015}, and is adequate to represent our experimentally measured diffusion kinetics. The fitted $D$ should therefore be interpreted as an effective poroelastic diffusivity that captures the dominant relaxation timescale around the free-swollen state. On this basis, we formulate the problem and boundary conditions directly in terms of $C$, rather than the more rigorous but less accessible chemical potential $\mu$.

\subsection{Free swelling of a thin sheet}
\label{secS:FS_Bulk}

We consider the diffusion of solvent from an infinitely large bath into a stress-free thin sheet of dry polymer with thickness $H_0$. By assuming the in-plane size is greatly larger than $H_0$, the solvent transport is predominantly one-dimensional through the thickness direction $z$, and is governed by the free swelling diffusion equation developed in \eqref{eq:diff_eq}. Let $z\in[-H_0/2,H_0/2]$ and $C(Z,t)$ be the solvent concentration.

\subsubsection{With fully drained boundary conditions}
The initial condition and boundary conditions (BCs) of this problem are classically set as:
\begin{subequations}\label{eq:S3:ICBC}
\begin{align}
C(Z,0) &= 0, \label{eq:S3:ICBCa}\\
C\!\left(\pm\frac{H_0}{2},t\right) &= C_s. \label{eq:S3:ICBCb}
\end{align}
\end{subequations}
where $C_s$ is the equilibrium (bath-imposed) concentration in the polymer at the surfaces. This classical boundary-value problem has the separated-variables solution~\cite{crank1979mathematics},
% Page 47, around equation 4.17
\begin{equation}
\frac{C(Z,t)}{C_s}=
1-\frac{4}{\pi}\sum_{n=0}^{\infty}
\frac{(-1)^n}{2n+1} 
\exp\!\left[- \frac{(2n+1)^2\pi^2 D t}{H_0^2}\right]\cos\!\big(\frac{(2n+1)\pi Z}{H_0}\big).
\label{eq:S5:C_Cs}
\end{equation}
From 3D particle tracking we obtain the local volume change $J=\det\mathbf F$, which is related to $C$ by incompressibility of polymer and solvent for a dry reference state, $J=1+\Omega C$,
% \label{eq:incompressible}
with solvent molar volume $\Omega$. Therefore, at a fixed depth $Z=\bar{Z}$, the normalized concentration is $C(\bar{Z},t)/C_s = (J(\bar{Z},t)-1)/(J_s-1)$, where $J_s=\lim_{t\to\infty}J(\bar{Z},t)$ is the fully swollen value. We estimate $D$ by nonlinear least-squares fits of \eqref{eq:S5:C_Cs}; in Fig.~\ref{figS:FS_sheet_Compare}A, we report the fit at the mid-thickness ($Z=0$), retaining the first four terms of the series ($n=0,\ldots,3$).

Integrating \eqref{eq:S5:C_Cs} across the thickness gives the classical series for the normalized uptake~\cite{uptake2015, crank1979mathematics, uptake2001},
\begin{equation}
\frac{W(t)}{W_\infty}
=1-\frac{8}{\pi^2}\sum_{n=0}^{\infty}\frac{1}{(2n+1)^2}
\exp\!\left[- \frac{(2n+1)^2\pi^2 D t}{H_0^2}\right],
\label{eq:S3:Mt_series}
\end{equation}
where $W(t)$ is the weight of the sample at time $t$, and $W_\infty$ is the equilibrium value.
For early times $(t\to 0^+)$, this reduces to, % ierfc(x) --> 0 when x--> inf
\begin{equation}
\frac{\Delta W(t)}{\Delta W_\infty}=\frac{N(t)}{N_\infty}
\approx \frac{4}{\sqrt{\pi}}\sqrt{\frac{D t}{H_0^2}},
\label{eq:S3:mass_fraction}
\end{equation}
and the early-time behavior follows a $\sqrt{t}$ law~\cite{crank1979mathematics, microchannel, uptake2001}. By fitting $\Delta W(t)/\Delta W_\infty$ to $\sqrt{t}/H_0$ with a linear function $\frac{\Delta W(t)}{\Delta W_\infty}\approx S \frac{\sqrt{t}}{H_0}$, the slope $S$ gives a direct estimate of $D = \pi S^2/16$; 
in the main text, we perform this linear fit within the early time $\sqrt{t}/H_0<\SI{0.1}{\second^{\frac{1}{2}}\micro m^{-1}}$.

\subsubsection{With limited-flux boundary conditions}
The fit to the solution with fully drained (Dirichlet) BCs largely overestimate the solvent concentration near the boundary, especially at early times, as shown in Fig.~\ref{figS:FS_sheet_BC}. This discrepancy indicates that solvent uptake is not instantaneous at the solvent--polymer interface; We therefore model the interface as a finite mass-transfer process and replace the Dirichlet BC by a Robin (mixed) BC,
\begin{equation}
-D\,\left.\frac{\partial C}{\partial n}\right|_{Z=\pm \frac{H_0}{2}}
= \kappa\Big(C_s - C(\pm \tfrac{H_0}{2},t)\Big),
\label{eq:robin_bc_FS}
\end{equation}
where $\kappa$ is an effective interfacial transfer coefficient and $n$ is the outward normal. Introducing the half-thickness $L=H_0/2$, the governing dimensionless group is the Biot number
\begin{equation}
\mathrm{Bi}=\frac{\kappa L}{D}.
\label{eq:biot}
\end{equation}
In the limit $\mathrm{Bi}\to\infty$, interfacial resistance vanishes and \eqref{eq:robin_bc_FS} reduces to the Dirichlet BC, $C(\pm L,t)\to C_s$, recovering \eqref{eq:S5:C_Cs}. Conversely, $\mathrm{Bi}\to 0$ approaches a no-flux boundary (Neumann) and uptake is strongly rate-limited by the interface.

% \paragraph{Series solution.}
Let $\theta(Z,t)=1-C(Z,t)/C_s$. The diffusion equation~\eqref{eq:diff_eq} with the initial condition~\eqref{eq:S3:ICBCa} and the Robin BC~\eqref{eq:robin_bc_FS} forms a standard Sturm--Liouville eigenvalue problem, which can be solved by separation of variables in textbook~\cite{strauss2007partial,hahn2012heat} as:
\begin{equation}
\frac{C(Z,t)}{C_s}
=
1-\sum_{n=1}^{\infty}A_n
\cos\!\Big(\mu_n\frac{|Z|}{L}\Big)\,
\exp\!\Big(-\mu_n^2\frac{Dt}{L^2}\Big),
\label{eq:FS_robin_C_series}
\end{equation}
where the dimensionless eigenvalues $\mu_n$ are the positive roots of $\mu_n\tan\mu_n=\mathrm{Bi}, n=1,2,\ldots,$ and the corresponding coefficients are $A_n=\dfrac{4\sin\mu_n}{2\mu_n+\sin(2\mu_n)}$.
In practice we truncate the series in \eqref{eq:FS_robin_C_series} to $N$ terms (here $N=5$), which is sufficient for the times and spatial resolution considered.
% \paragraph{Revise bulk uptake.}
The normalized uptake follows by averaging \eqref{eq:FS_robin_C_series} across the thickness:
\begin{equation}
\frac{W(t)}{W_\infty}
=\frac{1}{H_0}\int_{-L}^{L}\frac{C(Z,t)}{C_s}\,\mathrm{d}Z
=1-\sum_{n=1}^{\infty}B_n
\exp\!\Big(-\mu_n^2\frac{Dt}{L^2}\Big),
\label{eq:FS_robin_W_series}
\end{equation}
with $B_n=\dfrac{A_n}{\mu_n}\sin\mu_n$.

From 3D particle tracking we obtain $J(Z,t)=\det\mathbf F$, and use incompressibility $J=1+\Omega C$ to form the normalized concentration $C(Z,t)/C_s = \big(J(Z,t)-1\big)/\big(J_s-1\big)$, where $J_s=\lim_{t\to\infty}J(Z,t)$. For each silicone-oil viscosity, we fit the spatiotemporal dataset $C(Z,t)/C_s$ to \eqref{eq:FS_robin_C_series} by nonlinear least squares, treating $(D,\mathrm{Bi})$ as free parameters (equivalently, $(D,k)$). 
Across all viscosities we obtain $\mathrm{Bi}\approx 0.3$--$0.4$, indicating a finite interfacial transfer resistance and thus a flux-limited boundary; the fitted $\mathrm{Bi}$ and $D$ are summarized in Fig.~\ref{figS:FS_sheet_Compare}B and Fig.~\ref{fig:local-free-swelling}D, respectively. Under this limited-flux condition, the strong near-surface gradients predicted by the Dirichlet model are substantially reduced, and the resulting profiles agree with the measured $C(Z,t)/C_s$ at early times (Fig.~\ref{figS:FS_sheet_BC}). A comparison of the fitted $D$ from the Dirichlet and Robin analyses is plotted in Fig.~\ref{figS:FS_sheet_Compare}C. Finally, we reassess the bulk uptake by fitting the Robin uptake series \eqref{eq:FS_robin_W_series}, using $\mathrm{Bi}$ extracted from the local measurements (Fig.~\ref{figS:FS_sheet_Compare}B); the corresponding fits and the comparison of fitted $D$ are reported in Fig.~\ref{figS:FS_bulk_Compare}.

\subsection{Free swelling of a semi-infinite strip}
\label{secS:FS_strip}

We model the experiment as solvent uptake into a dry, stress-free strip along its length $y\ge 0$, with the end face at $y=0$ suddenly brought into contact with a large bath at $t=0^+$. Transport is one-dimensional along $y$ and governed by the free-swelling diffusion equation \eqref{eq:diff_eq} with a constant effective diffusivity $D$. The motion of the strip due to swelling is for simplicity. The initial and boundary condition in this experimental configuration can be expressed as 
\begin{equation}
\begin{aligned}
C(y,0) &= 0, \\
C(0,t>0) &= C_s, \\
\lim_{y\to\infty} C(y,t) &= 0,
\end{aligned}
\label{eq:S4:PDE}
\end{equation}
where $C_s$ is the bath-imposed surface concentration, which also equals the long-time equilibrium value at the contact surface. By the usual similarity reduction with $y/(2\sqrt{Dt})$, the concentration field is solved in~\cite{crank1979mathematics} as 
% Eq(3.13) Crank solved as (C-Cs)/(0-Cs) = erf,  here we used erfc which = 1-erf.
\begin{equation}
\frac{C(y,t)}{C_s} = \operatorname{erfc}\!\left(\frac{y}{2\sqrt{Dt}}\right).
\label{eq:S4:similarity}
\end{equation}
where $\operatorname{erfc}(\cdot)$ denotes the complementary error function. At a given concentration $C=C_\alpha=\alpha C_s$, the iso-concentration fronts $\{y_\alpha(t):  C=C_\alpha=\alpha C_s\}$ satisfies
$y_\alpha(t) = 2\sqrt{D}\operatorname{erfc}^{-1}(\alpha) \sqrt{t}$. 
By a linear fit of $Y_\alpha(t)$ versus $\sqrt{t}$ with slope $S_\alpha$, we obtain
\begin{equation}
D = \frac{S_\alpha^{ 2}}{ 4 [\operatorname{erfc}^{-1}(\alpha)]^{2}}.
\label{eq:S4:D_from_slope}
\end{equation}
In the following, we will provide two approaches to obtain the iso-concentration fronts $y_\alpha(t)$ which are necessary for the fit.

\subsubsection{Iso-concentration fronts from width measurements}\label{secS:Width_Based}
In the free-swelling experiment, a toluene-treated PDMS strip is hung with one end immersed in a large silicone-oil bath. Assuming isotropic swelling, the local volumetric change is $J(y,t)=\lambda^3(y,t)$, where $\lambda(y,t)$ is the in-plane linear swelling ratio obtained from the apparent width change by edge detection. Based on the assumption of incompressible solvent and polymer molecules, the concentration field is estimated as a function of $\lambda(y,t)$ as $C(y,t)= (J(y,t)-1)/\Omega = (\lambda^3(y,t)-1)/\Omega$. In the fully swollen state, $J_s=1+\Omega C_s=\lambda_s^3$, so the normalized concentration is
\begin{equation}
\alpha(y,t)\;=\;\frac{C(y,t)}{C_s}
= \frac{\lambda^3(y,t)-1}{\lambda_s^3-1}.
\label{eq:S4:lmd_C_Cs}
\end{equation}
At a given $\alpha$, the time-resolved iso-concentration fronts $y_\alpha(t)$ can therefore be retrieved.

\subsubsection{Intensity-based iso-concentration fronts via Beer--Lambert law with thickness correction}\label{secS:Beer_Lambert}
During this free swelling experiment, the silicone oil is labeled with an oil-based dye that co-migrates with the solvent; within the linear Beer–Lambert regime we assume $C_d(y,t)=\beta C(y,t)$ with constant proportionality $\beta$ between the concentration of the dye $C_d$ and the silicone oil $C$. We image in transmission through the sample thickness: before immersion a reference image provides the incident intensity field $I_0(y)$; after equilibration the bath is removed and excess surface solvent is gently blotted to obtain $I_s(y)$ in the fully swollen state.

The Beer--Lambert law~\cite{BeerLambert} gives
\begin{equation}
I(y,t)=I_0(y) e^{- \varepsilon C_d(y,t) l(y,t)},
\label{eq:S4:BL_I}
\end{equation}
where $\varepsilon$ is the dye molar absorptivity and $l(y,t)$ is the optical path length, which is corrected by the linear swelling ratio $\lambda(y,t)$ determined in Sec.~\ref{secS:Width_Based}, $l(y,t)=\lambda(y,t)l_0$ with the initial sample thickness $l_0$. Solving for $C$ with \eqref{eq:S4:BL_I},
\begin{equation}
C(y,t)=- \frac{1}{\varepsilon \beta l_0} \frac{1}{\lambda(y,t)} 
\ln\!\frac{I(y,t)}{I_0(y)}.
\label{eq:S4:BL_C_corrected}
\end{equation}
In the fully swollen state, we estimate $C_s= \langle C(y, t\to\infty) \rangle_y$ as an average along $y$ over the saturated region (where strip was below the bath surface). Dividing \eqref{eq:S4:BL_C_corrected} with $C_s$ eliminates the unknown optical constants and yields the normalized concentration:
\begin{equation}
\alpha(y,t)\;=\;\frac{C(y,t)}{C_s}
=
\frac{\displaystyle \frac{1}{\lambda(y,t)} \ln\!\frac{I_0(y)}{I(y,t)}}
{\displaystyle \Big\langle \frac{1}{\lambda_s(y)} \ln\!\frac{I_0(y)}{I_s(y)} \Big\rangle_y},
\label{eq:S4:BL_C_Cs}
\end{equation}
with all right-hand-side quantities measured from images and $\lambda(y,t)$ from Sec.~\ref{secS:Width_Based}. With the $\alpha(y,t)$ field, iso-concentration front can be resolved with time.

% \newpage
\section{Linear poroelastic diffusion in a fully swollen bent beam}
\label{secS:beam_solution}

We consider a fully-swollen, stress-free beam that is rapidly bent at $t=0$ to a prescribed curvature and then held. This initial swollen configuration serves as the reference. Let $(x,y,z)$ be a beam-fixed orthonormal basis with $x$ along the span $L_0$, $y$ through the thickness $H_0$, and $z$ across the height $B_0$. Under bending, mechanical fields and chemical potential vary predominantly along the thickness direction; accordingly, we model transport as one-dimensional along $y\in[0,H_0]$ at a midspan slice $x=L_0/2$.

We measure the local volumetric change, $J(y,t)=\det\mathbf{F}(y,t)$, relative to this swollen reference, and define the solvent concentration $C(y,t)$ as the moles of solvent per unit reference volume of the swollen beam, with homogeneous initial value $C_0$. Both the polymer network and the solvent are assumed incompressible, so any local volume change arises solely from solvent uptake or release~\cite{yoon2010, bouklas2012, hong2008theory}:
\begin{equation}
J - 1 = \Omega(C(y,t)-C_0).
\label{eq:S6:C_J}
\end{equation}
where $\Omega$ is the solvent molar volume and $J$ is measured from particle tracking \emph{with the free-swollen state as the reference}.

Following Biot's linear poroelastic framework as specialized to gels~\cite{hong2008theory, yoon2010, bouklas2012}, the local free energy density is taken quadratic in the deformation gradient $\mathbf{F}$ with an additional term coupling to solvent concentration $C$, as $W(\mathbf{F}, C)$. Accordingly, the chemical potential of solvent $\mu_s$, defined as $\mu_s = \partial W/\partial C$, is a function of both $C$ and $\mathbf{F}$.

\paragraph{Free-swollen reference state and rapid bending}
Before bending, the beam is fully swollen and stress-free in a homogeneous reference state,
\begin{equation}
C(y,t<0)=C_0,
\qquad \sigma_{ij}=0,
\qquad \mu_s\big(y,t<0\big)=\mu_s\big(C_0,\mathbf{F}=\mathbf{I}\big) = \mu_{\rm ext},
\label{eq:S6:ref_state}
\end{equation}
with $\mu_{\rm ext}$ the external chemical potential (i.e. that of the surrounding silicone oil in the free-swelling bath).

At $t=0$, we impose a rapid bending step: the displacement and deformation gradient jump to a bent configuration $\mathbf{F}^{\rm bend}(y)$ corresponding to the prescribed curvature, but there is no time for solvent migration. Thus, at $t=0^+$,
\begin{equation}
C(y,0^+) = C_0,\qquad \mathbf{F}(y,0^+) = \mathbf{F}^{\rm bend}(y).
\end{equation}
The bending-induced elastic deformation changes the elastic contribution to $\mu_s$, and therefore, $\mu_s$ becomes non-uniform:
\begin{equation}
\mu_s(y,0^+) 
= \mu_s\big(C(y,0^+), \mathbf{F}(y,0^+) \big)
= \mu_s\big(C_0,\mathbf{F}^{\rm bend}(y)\big).
\end{equation}
This inhomogeneous $\mu_{\rm mech}(y):=\mu_s(y,0^+)$ drives solvent migration for $t>0$.

% \paragraph{Linear poroelastic background and choice of variables.}

During the poroelastic solvent migration, for small incremental changes about a swollen, bent reference state $(C_0,\mathbf{F}^{\rm bend}(y))$, we expand $\mu_s$ to first order as
\begin{equation}
\mu_s(C,\mathbf{F}) \;\approx\; 
\mu_s\big(C_0,\mathbf{F}^{\rm bend}(y)\big)
+ \left.\frac{\partial \mu_s}{\partial C}\right|_{C_0,\mathbf{F}^{\rm bend}}
\big(C - C_0\big)
+ \left.\frac{\partial \mu_s}{\partial \mathbf{F}}\right|_{C_0,\mathbf{F}^{\rm bend}}
:\big(\mathbf{F} - \mathbf{F}^{\rm bend}(y)\big),
\label{eq:S6:mu_Taylor}
\end{equation}
where $:$ denotes the double contraction of tensors, and all partial derivatives are evaluated at $(C_0,\mathbf{F}^{\rm bend}(y))$.

Because mechanical equilibrium is restored much faster than solvent diffuses, 
the deformation at each $y$ is slaved to the local concentration through 
a quasi-static relation $\mathbf{F} = \mathbf{F}(C,y)$. 
Linearizing this mechanical response around the reference state $C_0$ gives
\begin{equation}
\mathbf{F}(y,t) - \mathbf{F}^{\rm bend}(y)
\;\approx\;
\left.\frac{\partial \mathbf{F}}{\partial C}\right|_{C_0}\big(C(y,t) - C_0\big),
\end{equation}
where the derivative is evaluated along this quasi-static equilibrium path.
Substituting this into \eqref{eq:S6:mu_Taylor} yields
\begin{equation}
\mu_s(C,\mathbf{F}) \;\approx\;
\mu_{\rm mech}(y) + K_\mu \big(C - C_0\big),
\label{eq:S6:mu_linear_C0}
\end{equation}
with
\begin{equation}
K_\mu :=
\left.\frac{\partial \mu_s}{\partial C}\right|_{C_0,\mathbf{F}^{\rm bend}}
+ \left.\frac{\partial \mu_s}{\partial \mathbf{F}}\right|_{C_0,\mathbf{F}^{\rm bend}}
:\left.\frac{\partial \mathbf{F}}{\partial C}\right|_{C_0}.
\label{eq:S6:chi_eff}
\end{equation}
Thus, the chemical potential is linearized with respect to concentration around the bent configuration at $C_0$, and $K_\mu$ acts as an effective osmotic modulus that incorporates both the direct and deformation-mediated sensitivities of $\mu_s$ to changes in $C$. For the moderate concentration perturbations relevant to our experiments, a first-order Taylor expansion (\eqref{eq:S6:mu_linear_C0}) is expected to be accurate, supported by the close agreement between linear and nonlinear poroelastic prediction for swelling reported in~\cite{bouklas2012}.

The solvent flux is taken to obey Darcy's law in $\mu_s$,
\begin{equation}
q = -M \partial_y \mu_s,
\label{eq:S6:Darcy}
\end{equation}
where $q(y,t)$ is the one-dimensional solvent flux and $M$ is the isotropic mobility. The mass balance reads
\begin{equation}
\partial_t C + \partial_y q = 0.
\label{eq:S6:mass_balance}
\end{equation}
Define the deviation of concentration from the free-swollen reference as
\begin{equation}
\delta C(y,t) := C(y,t) - C_0.
\label{eq:S6:DeltaC_def}
\end{equation}
Inserting \eqref{eq:S6:mu_linear_C0} into \eqref{eq:S6:Darcy} and \eqref{eq:S6:mass_balance} yields
\begin{equation}
q = -M\Big[\partial_y \mu_{\rm mech}(y) + K_\mu \partial_y \delta C\Big],
\label{eq:S6:Jc_M_C}
\end{equation}
and
\begin{equation}
\partial_t \delta C = M \partial_y\!\big[\partial_y \mu_{\rm mech}(y) + K_\mu \partial_y \delta C\big].
\label{eq:S6:mass_balance2}
\end{equation}
Defining the effective diffusivity
\footnote{We note that the diffusion coefficient $D$ introduced in \eqref{eq:S6:D_def} is obtained by linearizing the solvent chemical potential around the free-swollen reference state and combining the mobility $M$ with the osmotic modulus $K_\mu$. This definition is algebraically different from classical linear poroelastic formulations~\cite{hong2008theory, yoon2010, bouklas2012}, where $D$ is expressed in terms of permeability, drained elastic moduli, and poroelastic coupling coefficients. Nevertheless, both quantities describe the same physical mechanism---solvent diffusion that relaxes chemical-potential gradients around the reference state. Therefore, they are physically equivalent, and set the same characteristic poroelastic timescale $\tau \sim L^2/D$.}
\begin{equation}
D := M K_\mu,
\label{eq:S6:D_def}
\end{equation}
we obtain a diffusion equation for $\delta C$ from \eqref{eq:S6:mass_balance2}:
\begin{equation}
\partial_t \delta C(y,t)
= D \partial_{yy}\delta C(y,t)
+ f(y),
\label{eq:S6:DeltaC_forced}
\end{equation}
where $f(y) := M \partial_{yy}\mu_{\rm mech}(y)$ is the bending-induced contribution to the chemical-potential field. In this form, the equation makes explicit that bending imposes a fixed spatial pattern $\mu_{\rm mech}(y)$ whose curvature acts as a distributed source driving solvent redistribution toward a new non-uniform equilibrium profile.

\paragraph{Steady state and reduction to a homogeneous problem}

At long-time limit ($t\to\infty$), The steady state $\delta C_\infty(y)$  satisfies
\begin{equation}
0 = D \partial_{yy}\delta C_\infty(y) + f(y).
\label{eq:S6:DeltaC_steady}
\end{equation}
This steady solution corresponds to the eventual bent poroelastic equilibrium in which the total chemical potential $\mu_s(y,\infty)$ becomes uniform and equal to $\mu_\infty$, while stresses satisfy mechanical equilibrium.

To homogenize \eqref{eq:S6:DeltaC_forced}, we define the concentration perturbation relative to this bent equilibrium profile as:
\begin{equation}
c(y,t) := C(y,t) - C_\infty(y)
= \delta C(y,t) - \delta C_\infty(y),
\label{eq:S6:c_def}
\end{equation}
with $C_\infty(y) = C_0 + \delta C_\infty(y)$ at $t\to\infty$. Using the incompressible condition \eqref{eq:S6:C_J}, $c(y,t)$ can be measured as
\begin{equation}
c(y,t)
= \frac{1}{\Omega} \big[ J(y,t)-J_\infty(y) \big],
\label{eq:S6:c_J_relation}
\end{equation}
where $J_\infty(y)$ is the measured Jacobian in the final bent equilibrium state. Subtracting \eqref{eq:S6:DeltaC_steady} from \eqref{eq:S6:DeltaC_forced}, we find
\begin{equation}
\partial_t c(y,t)
= D \partial_{yy}c(y,t).
\label{eq:S6:diff_c}
\end{equation}
Thus, by shifting from $\delta C = C - C_0$ to $c(y,t) = C(y,t) - C_\infty(y)$, the bend-induced source term is absorbed into the definition of the equilibrium profile $C_\infty(y)$. In solving for $c$, we neglect the $y$-dependence of the osmotic modulus $K_\mu(y)$, so that the dynamics are governed by a homogeneous diffusion equation with a constant effective diffusivity $D$.

\paragraph{Initial condition for the homogeneous problem}
The initial condition for $c$ measures how far the instantaneously bent beam is from its final bent equilibrium:
\begin{equation}
c(y,0) = C_0 - C_\infty(y),
\label{eq:S6:IC_c}
\end{equation}
and in terms of measurable $J(y,t)$,
\begin{equation}
c_0(y)
= \frac{1}{\Omega} \big[ J(y,0)-J_\infty(y) \big]
\propto 1 - J_\infty(y),
\label{eq:S6:c0_J_relation}
\end{equation}
with $J(y,0)=J_0=1$ immediately after bending. 

\subsection{Solution with Neumann--Neumann boundary conditions}

In addition to the initial condition, we must specify boundary conditions (BCs) at the beam surfaces to solve~\eqref{eq:S6:diff_c}. Because both the tensile face ($y=H_0$) and the compressive face ($y=0$) are immersed a water--glycerol mixture, which is immiscible with with the silicone-oil solvent or the PDMS network, it is natural to set both faces as effectively sealed and treat them with \emph{no-flux} Neumann BCs,
\begin{equation}
q(0,t)=0,\qquad q(H_0,t)=0.
\label{eq:S6:Jc_t}
\end{equation}
% In the long-time equilibrium state $t\to\infty$, we have
% \begin{equation}
% q_{\infty}(0)=0,\qquad q_{\infty}(H_0)=0.
% \label{eq:S6:Jc_inf}
% \end{equation}
Using Darcy's law~\eqref{eq:S6:Darcy} together with the linearized
constitutive relation~\eqref{eq:S6:Jc_M_C}, and introducing the perturbation $c(y,t)$ defined in~\eqref{eq:S6:c_def}, the no-flux condition at each face reduces to homogeneous Neumann boundary conditions for $c$,
\begin{equation}
\partial_y c(0,t)=0,\qquad \partial_y c(H_0,t)=0.
\label{eq:S6:BC_Neumann_c}
\end{equation}
Together with governing equation~\eqref{eq:S6:diff_c} and initial condition~\eqref{eq:S6:IC_c}, these equations define a classical 1D linear diffusion problem on $0<y<H_0$ with Neumann–Neumann (NN) BCs.

\begin{figure}[H]
\centering
\includegraphics[width=0.9\linewidth,trim={0 0 0 10mm},clip]{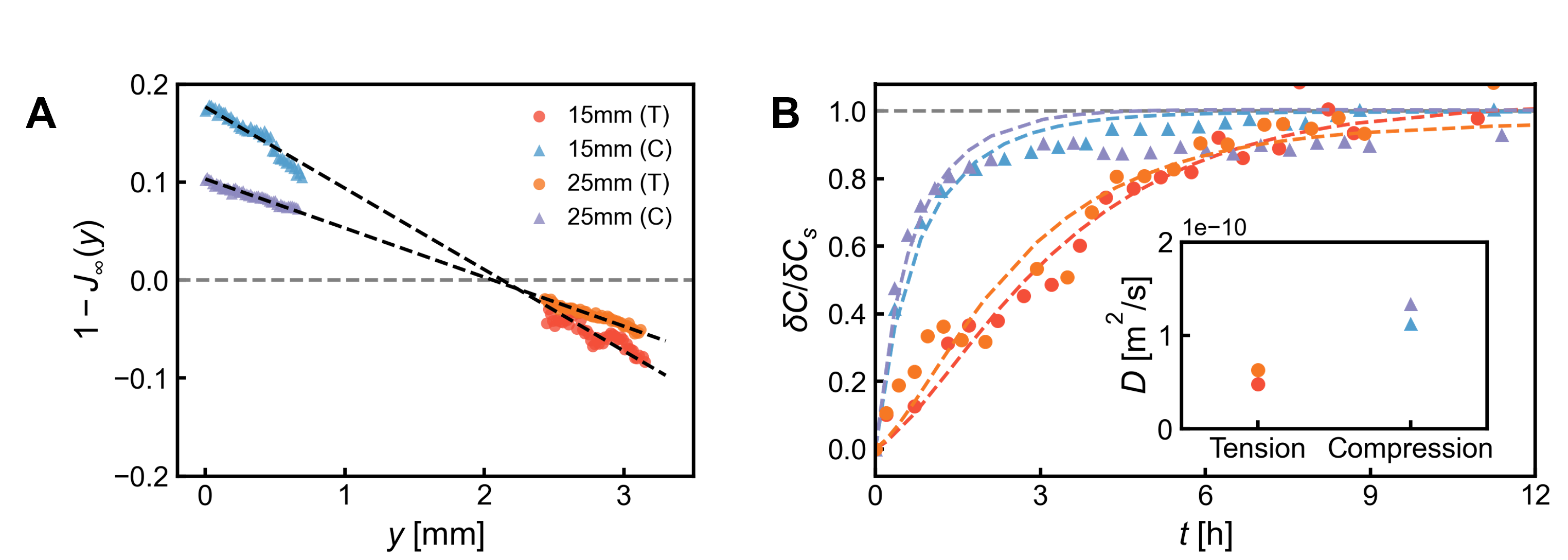}
\caption{(A) Profiles of the volumetric stretch $J_\infty(y)$ measured at midspan in the last frame for fully-swollen \SI{5}{\centi P} beams with spans $\SI{15}{mm}$ and $\SI{25}{mm}$ on the tensile (T, circles) and compressive (C, triangles) sides. Symbols are experimental data; black dashed lines are linear least-squares fits $J_\infty^{\rm fit}(y)$ used to construct the initial profile $c_0(y)$ in the diffusion model. The horizontal grey dashed line marks $J_\infty=1$, corresponding to no net volume change. The nonzero thickness-integral of $1-J_\infty(y)$ indicates a net solvent uptake or loss, ruling out strictly sealed Neumann--Neumann boundary conditions. 
(B) Normalized concentration change $\delta C/\delta C_s = 1-\theta$ measured at the center of the volume of interest as a function of time for the same beams and sides (markers). Colored dashed curves are Neumann--Robin predictions with best-fit $D$ and $\mathrm{Bi}$. All datasets are well described by the Neumann--Robin model and yield $\mathrm{Bi}\approx 10^9 \gg 1$, indicating a strongly drained compressive surface and motivating the Neumann--Dirichlet boundary conditions adopted in the main text.}
\label{figS:BC_Check}
\end{figure}

Integrating \eqref{eq:S6:diff_c} over the thickness and using \eqref{eq:S6:BC_Neumann_c} gives
\begin{equation}
\frac{{\rm d}}{{\rm d}t}\int_0^{H_0} c(y,t) {\rm d}y
= D\Big[\partial_y c(H_0,t) - \partial_y c(0,t)\Big] = 0,
\end{equation}
so the integration of $c$ in thickness is conserved in time. By construction, $c(y,t)\to 0$ as $t\to\infty$, hence this conserved quantity must vanish and the initial profile must satisfy
\begin{equation}
\int_0^{H_0} c_0(y) {\rm d}y = \int_0^{H_0} c(y,t\to\infty) {\rm d}y =0.
\label{eq:S6:NN_integral_constraint}
\end{equation}
Based on the assumption of incompressibility, we show that $c_0(y)$ is proportional to $1 - J_\infty(y)$ in~\eqref{eq:S6:c0_J_relation}, where $J_\infty(y)$ is the measured long-time Jacobian profile. Under strict NN BCs, \eqref{eq:S6:NN_integral_constraint} therefore requires
\begin{equation}
\int_0^{H_0} \big[1 - J_\infty(y)\big] {\rm d}y = 0.
\label{eq:S6:NN_qonstraint}
\end{equation}
However, while $1-J_\infty(y)$ is measured very linear in $y$, this integral, calculated with the linear least-squares fit $1-J^{\rm fit}_\infty(y)$, yields clearly non-zero, as shown in~Fig.~\ref{figS:BC_Check}A. \emph{This demonstrates that a perfectly sealed beam (NN BCs) is incompatible with the measured final swollen state}, and the beam must have exchanged solvent beyond this 1D diffusion~\cite{scherer2009bending} or through the boundaries.
% ; in other words, there must be net solvent transport, either laterally along the beam~\cite{scherer2009bending} (1D diffusion $\to$ 2D diffusion) or by exchange with the adjacent micro-environment, as shown in Fig.~\ref{figS:bending_BC}.

\subsection{Solution with Neumann--Robin boundary conditions}\label{secsss:beam_NR_BC}

\begin{figure}[H]
\centering
\includegraphics[width=0.5\linewidth]{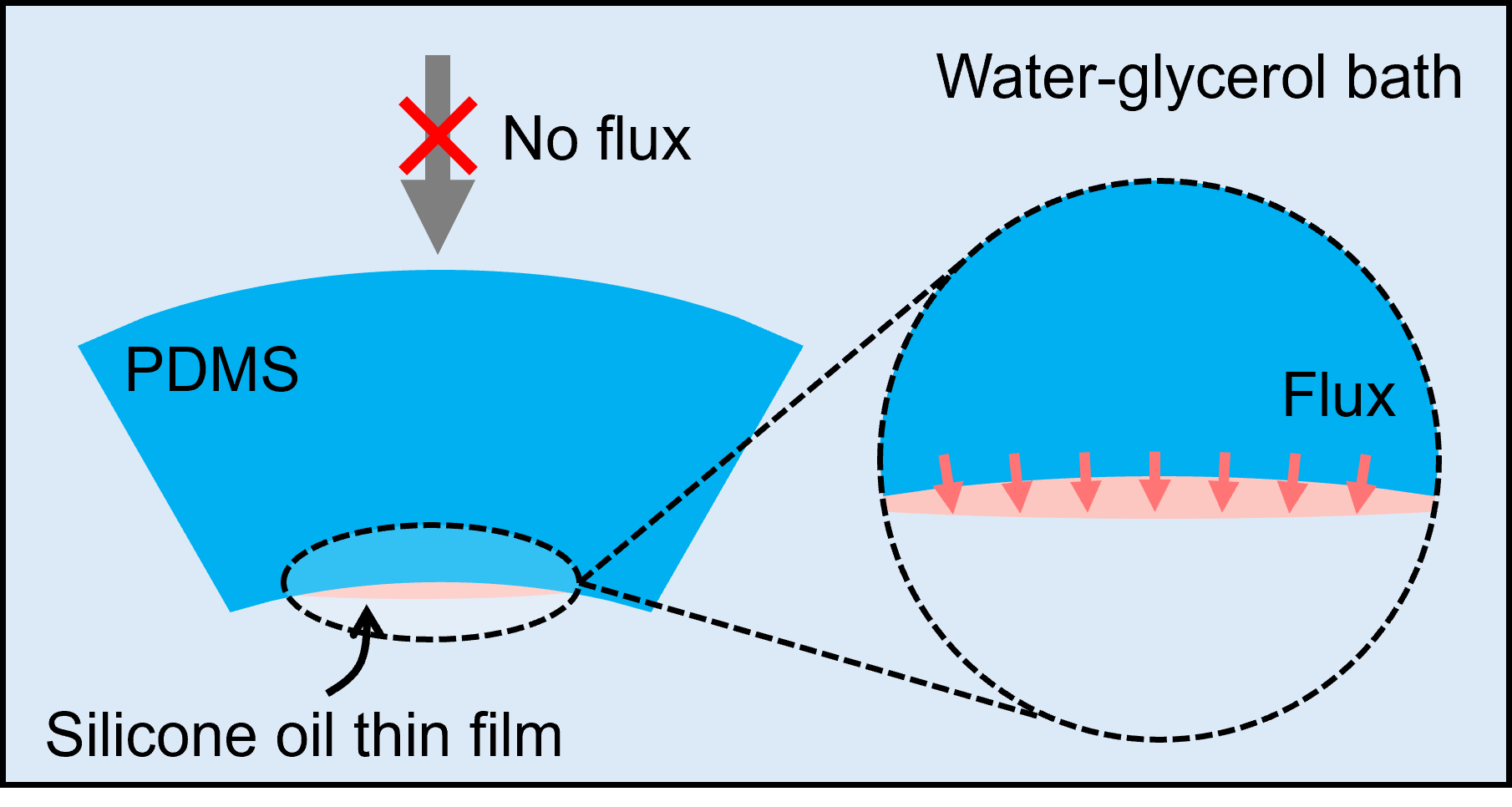}
\caption{Schematic of the interfacial boundary conditions for a bent PDMS beam immersed in a water--glycerol bath (mechanical forces and moments omitted for clarity). The water--glycerol is immiscible with either the silicone-oil solvent or the PDMS; therefore, upon bending, the tensile-side surface in direct contact with the bath cannot take up solvent and behaves as an effective no-flux boundary. In contrast, residual silicone oil due to incomplete removal leaves a thin interfacial film on the compressive face. This thin layer of silicone oil remains in diffusive contact with the internal solvent, enabling solvent flux across the interface and thereby rendering the compressive surface effectively drained.
}
\label{figS:bending_BC}
\end{figure}

The inconsistency from the NN solution is rationalized by assuming a thin film of residual silicone oil on the compressive surface, as shown in Fig.~\ref{figS:bending_BC}. Owing to the strong compatibility of silicone oils with PDMS and their poor removability, it is difficult to completely clean the silicone oil from the beam surfaces. As a result, a thin interfacial film of residual silicone oil persists on the compressive face and can diffusively exchange with the internal silicone-oil solvent. Therefore, we relax the BC at the compressive surface, and allow \emph{finite} solvent exchange by imposing a \emph{limited-flux} Robin BC at $y=0$,
\begin{equation}
q(0,t) = \kappa \big[\mu_s(0,t)-\mu_{\rm ext}\big],
\label{eq:S6:BC_Robin_mu}
\end{equation}
with $\kappa$ an interfacial mass-transfer coefficient and $\mu_{\rm ext}$ the chemical potential of the external silicone oil. On the tensile surface, \emph{no-flux} BC is retained,
\begin{equation}
q(H_0,t)=0 \quad\Longleftrightarrow\quad \partial_y c(H_0,t)=0.
\label{eq:S6:BC_NR_H0}
\end{equation}
Using the same technique in obtaining ~\eqref{eq:S6:BC_Neumann_c}, \eqref{eq:S6:BC_Robin_mu} becomes
\begin{equation}
- D \partial_y c(0,t) = \kappa c(0,t),
\label{eq:S6:BC_Robin_c}
\end{equation}
and the dimensionless mass-transfer Biot number~\cite{Mass_transfer}, is defined as
\begin{equation}
\mathrm{Bi} := \frac{\kappa H_0}{D},
\label{eq:S6:Da_def}
\end{equation}
which provides a one-parameter family of BCs interpolating between a Neumann--Neumann limit ($\mathrm{Bi}\to 0$) and a Neumann--Dirichlet (fully-drained) limit ($\mathrm{Bi}\to\infty$).

With the governing equation~\eqref{eq:S6:diff_c}, initial condition~\eqref{eq:S6:IC_c}, no-flux BC at the tensile face~\eqref{eq:S6:BC_NR_H0}, and limited-flux BC at the compressive face~\eqref{eq:S6:BC_Robin_c}, the system constitutes a standard Sturm--Liouville eigenvalue problem. 1D diffusion problems with various combinations of Dirichlet, Neumann, and Robin boundary conditions have been solved by separation of variables in textbooks~\cite{strauss2007partial, hahn2012heat}. Solving the present problem gives the general series 
solution:
\begin{equation}
c(y,t) = \sum_{n=0}^{\infty} A_n(\mathrm{Bi})\,\Phi_n(y;\mathrm{Bi})\,\exp\!\big(-D \lambda_n^2 t\big).
\label{eq:S6:c_series_NR}
\end{equation}
The spatial eigenfunctions, $\Phi_n(y)$, are determined by imposing the boundary conditions on the spatial component, leading to the transcendental equation for the admissible eigenvalues $\zeta_n = \lambda_n H_0$:
\begin{equation}
\zeta_n \tan\zeta_n + \mathrm{Bi} = 0,
\label{eq:S6:Robin_eig}
\end{equation}
with corresponding eigenfunctions
\begin{equation}
\Phi_n(y) = \cos(\lambda_n y) + \tan\zeta_n\,\sin(\lambda_n y).
\end{equation}
Due to the properties of the regular Sturm--Liouville problem, the eigenfunctions are orthogonal. Consequently, the coefficients $A_n(\mathrm{Bi})$ are determined by projecting the initial concentration profile $c_0(y)$ onto the basis set $\{\Phi_n\}$:
\begin{equation}
A_n(\mathrm{Bi})
= \frac{\displaystyle \int_0^{H_0} c_0(y)\,\Phi_n(y;\mathrm{Bi})\,\mathrm{d}y}
       {\displaystyle \int_0^{H_0} \big[\Phi_n(y;\mathrm{Bi})\big]^2\,\mathrm{d}y}.
\label{eq:S6:Bn_projection}
\end{equation}
Thus, for each trial value of $\mathrm{Bi}$, the spatial weights $A_n(\mathrm{Bi})$ are fully determined by the measured $c_0(y)$.

Experimentally, we do not measure the concentration field $C(y,t)$ or $c(y,t)$ directly, but instead the local Jacobian $J(y,t)$ from 3D particle tracking. Using incompressibility, $c(y,t)$ and $J(y,t)$ are related by the affine mapping \eqref{eq:S6:c_J_relation}, but the solvent molar mass $\Omega$ is still involved. To eliminate any uncertain scale and focus on the shape and timescale of the relaxation, we work with a normalized, dimensionless signal at a given measurement location $y=\bar{Y}$,
\begin{equation}
\theta(\bar{Y},t)
:= \frac{J(\bar{Y},t)-J_\infty(\bar{Y})}{J(\bar{Y},0)-J_\infty(\bar{Y})}= \frac{c(\bar{Y},t)}{c(\bar{Y},0)}.
\label{eq:S6:theta_def}
\end{equation}
Note that in the main text, the normalized concentration change $\delta C(\bar{Y},t)/\delta C_s(\bar{Y})$ is plotted (e.g., Fig.~\ref{figS:bending_BC}B), corresponding to $1-\theta(\bar{Y},t)$. Substituting the modal solution \eqref{eq:S6:c_series_NR} into \eqref{eq:S6:theta_def} yields the theoretical prediction used to fit the experimental data:
\begin{equation}
\theta(\bar{Y},t)
= \frac{\displaystyle\sum_{n=0}^{\infty} A_n\,\Phi_n(\bar{Y})\,e^{-D\lambda_n^2 t}}
{\displaystyle\sum_{n=0}^{\infty} A_n\,\Phi_n(\bar{Y})},
\label{eq:S6:theta_series_NR}
\end{equation}
which is the theoretical prediction that we use to fit the measured $\theta(t)$.

In practice, we estimate $D$ and $\mathrm{Bi}$ from the measured $(t,\theta(\bar{Y}, t))$ trace at a given location $\bar{Y}$, treating $D$ and $\mathrm{Bi}$ as fitting parameters. The procedure is as follows. We first choose a trial value of $\mathrm{Bi}$ and solve the eigenvalue equation~\eqref{eq:S6:Robin_eig} numerically (using a Brent root-finding algorithm~\cite{2020SciPy}) to obtain a discrete set of eigenvalues $\{\lambda_n(\mathrm{Bi})\}$ and the associated eigenfunctions $\{\Phi_n(y;\mathrm{Bi})\}$. The corresponding modal amplitudes $A_n(\mathrm{Bi})$ are then determined from the measured initial profile $c_0(y)$ via the projection formula~\eqref{eq:S6:Bn_projection}. Once $\{\lambda_n(\mathrm{Bi}),\Phi_n(y;\mathrm{Bi}),A_n(\mathrm{Bi})\}$ are known for a given $\mathrm{Bi}$, the diffusivity $D$ enters only through the decay rates $D\lambda_n^2$ in the time exponentials. Using \eqref{eq:S6:theta_series_NR}, for each pair $(D,\mathrm{Bi})$ we compute the theoretical prediction $\theta_{\rm th}(\bar{Y},t;D,\mathrm{Bi})$ at the experimental times and at the measurement location $y=\bar{Y}$. The best-fit values of $D$ and $\mathrm{Bi}$ are obtained by minimizing the least-squares misfit between $\theta_{\rm th}(\bar{Y},t;D,\mathrm{Bi})$ and the measured $\theta(\bar{Y},t)$. 

As shown in Fig.~\ref{figS:bending_BC}B, the NR solution~\eqref{eq:S6:theta_series_NR} predicts well with the experimentally measurement, and the best-fit parameters consistently satisfy $\mathrm{Bi}\gg 1$ for all datasets. \textbf{Such large Bi numbers correspond to a highly diffusive compressive surface, which equilibrates rapidly with the surroundings, and therefore, behaves as an effectively drained boundary.} Motivated by this observation, in the main text, we treat the compressive surface as fully drained (Dirichlet) and retain a no-flux (Neumann) condition at the tensile surface, so that $D$ becomes the only fitting parameter. The solution to this resulting Neumann--Dirichlet problem is derived briefly in the following.

\subsection{Solution with Neumann--Dirichlet boundary conditions}
In the drained limit $\mathrm{Bi}\gg 1$, the Robin BC at the compressive face reduces to an effective Dirichlet condition, $c(0,t)=0$. Together with the no-flux condition at $y=H_0$, this defines a standard Neumann--Dirichlet (ND) eigenvalue problem. The resulting eigenfunctions and eigenvalues are solved analytically~\cite{hahn2012heat}:
\begin{equation}\Phi_n(y) = \sin(\lambda_n y), \qquad \lambda_n = \frac{(n+\tfrac{1}{2})\pi}{H_0}, \quad n=0,1,2,\dots\label{eq:S6:eigenfun_ND}\end{equation}The series solution for $c(y,t)$ and the projection coefficients $A_n$ follow the same forms derived in the Neumann--Robin analysis (\eqref{eq:S6:c_series_NR} and \eqref{eq:S6:Bn_projection}, respectively), using the basis set \eqref{eq:S6:eigenfun_ND}. Consequently, the theoretical prediction for the normalized measurement becomes:   
\begin{equation}
\theta(\bar{Y},t)
= \frac{\displaystyle\sum_{n=0}^{\infty} A_n\,\Phi_n(\bar{Y})\,e^{-D(\lambda_n)^2 t}}%
{\displaystyle\sum_{n=0}^{\infty} A_n\,\Phi_n(\bar{Y})}.
\label{eq:S6:theta_series_ND}
\end{equation}

In this ND formulation, the shape of the initial profile $c_0(y)$ and the modal weights $A_n$ are fixed once and for all from the measured $J_\infty(y)$ and the beam geometry, while the diffusivity $D$ is the only adjustable parameter. $D$ is obtained by minimizing the least-squares misfit between the prediction $\theta_{\rm th}(\bar{Y},t_i;D)$ and experimental measurement $\theta_{\rm exp}(\bar{Y},t_i)$ at evaluation point $y=\bar{Y}$ and experimental times $t_i$. The best-fit $D$ and corresponding predictions are provided in the main text.

% \newpage
\section{Swelling equilibrium of a crosslinked network in pure solvent}\label{secS:FR_equil}

We consider a crosslinked polymer network exchanging solvent with a large bath of pure solvent at fixed temperature and pressure. The mixture is incompressible and swells isotropically, with
\[
J=\frac{V}{V_0},\qquad \phi=\frac{1}{J},\qquad \lambda=J^{1/3}=\phi^{-1/3}.
\]
Here $V_0$ is the dry polymer volume, $V$ is the current swollen volume of the gel, so that $J$ is the volumetric swelling ratio, $\phi$ is the polymer volume fraction, and $\lambda$ is the isotropic linear swelling ratio.

Following the standard Flory--Huggins lattice treatment of polymer mixtures~\cite{rubinstein2003}, we model the gel on a lattice of equal-volume sites, with $L$ denoting the total moles of lattice sites, $L = N_s n_s+N_p n_p$, where $n_s$ and $n_p$ are the moles of solvent molecules and network strands; $N_s$ and $N_p$ are the numbers of segments per solvent molecule and per strand. The corresponding site (volume) fractions are
\[
\phi_s=\frac{N_s n_s}{L}=1-\phi,\qquad
\phi_p=\phi=\frac{N_p n_p}{L},\qquad
\phi_s+\phi_p=1.
\]
The solvent molar volume is $V_s = N_s v_{\mathrm{seg}} = M_s/\rho_s$, with $v_{\mathrm{seg}}$ the segment molar volume, $M_s$ the solvent molar mass, and $\rho_s$ the solvent density.
% The network elastic density per dry volume is parameterized by $\nu_e=G_{\mathrm{dry}}/(RT)$. The network limit is taken as $N_p\to\infty$ at fixed $\phi$, so that polymer translational entropy can be neglected.

\medskip
\paragraph{Mixing free energy and solvent chemical potential.}
With Flory--Huggins interaction parameter $\chi$, the mixing free-energy density per mole of lattice sites~\cite{rubinstein2003} is %eq4.23
\begin{equation}
f_{\mathrm{mix}} = RT\left[\frac{\phi_s}{N_s}\ln\phi_s \;+\; \frac{\phi_p}{N_p}\ln\phi_p \;+\; \chi \phi_s\phi_p\right].
\label{eq:S_FR_eq:f_mix}
\end{equation}
Using $\phi_s/N_s=n_s/L$ and $\phi_p/N_p=n_p/L$, the total mixing free energy is
\begin{equation}
F_{\mathrm{mix}}=RT\left[n_s\ln\phi_s \;+\; n_p\ln\phi_p \;+\; \chi \frac{N_sN_p n_sn_p}{L}\right].
\label{eq:S_FR_eq:F_mix}
\end{equation}
The solvent mixing chemical potential relative to pure solvent, $\mu_s^{\mathrm{mix}}-\mu_s^{\mathrm{pure}}$, is obtained by differentiating $F_{\mathrm{mix}}$ with respect to the number of solvent molecules $n_s$~\cite{flory1953}. In the network limit of a crosslinked polymer, $N_p \to \infty$ at fixed composition~\cite{rubinstein2003}%discussion following eq4.12 N is large,....
, the contribution from the polymer translational entropy vanishes. Carrying out this differentiation yields
\begin{equation}
% Flory, Principles of Polymer Chemisty 1953, Ch. XII, Eq26
\mu_s^{\mathrm{mix}}-\mu_s^{\mathrm{pure}}
=
RT\left[
\ln(1-\phi)\;+\;\phi\;+\;\chi N_s \phi^{2}
\right].
\label{eq:S_FR_eq:mu_mix}
\end{equation}
In our notation, $\chi$ is defined per solvent segment, so that the effective Flory--Huggins interaction parameter for a solvent molecule containing $N_s$ segments is $\chi N_s$, which tunes the interaction between silicone oils of different chain length and the network. For a small-molecule solvent with $N_s=1$, $\chi N_s$ reduces to the standard $\chi$ used in the Flory--Rehner equation.

\medskip
\paragraph{Elastic contribution.}
Let $G_{\mathrm{dry}}$ be the dry shear modulus and $W_{\mathrm{el}}$ the elastic strain-energy density per dry volume. For an incompressible mixture, the elastic contribution to the solvent chemical potential is $\mu_s^{\mathrm{el}}=V_s (\partial W_{\mathrm{el}}/\partial J)$.
We consider two network models: the classic Gaussian model~\cite{flory1953} and the Gent model with chain-locking parameter $J_m$~\cite{gent1996}. Here, we explicitly assume the swelling is \emph{isotropic}, characterized by uniform stretch ratios $\lambda = J^{1/3}$. This kinematics is distinct from \emph{uniaxial} extension. Using the isotropic relation, the first invariant becomes $I_1=3J^{2/3}=3\phi^{-2/3}$, and the energy densities are
\begin{equation}
\begin{aligned}
% flory1953, Page 578, XIII-3a, Eq35
W_{\mathrm{el}}^{\mathrm{G}} &= \frac{G_{\mathrm{dry}}}{2} \left[ (I_1-3) - \ln J \right],\\
W_{\mathrm{el}}^{\mathrm{Gent}} &= \frac{G_{\mathrm{dry}}}{2} \left[ -J_m\ln\!\Big(1-\frac{I_1-3}{J_m}\Big) - \ln J \right].
\end{aligned}
\label{eq:S_FR_eq:Wel}
\end{equation}
It is noted that the logarithmic term originates from the entropy of deformation and its prefactor can be chosen other than unity~\cite{ okumura2018gentAB, zhou2020hydrolysis, Suo2014gent}. For example, from a continuum mechanics perspective, the prefactor of the $\ln J$ term is often set to $2$ to ensure a zero-stress condition in the dry reference state~\cite{hong2010, Suo2014gent}. Here, however, we adhere to the classic Flory--Rehner (FR) network derivation to maintain consistency with standard polymer physics literature~\cite{flory1953, treloar1975physics, quesada2011gel}. Differentiating these with respect to swelling $J$ yields the corresponding chemical potentials:
\begin{equation}
\begin{aligned}
\mu_s^{\mathrm{el,G}} &= G_{\mathrm{dry}} V_s\big(\phi^{1/3}-\tfrac{1}{2}\phi\big),\\
\mu_s^{\mathrm{el,Gent}} &= G_{\mathrm{dry}} V_s\left[\frac{\phi^{1/3}}{ 1-\frac{3}{J_m}\big(\phi^{-2/3}-1\big) }-\tfrac{1}{2}\phi\right].
\end{aligned}
\label{eq:S_FR_eq:mu_el}
\end{equation}
For the Gent model, swelling is physically limited to $\phi \ge (1+J_m/3)^{-3/2}$.

\medskip
\paragraph{Swelling equilibrium.}
Equilibrium with a pure solvent bath requires $\mu_s^{\mathrm{mix}}-\mu_s^{\mathrm{pure}}+\mu_s^{\mathrm{el}}=0$. Substituting the mixing term \eqref{eq:S_FR_eq:mu_mix} and the elastic terms \eqref{eq:S_FR_eq:mu_el}, and defining the effective chain molar density $\nu_e = G_{\mathrm{dry}}/RT$, we obtain the dimensionless equilibrium conditions.
For a Gaussian network, this gives the classic FR equation:
\begin{equation}
\ln(1-\phi)+\phi+\chi N_s \phi^2
+\nu_e V_s\big(\phi^{1/3}-\tfrac{1}{2}\phi\big)=0.
\label{eq:S_FR_eq:FR_balance}
\end{equation}
For a Gent network, the finite extensibility modifies the elastic term:
\begin{equation}
\ln(1-\phi)+\phi+\chi N_s \phi^2
+\nu_e V_s\!\left[
\frac{\phi^{1/3}}{ 1-\frac{3}{J_m}\big(\phi^{-2/3}-1\big) }-\tfrac{1}{2}\phi
\right]=0.
\label{eq:S_FR_eq:FRGent_balance}
\end{equation}

In \eqref{eq:S_FR_eq:FR_balance} and \eqref{eq:S_FR_eq:FRGent_balance}, we set $\chi=1\times10^{-3}$ and take $\nu_e$ from the small-strain shear modulus. Since the mixture is nearly athermal, the interaction parameter plays a negligible role, and therefore, the equilibrium $\phi$ is predominantly governed by the silicone oil molar volume $V_s$ (determined by the oil chain length). The FR equilibrium swelling ratio $Q=1/\phi$ is obtained by solving \eqref{eq:S_FR_eq:FR_balance} for $\phi$ with a bracketing bisection scheme; for a given solvent ($M_s$, $V_s$, $N_s$), the equation is monotone and admits a unique root. For the FR–Gent model, we examined two calibrations of $J_m$: (i) fitting the measured $\phi$ gives $J_m\approx2.4$ with an excellent match; (ii) an independent Gent fit to the uniaxial tensile data (Fig.~\ref{figS:tensile}B) yields $J_m\approx12$. Using this tensile-derived $J_m$, we then solved \eqref{eq:S_FR_eq:FRGent_balance} by bisection to predict $\phi$ as a function of $M_s$.

\section{Visualization of solvent diffusion during free-swelling}\label{secS:FS_Visual}

\begin{figure}[!htbp]
\centering
\includegraphics[width=0.9\linewidth,trim={0 0 0 2mm},clip]{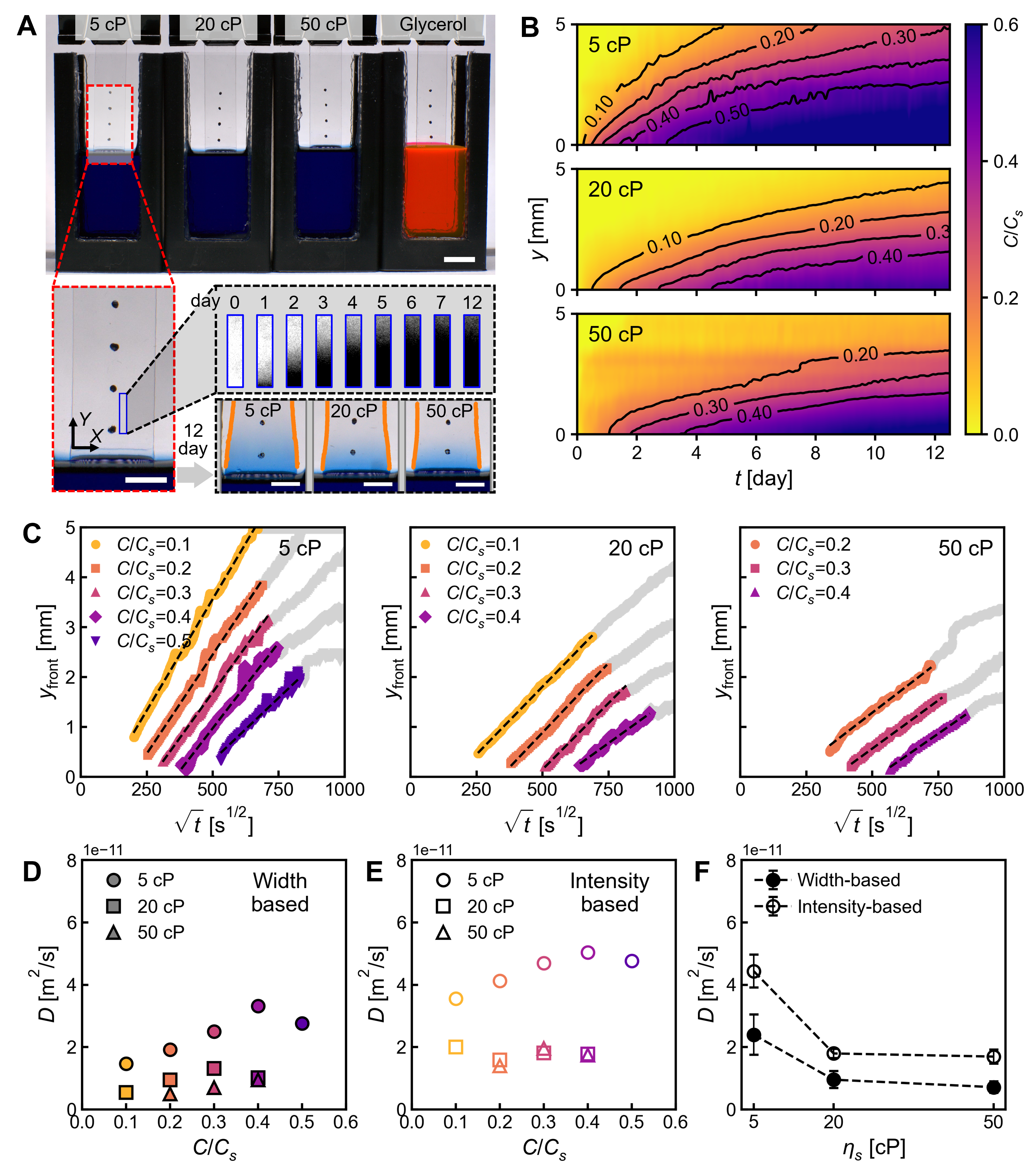}
\caption{Visualizing the diffusion of silicone oils in TT PDMS strips. (A) Experimental configuration. TT PDMS strips (\SI{60}{\mm} long, \SI{10}{\mm} wide, and \SI{1}{\mm} thick) are clamped vertically and their lower ends are immersed in dyed silicone oils of viscosity $\eta_s=\numlist{5;20;50}\,\unit{\centi P}$; dyed glycerol serves as a non-swelling reference. Black fiducial dots mark fixed positions along the strip. Lower left: zoom-in view at $t=0^+$ immediately after immersion in $\eta_s=\SI{5}{\centi P}$ silicone oil. 
Middle right ($\eta_s=\SI{5}{\centi P}$ example): a $\SI{1}{\mm}\times\SI{5}{\mm}$ rectangular region-of-interest (ROI) near the immersed edge is used to visualize the spatiotemporal evolution of the advance of the dyed-solvent front (contrast adjusted for better visualization). 
Lower right: zoom-in views after 12 days swelling for the three $\eta_s$. The orange curves denotes the detected edge which are used to measure the linear swelling ratio $\lambda(y,t)$. Scale bars: \SI{10}{\mm} (top) and \SI{5}{\mm} (insets).
(B) Across the ROI for three viscosities, $\lambda(y,t)$ is measured and converted to the solvent concentration $C(y,t)$ and normalized by the equilibrium solvent concentration $C_s$; iso-contours of $C/C_s$ are overlaid.
(C) Normalized concentration profiles are extracted from (B), and the front position $y_\mathrm{front}$, defined by a given iso-concentration level, increases linearly with $\sqrt{t}$, consistent with Fickian diffusion-type imbibition. A linear fit of $y_\mathrm{front}$ versus $\sqrt{t}$ yields the effective diffusion coefficient according to \eqref{eq:S4:D_from_slope}. 
(D) Diffusion coefficients obtained from the linear fit of $y_\mathrm{front}$ versus $\sqrt{t}$ in (C).
% Despite a small variance at \SI{5}{\centi P}, the values of $D$ estimated at different iso-concentration levels are consistent for all three viscosities. 
(E) The normalized concentration $C/C_s$ is also estimated from the image intensity in Fig.~\ref{figS:visual}. Using the same fitting protocol, the corresponding diffusion coefficients are obtained and plotted as a function of $C/C_s$. 
(F) Summary of diffusivities $D$ obtained by both methods for the three viscosities.}
\label{figS:free_swelling_visual}
\end{figure}

\begin{figure}[!htbp]
\centering
\includegraphics[width=0.9\linewidth]{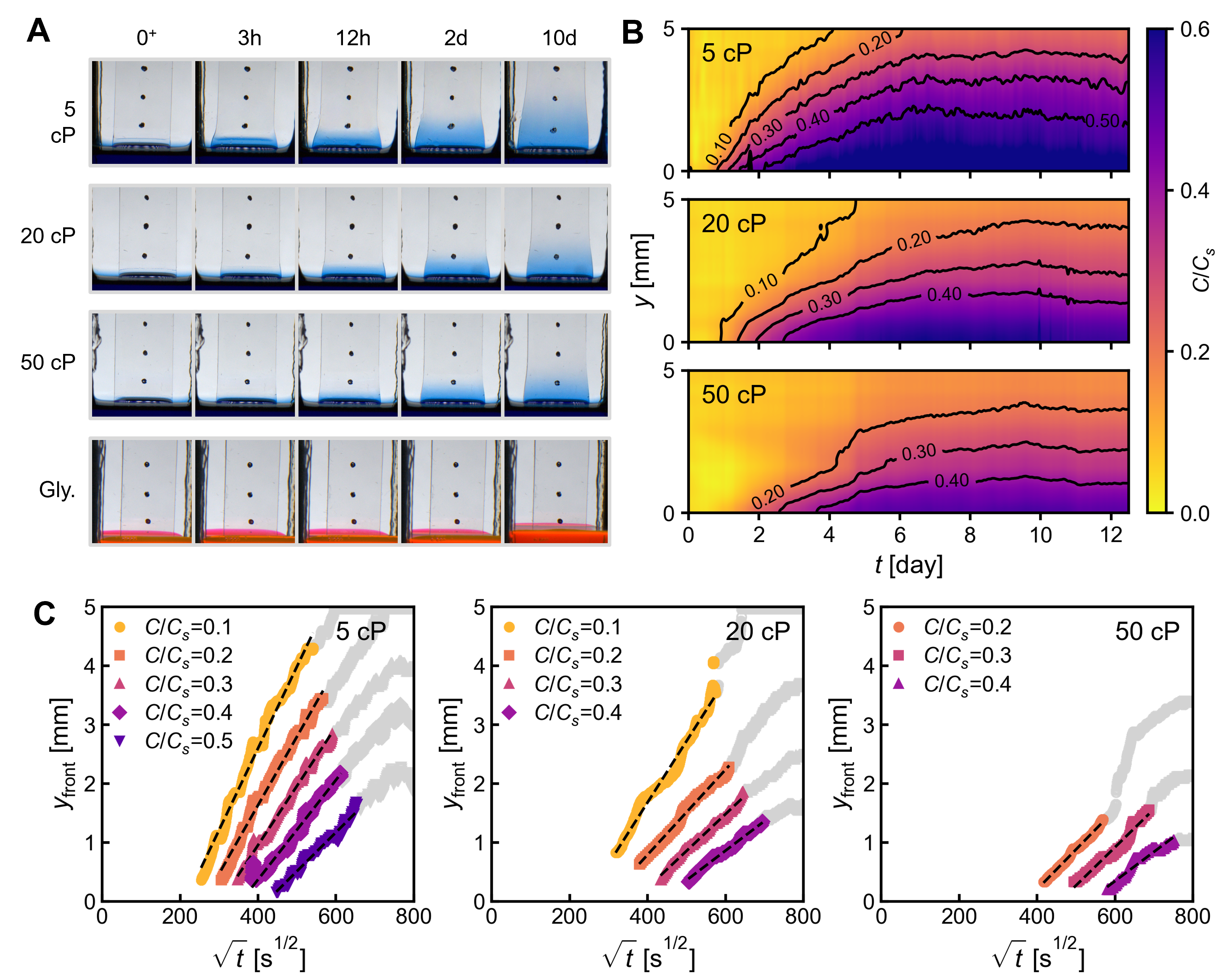}
\caption{Visualizing free-swelling of TT PDMS strips and extracting front kinetics. 
(A) Time-lapse photographs of strips with their lower ends immersed in dyed silicone oils of viscosity $\eta_s=\numlist{5;20;50}\,\unit{\centi P}$; dyed glycerol is shown as a non-swelling control. Black dots are fixed fiducials.
(B) For each $\eta_s$, same ROI in Fig.~\ref{figS:free_swelling_visual}A at the immersed edge is converted to a spatiotemporal map of normalized solvent concentration $C/C_s$; iso-contours of $C/C_s$ are overlaid. 
(C) Front position $y_\mathrm{front}$ defined by the same iso-concentration levels plotted against $\sqrt{t}$ for $\eta_s=\numlist{5;20;50}\,\unit{\centi P}$. The near-linear dependence confirms Fickian diffusion-type imbibition; slopes of linear fits give effective diffusivities and are reported in Fig.~\ref{figS:free_swelling_visual}E.
}
\label{figS:visual}
\end{figure}

To directly visualize and measure diffusion during free swelling, TT strip samples were clamped vertically and partially immersed in silicone oils dyed with $0.0005 \text{wt}\%$ trace dye, as shown in Fig.~\ref{figS:free_swelling_visual}A. The propagation of the dyed-oil front, representing the diffusion process, was visualized as a spatial intensity gradient in the transmission images and recorded as a time series (Fig.~\ref{figS:visual}A). To avoid meniscus artifacts and approximate one-dimensional transport, we selected a region of interest ($5 \unit{mm}\leq y \leq 10 \unit{mm}$) above the oil-sample interface, and analyzed the diffusion by two independent approaches (see Sec.~\ref{secS:FS_strip}): (i) edge segmentation to obtain the local strip width $w(y,t)$ and hence the linear swelling $\lambda(y,t)={w(y,t)}/{w_0(y)}$ and the solvent concentration; and (ii) conversion of transmitted-light intensity to dye (solvent) concentration via a Beer–Lambert calibration. Quantifying their time evolution produces the spatial-temporal normalized concentration field $C/C_s$ for oil viscosities $\eta_s=\numlist{5;20;50}\,\unit{\centi P}$, as shown in Fig.~\ref{figS:free_swelling_visual}B for the width-based measurement, and in Fig.~\ref{figS:visual}B for the intensity-based measurement.

From the normalized concentration fields, we extracted iso-contours of $C/C_s$ and plotted their position against $\sqrt{t}$ in Fig.~\ref{figS:free_swelling_visual}C. The clear linear relationship between the diffusion front position and $\sqrt{t}$ confirms that the swelling process follows Fickian diffusion dynamics; linear fits to the early-time regime of these plots yield effective diffusion coefficients $D$ for each iso-concentration level and each oil (Fig.~\ref{figS:free_swelling_visual}D). A parallel analysis was conducted using the intensity-based measurements (Fig.~\ref{figS:visual}B-C), and the resulting values of $D$ are provided in Fig.~\ref{figS:free_swelling_visual}E. 

Both methods yield effective diffusion coefficients $D$ that fall within the same order of magnitude and show identical trends (Fig.~\ref{figS:free_swelling_visual}D-E): $D$ is larger for lower-viscosity oils, and at a given viscosity, subtle variation exists across concentration levels, particularly for $\eta_s=5 \unit{cP}$. This variation is consistent with free-volume theory, where the ingress of solvent increases the network’s fractional free volume, lowering segmental friction and thereby increasing diffusivity~\cite{vrentas_duda_FVT, FVT, concentration_dependent_D}. Notably, the intensity-based method produces systematically higher $D$ values, which could result from potential nonlinearity in the dye concentration-intensity relationship or the dye’s non-covalent association with the oil. Despite this minor offset, $D$ values from both methods across all concentrations, when summarized versus $\eta_s$ (Fig.~\ref{figS:free_swelling_visual}F), are consistent with independent mass-uptake measurements in Fig.~\ref{fig:bulk_swelling}(E), confirming the robustness of our results.

%%% Add this line AFTER all your figures and tables
\FloatBarrier

% \movie{Propagation of dyed silicone-oil fronts into TT-PDMS strips}\label{movieS1}

\end{document}